\documentstyle[12pt,aaspp4]{article}

\newcommand\about     {\hbox{$\sim$}}

\makeatother
\def\eq#1{\begin{equation} #1 \end{equation}}
\def\eqarray#1{\begin{eqnarray} #1 \end{eqnarray}}

\def\non    {\nonumber \\}

\def\case#1/#2{\hbox{$\frac{#1}{#2}$}}
\def\about  {\hbox{$\sim$}}

\def\g                {\hbox{$g^*$}}
\def\r                {\hbox{$r^*$}}
\def\i                {\hbox{$i^*$}}
\def\z                {\hbox{$z^*$}}
\def\ug               {\hbox{$u^*-g^*$}}
\def\gr               {\hbox{$g^*-r^*$}}
\def\ri               {\hbox{$r^*-i^*$}}
\def\iz               {\hbox{$i^*-z^*$}}

\def\a                {\hbox{$a^*$}}
\def\dd               {\hbox{deg/day}}
\def\mic              {\hbox{$\mu{\rm m}$}}

\def\comment#1        {\tt #1}

\begin{document}

\rightline{\it Submitted to AJ.}

\title{                 Solar System Objects
               Observed in the Sloan Digital Sky Survey
                       Commissioning Data$^1$}

\author{
\v{Z}eljko Ivezi\'{c}\altaffilmark{\ref{Princeton}},
Serge Tabachnik\altaffilmark{\ref{Princeton}},
Roman Rafikov\altaffilmark{\ref{Princeton}},
Robert H. Lupton\altaffilmark{\ref{Princeton}},
Tom Quinn\altaffilmark{\ref{Washington}},
Mark Hammergren\altaffilmark{\ref{LLNL}},
Laurent Eyer\altaffilmark{\ref{Princeton}},
Jennifer Chu\altaffilmark{\ref{Princeton},\ref{Ramapo}},
John C. Armstrong\altaffilmark{\ref{Washington}},
Xiaohui Fan\altaffilmark{\ref{Princeton}},
Kristian Finlator\altaffilmark{\ref{Princeton}},
Tom R. Geballe\altaffilmark{\ref{UKIRT}},
James E. Gunn\altaffilmark{\ref{Princeton}},
Gregory S. Hennessy\altaffilmark{\ref{USNO1}},
Gillian R. Knapp\altaffilmark{\ref{Princeton}},
Sandy K. Leggett\altaffilmark{\ref{UKIRT}},
Jeffrey A. Munn\altaffilmark{\ref{USNO2}},
Jeffrey R. Pier\altaffilmark{\ref{USNO2}},
Constance M. Rockosi\altaffilmark{\ref{Chicago}},
Donald P. Schneider\altaffilmark{\ref{PennState}},
Michael A. Strauss\altaffilmark{\ref{Princeton}},
Brian Yanny\altaffilmark{\ref{FNAL}},
Jonathan Brinkmann\altaffilmark{\ref{APO}},
Istv\'an Csabai\altaffilmark{\ref{JHU},\ref{Eotvos}},
Robert B. Hindsley\altaffilmark{\ref{RSD}},
Stephen Kent\altaffilmark{\ref{FNAL}},
Bruce Margon\altaffilmark{\ref{Washington}},
Timothy A. McKay\altaffilmark{\ref{Michigan}},
J. Allyn Smith\altaffilmark{\ref{Wyoming}},
Patrick Waddel\altaffilmark{\ref{SOFIA}},
Donald G. York\altaffilmark{\ref{Chicago}}
(for the SDSS Collaboration)
}

\altaffiltext{1}{Based on observations obtained with the
Sloan Digital Sky Survey.}
\newcounter{address}
\setcounter{address}{2}
\altaffiltext{\theaddress}{Princeton University Observatory, Princeton, NJ 08544
\label{Princeton}}
\addtocounter{address}{1}
\altaffiltext{\theaddress}{University of Washington, Dept. of Astronomy,
Box 351580, Seattle, WA 98195
\label{Washington}}
\addtocounter{address}{1}
\altaffiltext{\theaddress}{Institute of Geophysics and Planetary Physics,
Lawrence Livermore National Laboratory, 7000 East Avenue, L-413, Livermore, CA 94550
\label{LLNL}}
\addtocounter{address}{1}
\altaffiltext{\theaddress}{Ramapo High School, 400 Viola Road, Spring Valley, NY 10977
\label{Ramapo}}
\addtocounter{address}{1}
\altaffiltext{\theaddress}{United Kingdom Infrared Telescope, Joint Astronomy Centre,
660 North A'ohoku Place, University Park, Hilo, HI 96720
\label{UKIRT}}
\addtocounter{address}{1}
\altaffiltext{\theaddress}{Department of Physics, University of Michigan, 
500 East University, Ann Arbor, MI, 48109
\label{Michigan}}
\addtocounter{address}{1}
\altaffiltext{\theaddress}{U.S. Naval Observatory,
Washington, DC  20392-5420
\label{USNO1}}
\addtocounter{address}{1}
\altaffiltext{\theaddress}{U.S. Naval Observatory,
Flagstaff Station, P.O. Box 1149, Flagstaff, AZ 86002
\label{USNO2}}
\addtocounter{address}{1}
\altaffiltext{\theaddress}{University of Chicago, Astronomy \& Astrophysics
Center, 5640 S. Ellis Ave., Chicago, IL 60637
\label{Chicago}}
\addtocounter{address}{1}
\altaffiltext{\theaddress}{Dept. of Astronomy and Astrophysics,
The Pennsylvania State University,
University Park, PA 16802
\label{PennState}}
\addtocounter{address}{1}
\altaffiltext{\theaddress}{Fermi National Accelerator Laboratory,
P.O. Box 500, Batavia, IL 60510
\label{FNAL}}
\addtocounter{address}{1}
\altaffiltext{\theaddress}{Apache Point Observatory,
2001 Apache Point Road, P.O. Box 59, Sunspot, NM 88349-0059
\label{APO}}
\addtocounter{address}{1}
\altaffiltext{\theaddress}{Dept. of Physics and Astronomy,
The Johns Hopkins University, 3701 San Martin Drive, Baltimore, MD 21218
\label{JHU}}
\addtocounter{address}{1}
\altaffiltext{\theaddress}{Dept. of Physics of Complex Systems,
E\"otv\"os University, P\'azm\'any P\'eter s\'et\'any 1/A, Budapest, H-1117, Hungary
\label{Eotvos}}
\addtocounter{address}{1}
\altaffiltext{\theaddress}{Remote Sensing Division, Code 7215, Naval Research
Laboratory, 4555, Overlook Ave. SW, Washington, DC 20375
\label{RSD}}
\addtocounter{address}{1}
\altaffiltext{\theaddress}{Department of Physics \& Astronomy,
P.O. Box 3905, University of Wyoming Laramie, WY 82071-3905
\label{Wyoming}}
\addtocounter{address}{1}
\altaffiltext{\theaddress}{USRA -- SOFIA, NASA Ames Research Center,
MS 144-2, Moffett Field, CA 94035-1000
\label{SOFIA}}

\begin{abstract}

We discuss measurements of the properties of $\about$10,000 asteroids
detected in 500 deg$^2$ of sky in the Sloan Digital Sky Survey (SDSS)
commissioning data. The moving objects are detected in the magnitude
range 14 $<$ \r $<$ 21.5, with a baseline of \about 5 minutes, resulting
in typical velocity errors of \about 3\%. Extensive tests show that the
sample is at least 98\% complete, with the contamination rate of less
than 3\%.

We find that the size distribution of asteroids resembles a broken
power-law, independent of the heliocentric distance: $D^{-2.3}$ for
0.4 km $\la D \la$ 5 km, and $D^{-4}$ for 5 km $\la D \la$ 40 km.
As a consequence of this break, the number of asteroids with \r $<$ 21.5
is ten times smaller than predicted by extrapolating the power-law relation
observed for brighter asteroids (\r $\la$ 18). The observed counts imply
that there are about 530,000 objects with $D>1$ km in the asteroid belt,
or about four times less than previous estimates. We predict that by
its completion SDSS will obtain about 100,000 near simultaneous five-band
measurements for a subset drawn from 280,000 asteroids brighter than
\r $<$ 21.5 at opposition. Only about a third of these asteroids
have been previously observed, and usually in just one band.

The distribution of main belt asteroids in the 4-dimensional SDSS color
space is bimodal, and the two groups can be associated with S (rocky)
and C (carbonaceous) asteroids. A strong bimodality is also seen in the
heliocentric distribution of asteroids and suggests the existence of
two distinct belts: the inner rocky belt, about 1 AU wide (FWHM) and
centered at $R$ \about 2.8 AU, and the outer carbonaceous belt, about
0.5 AU wide and centered at $R$ \about 3.2 AU. The median color of each
class becomes bluer by about 0.03 mag AU$^{-1}$ as the heliocentric distance
increases. The observed number ratio of S and C asteroids in a sample
with \r $<$ 21.5 is 1.5:1, while in a sample limited by absolute magnitude
it changes from 4:1 at 2 AU, to 1:3 at 3.5 AU. In a size-limited sample
with $D > 1$ km, the number ratio of S and C asteroids in the entire main
belt is 1:2.3.

The colors of Hungarias, Mars crossers, and near-Earth objects, selected
by their velocity vectors, are more similar to the C-type than to S-type
asteroids, suggesting that they originate in the outer belt. In about 100
deg$^2$ of sky along the Celestial Equator observed twice two days apart,
we find one plausible Kuiper Belt Object (KBO) candidate, in agreement with
the expected KBO surface density. The colors of the KBO candidate are
significantly redder than the asteroid colors, in agreement with colors of
known KBOs. We explore the possibility that SDSS data can be used to search
for very red, previously uncatalogued asteroids observed by 2MASS, by
extracting objects without SDSS counterparts. We do not find evidence for
a significant population of such objects; their contribution is no more than
10\% of the asteroid population.

\end{abstract}


\section{                              Introduction                            }

The scientific relevance of the small bodies in our solar system ranges from
fundamental questions about their origins to pragmatic societal concerns about
the frequency of asteroid impacts on the Earth. A proper inventory of these
objects requires a survey with

\begin{enumerate}
\item
Large sky coverage.
\item
Faint limiting magnitude.
\item
Uniform and well-defined detection limits in magnitude and proper motion.
\item
Accurate multicolor photometry for taxonomy.
\item
Sufficient multi-epoch observations or follow-up observations to determine
the orbits of all, or at least of particularly interesting objects.
\end{enumerate}

The Sloan Digital Sky Survey (SDSS, York {\em et al.} 2000), which was primarily
designed for studies of extragalactic objects, satisfies all of the above
requirements, except the last one. Although the SDSS cannot be used
to determine the orbits of detected moving objects (except in the special case of the
Kuiper Belt Objects discussed in \S 8), its accurate and deep near simultaneous
five-color photometry, ability to detect the motion of objects moving faster than \about
0.03 deg day$^{-1}$, and large sky coverage, can be efficiently used for studying
small solar system objects.  For example, the largest multi-color asteroid survey to
date is the Eight
Color Asteroid Survey (ECAS, Zellner, Tholen \& Tedesco 1985), in which 589 asteroids
were observed in eight different passbands. The five-color SDSS photometry spans
roughly the same wavelength range as the ECAS passbands, and will be available for
about 100,000 asteroids to a limit about seven magnitudes fainter than ECAS.
This represents an increase in the number of observed objects with accurate
multi-color photometry by more than two orders of magnitude. The only other survey
with depth and number of observed objects comparable to SDSS is Spacewatch II
(Scotti, Gehrels \& Rabinowitz 1991), which, however, does not provide any color
information\footnote{For more details on the Spacewatch project see
http://pirlwww.lpl.arizona.edu/spacewatch}.
Color information is vital in e.g. determining the asteroid size distribution,
which is considered to be the ``planetary holy grail'' by Jedicke \& Metcalfe
(1998), because the colors can be used to distinguish different types of asteroids
and thus avoid significant ambiguities (Muiononen, Bowell \& Lumme 1995). For
informative reviews of asteroid research we refer the reader to Gehrels (1979) and
Binzel (1989).

This paper presents some of the early solar system science from the SDSS
commissioning data. These data cover about 5\% of the total sky area to be
observed by the survey completion (in about 5 years). Section 2 describes the
SDSS, its capabilities for moving object detection, and the detection
algorithm implemented in the photometric processing pipeline. The data
and the accuracy of measured parameters are described in Section 3.
The colors of detected objects are discussed in Section 4, and their proper
motions in Section 5. The distributions of heliocentric distances and sizes
of main belt asteroids are discussed in Section 6. We describe the use of SDSS
data for finding asteroids in 2MASS data in Section 7, the search for Kuiper
Belt Objects using multi-epoch SDSS data in Section 8, and discuss the
results in Section 9.

\section{               Solar System Objects in SDSS          }

\subsection{                 SDSS Imaging Data        }

The SDSS is a digital photometric and spectroscopic survey which will
cover 10,000 deg$^2$ of the Celestial Sphere in the North Galactic cap and
produce a smaller ($\sim$ 225 deg$^2$) but much deeper survey in the
Southern Galactic hemisphere (York {\em et al.} 2000\footnote{See also
http://www.astro.princeton.edu/PBOOK/welcome.htm} and references therein).
The survey sky coverage will result in photometric measurements for about 
50 million stars and a similar number of galaxies. The
flux densities of detected objects are measured almost simultaneously in five bands
($u'$, $g'$, $r'$, $i'$, and $z'$; \cite{F96}) with effective wavelengths of 3561~\AA,
4676~\AA, 6176~\AA, 7494~\AA, and 8873~\AA, 95\% complete\footnote{These
values are determined by comparing multiple scans of the same area obtained
during the commissioning year. Typical seeing in these observations was
1.5$\pm$0.1 arcsec.} for point sources to limiting magnitudes of 22.1, 22.4, 22.1, 21.2,
and 20.3 in the North Galactic cap\footnote{We refer to the measured magnitudes in
this paper as $u^*, g^*, r^*, i^*,$ and $z^*$ because the absolute calibration
of the SDSS photometric system (dependent on a network of standard stars)
is still uncertain at the $\sim 0.03^m$ level. The SDSS filters themselves are
referred to as $u', g', r', i',$ and $z'$. All magnitudes are given on the
AB$_\nu$ system (Oke \& Gunn 1983, for additional discussion regarding the SDSS
photometric system see \cite{F96}, Fan 1999, and Fan {\em et al.} 2001a).}.
Astrometric positions are accurate to about 0.1 arcsec
per coordinate (rms) for sources brighter than 20.5$^m$ (Pier {\em et al.} 2001),
and the morphological information from the images allows robust star-galaxy
separation to $\sim$ 21.5$^m$ (Lupton {\em et al.} 2001).

\begin{figure}
\plotfiddle{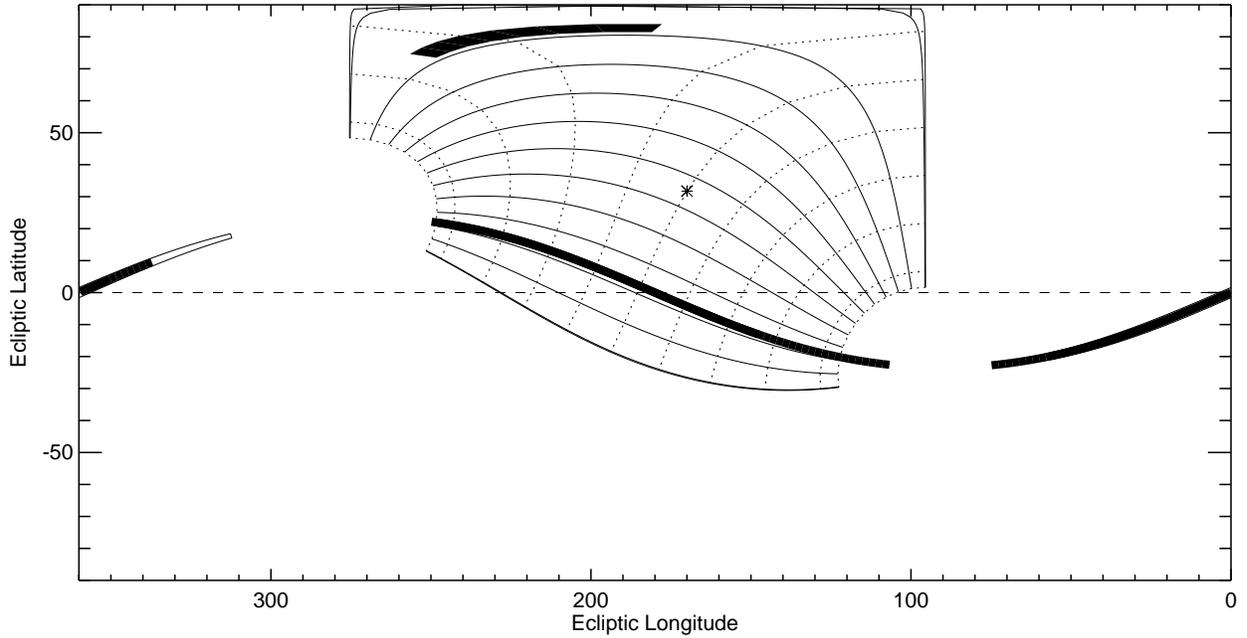}{10cm}{90}{67}{67}{280}{-80}
\caption{The SDSS footprint in ecliptic coordinates. The survey avoids
the Galactic plane and is limited to the area with Galactic latitude
$b > 30^\circ$ (approximately). The survey is performed by scanning along great
circles indicated by solid lines. The position of the North Galactic Pole is
indicated by an asterisk. The disconnected stripe that crosses $\lambda = 0^\circ$
is the Southern strip. The shaded regions represent areas analyzed in this work.
\label{SDSSfoot}
}
\end{figure}

The SDSS footprint in ecliptic coordinates is shown in Figure \ref{SDSSfoot}. The
survey avoids the Galactic plane, which is of limited use for solar system
surveys anyway because of the dense stellar background. The survey is performed
by scanning along great circles, indicated by solid lines. There are several important
areas for solar system science. The obvious area is the coverage of the Ecliptic from
$\lambda = 100^\circ$ to $\lambda = 225^\circ$. Less obvious is the area around
$\lambda = 100^\circ$ at one end of the great circle scans. Here, there will be
significant convergence of the scans, so about half of the sky is scanned twice in
the course of the survey. The third region, where the southern strip crosses
the Ecliptic ($\lambda = 0^\circ$), will be scanned several dozen times, and 
will be useful for studying the inclination and ecliptic latitude distributions 
of detected objects. The shaded regions represent areas analyzed in this work, 
and are described in more detail in Section 3.1.

\subsection{      The SDSS Sensitivity for Detecting Moving Objects    }

The SDSS camera (Gunn {\em et al.} 1998) detects objects in the order
$r'-i'-u'-z'-g'$, with detections in two successive bands separated in time
by 72 seconds. The mean astrometric accuracy for band-to-band transformations
is 0.040 arcsec per coordinate\footnote{This accuracy corresponds to moving
objects. The relative astrometric accuracy for stars is around 0.025 arcsec.}
(Pier {\em et al.} 2001). With this accuracy,
an $8\sigma$ moving object detection between the $r'$ and $g'$ bands corresponds
to an angular motion of \about0.025 \dd\ (or 3.8 arcsec/hr). This limit
corresponds to the Earth reflex motion for an object at a distance of $\about$
34 AU (i.e. the distance of Neptune) and shows that all types of asteroid
(including Trojans) can be readily detected (assuming an object at opposition).
With this level of accuracy it seems that the motion of Kuiper Belt Objects
(KBOs), which are mostly found beyond Neptune's orbit, could be detected at a
significance level better than \about6$\sigma$. Unfortunately, the distribution
of the astrometric errors (which include contributions from centroiding and
band-to-band transformations) is not strictly Gaussian, and the tests show
that the KBOs would be effectively detected at only \about3$\sigma$ level
(a typical KBO would move about 0.3 arcsec between the $r'$ and $g'$
exposures). Since the stellar density at the faint magnitudes probed by
SDSS is more than 10$^5$ times larger than the expected KBO density ($\about$
0.05-0.1 deg$^{-2}$ for \r $\la$ 22, Jewitt 1999), the false candidates would
preclude the routine detection of KBOs.

The upper limit on angular motion for detecting moving objects in a single scan with
the present software is about 1 \dd. This value is determined by the upper limit on
the angular distance between detections of an object in two different bands, for them
to be classified by the SDSS software as a single object. While somewhat seeing dependent
(see the next section), it is sufficiently high not to impose any practical limitation
for the detection of main belt asteroids.

\begin{figure}
\plotfiddle{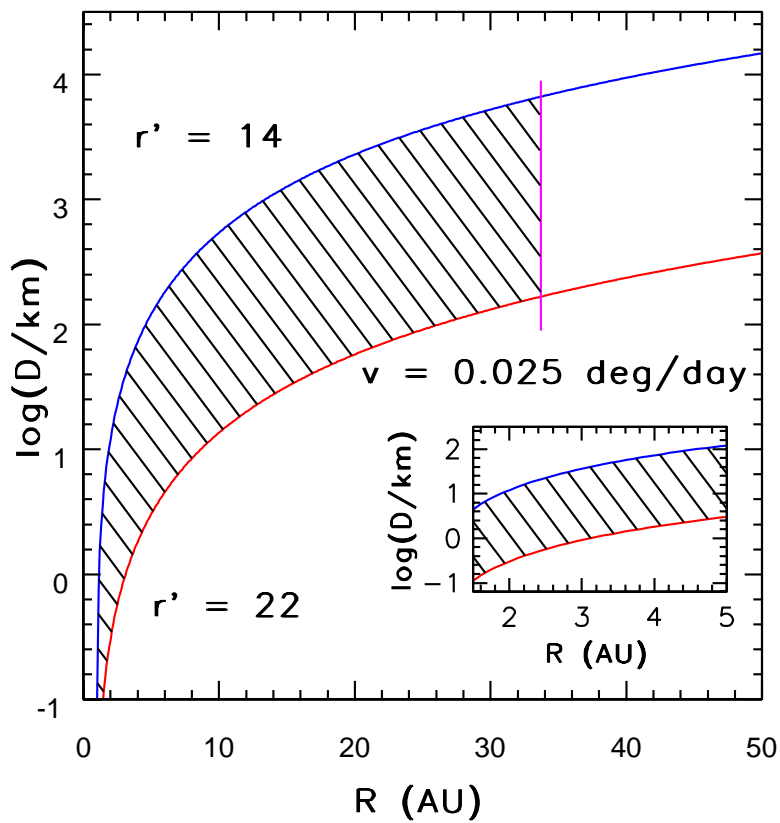}{10cm}{0}{80}{80}{-280}{-230}
\caption{The SDSS sensitivity for finding moving objects in a single scan.
The objects are assumed to be in circular orbits, have albedo of 0.1, and
are observed at opposition. The upper limit on the object size is set by the
image saturation, and the lower limit is set by the limiting magnitude for
the survey (\r \about 22). The upper limit on the heliocentric distance (or
equivalently, the lower limit on the detectable motion) is set by the astrometric
accuracy.
\label{window}
}
\end{figure}

The SDSS coverage of the size -- distance plane for objects on circular orbits observed
close to the antisolar point is shown in Figure \ref{window}. The two curved lines represent
the SDSS CCD saturation limit, \r=14, and the faint limit \r=22. They lines are determined
from (e.g. Jewitt 1999)
\eq{
\label{distmod}
                \r = \r(1,0) + 5 \log[R(R-1)],
}
where the heliocentric distance, $R$, is measured in AU, and $\r(1,0)$ is the ``absolute
magnitude'', the magnitude that an asteroid would have at a distance of 1 AU from
the Sun and from the Earth, viewed at zero phase angle. This is an impossible
configuration, of course, but the definition is motivated by desire to separate
asteroid physical characteristics from the observing configuration. The absolute magnitude
is given by
\eq{
\label{absmag}
             \r(1,0) = 17.9 - 2.5 \log({p \over 0.1}) - 5 \log(D / 1 {\rm km}),
}
where $p$ is the albedo, and the object diameter is $D$ (assuming a spherical asteroid).
The constant 17.9 was derived by assuming that the apparent magnitude of the Sun in the
$r'$ band is -26.95, obtained from $V_\odot$ = -26.75 and $(B-V)_\odot = 0.65$ (Allen 1973)
with the aid of photometric transformations from Fukugita {\em et al.} (1996). This
constant in an equivalent expression for the IAU-recommended (see e.g. Bowel {\em et al.}
1989) asteroid $H$ magnitude (not to be confused with the near-infrared $H$ magnitude
at \about 1.65 $\mic$) is 18.1. The plotted curves can be shifted vertically by
changing the albedo (we assumed $p=0.1$). The vertical line corresponds to an angular
motion of 0.025 \dd, approximately transformed into distance by using (e.g. Jewitt 1999)
\eq{
\label{veq}
                       v = { 0.986 \over R + \sqrt{R}} \, \dd.
}

The shaded region shows the part of the $D$--$R$ plane that can be explored with
single SDSS scans. SDSS can detect main belt asteroids with radii larger
than $\about$ 100 m, while asteroids larger than $\about$ 10-100 km, depending on
their distance, saturate the detectors. The upper limit on the heliocentric distance
at which motion can be detected in a single observation is $\about$ 34 AU; at
that distance SDSS can detect objects larger than 100 km (although the practical
limit on heliocentric distance is somewhat smaller due to confusion with
stationary objects, as argued above).

Without an improvement by about a factor of 2 in relative astrometry,
single SDSS scans cannot be used to efficiently detect KBOs\footnote{The SDSS astrometric
accuracy is limited by anomalous refraction and atmospheric motions. We are currently
investigating the possibility of increasing the astrometric accuracy by improving
centroiding algorithm in the photometric pipeline, and by post-processing data with
the aid of more sophisticated astrometric models than used in the automatic survey
processing.}. However, multiple scans of the same area can be used to search for KBOs,
and we describe such an effort in Section 8. The utilization of two-epoch data
obtained on time scales of up to several days allows the detection of moving objects
at distances as large as 100 AU. For example, at $R$ = 50 AU SDSS can detect objects
larger than $\about$ 200 km.

\subsection{       Detection of Moving Objects in SDSS Data     }
\label{DMflag}

The SDSS photometric pipeline (Lupton {\em et al.} 2001) automatically
flags all objects whose position offsets between the detected bands are consistent
with motion. Without this special treatment, main belt and faster
asteroids would be deblended\footnote{Here ``deblending" means separating
complex sources with many peaks into individual, presumably single-peaked,
components.} into one very red and one very blue object,
producing false candidates for objects with non-stellar colors.
In particular, the candidate quasars selected for SDSS spectroscopic observations
would be significantly contaminated because they are recognized by their
non-stellar colors (Richards {\em et al.} 2001).

As discussed in the previous section, objects are detected in the order
$r' - i' - u'- z' - g'$, with detections in two successive bands separated in time
by 72 seconds. The images of objects moving slower than about 0.5 \dd\ (about
1 arcsec during the exposure in a single band) are indistinguishable from stellar;
only extremely fast near-Earth objects are expected to be extended along the
motion vector. After finding objects and measuring their peaks in each band, they
are merged together by constructing the union of all pixels belonging to them.
The five lists of peaks, one for each band, are then searched for peaks that appear
at a given position (within 2$\sigma$) in only one band (if an object is detected
at the same position in at least two bands it cannot be moving). When all the
single-band peaks have been found, if there are detections in at least three bands,
the algorithm fits for two components of proper motion; if the $\chi^2$ is acceptable
($<$3 per degree of freedom), and the resulting motion is sufficiently different
from zero ($> 2\sigma$), the object is declared moving\footnote{Note that, with
the available positional accuracy of $\about$ 0.040 arcsec per coordinate, the proper
motion during a few minutes is not sufficiently non-linear to obtain a useful
orbital solution.}.

For the traditional detection methods based on asteroid trails, the detection
efficiency decreases with the speed of the asteroid at a given magnitude because
the trail becomes fainter (e.g. Jedicke \& Metcalfe, 1998). Any quantitative analysis
needs to account for this effect, and the correction is difficult to determine. One
of the advantages of SDSS is that the detection efficiency does not depend on the asteroid
rate of motion within the relevant range\footnote{Objects that move faster than
$\about$0.5 \dd\ may be classified by software as separate objects. They could
be searched for at the database level as single band detections. Such analysis
will be presented in a separate publication.} (most asteroids move slower than
0.5 \dd).

\section{    The Search for Moving Objects in SDSS Commissioning Data }

\subsection{          The Selection of Moving Objects           }

We utilize a portion of SDSS imaging data from six commissioning runs
(numbered 94, 125, 752, 756, 745, and 1336). The first four runs cover
481.6 deg$^2$ of sky along the Celestial Equator ($-1.27^\circ \la {\rm Dec}
\la 1.27^\circ$) with the ecliptic latitude, $\beta$, ranging from -14$^\circ$
to 20$^\circ$, and the last run covers
16.2 deg$^2$ of sky with $73^\circ < \beta < 86^\circ$. The fifth run (745)
covers roughly the same sky region as run 756, and was obtained 1.99 days earlier.
This two-epoch data set is used to determine the completeness of the asteroid
sample and to search for  KBOs. The data were taken during the Fall of 1998,
the Spring of 1999, and the Spring of 2000. More details about these runs are
given in Table 1. The data in each run were obtained in six parallel
scanlines\footnote{See also
http://www.astro.princeton.edu/PBOOK/strategy/strategy.htm}, each 13.5 arcmin wide
(the six scanlines from adjacent runs are interleaved to make a filled stripe).
The seeing in all runs was variable between 1 and 2 arcsec (FWHM) with the median
values ranging from 1.3 to 1.7 arcsec.

\begin{figure}
\plotfiddle{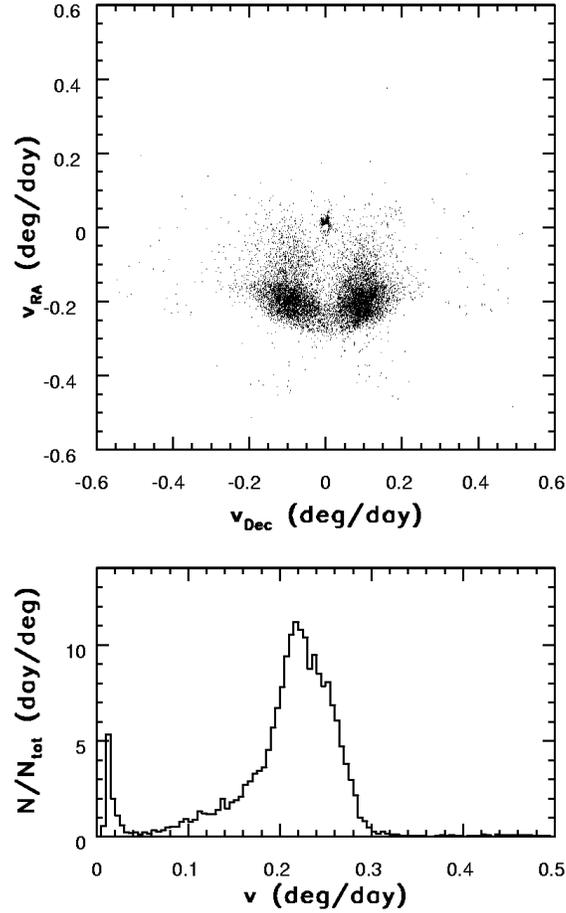}{10cm}{0}{60}{60}{-200}{-30}
\caption{The upper panel shows the velocity distribution in equatorial coordinates for
11,216 moving object candidates (see text). The lower panel shows the velocity histogram
for the same objects. Objects with $v < 0.03 \, \dd$ are likely to be spurious.
\label{vRAvDec}
}
\end{figure}

We first select all point sources which are flagged as moving (see \S \ref{DMflag})
and are brighter than \r=21.5. We choose a more conservative flux limit that discussed
above because the accuracy of the star-galaxy separation algorithm is not yet fully
characterized at fainter levels (we note that repeated commissioning scans imply that
significantly less than 5\% of sources with \r \about 21.5 are misclassified). There
are 11,216 such moving object
candidates in the analyzed area, and the upper panel in Figure \ref{vRAvDec} shows their
velocity distribution in equatorial coordinates (for the four equatorial runs which
dominate the sample, the $v_{RA}$ velocity component is parallel to the scanning direction).
There are two obvious concentrations of objects whose velocities satisfy
$v_{Dec} = \pm 0.43 v_{RA}$. The angle between the velocity vector and the Celestial
Equator is $\about$ 23 deg (i.e. arctan(0.43)) because the velocity vector is primarily
determined by the Earth reflex motion, and the asteroid density is maximal at the
intersection of the Celestial Equator and the Ecliptic. There are two peaks because
the sample includes both Spring and Fall data.

The candidates close to the origin ($v \la 0.03$ \dd, see the lower panel in Figure
\ref{vRAvDec}) are probably spurious detections since this velocity range is roughly
the same as the expected sensitivity for detecting moving objects (see \S 2.2).
Furthermore, we find that the distribution of these candidates is roughly circularly
symmetric around the origin, which further reinforces the conclusion that they are
not real detections. Consequently, in the remaining analysis we consider only the
10,678 moving object candidates with $v > 0.03$ \dd, which for simplicity we will
call asteroids hereafter.

\begin{figure}
\plotfiddle{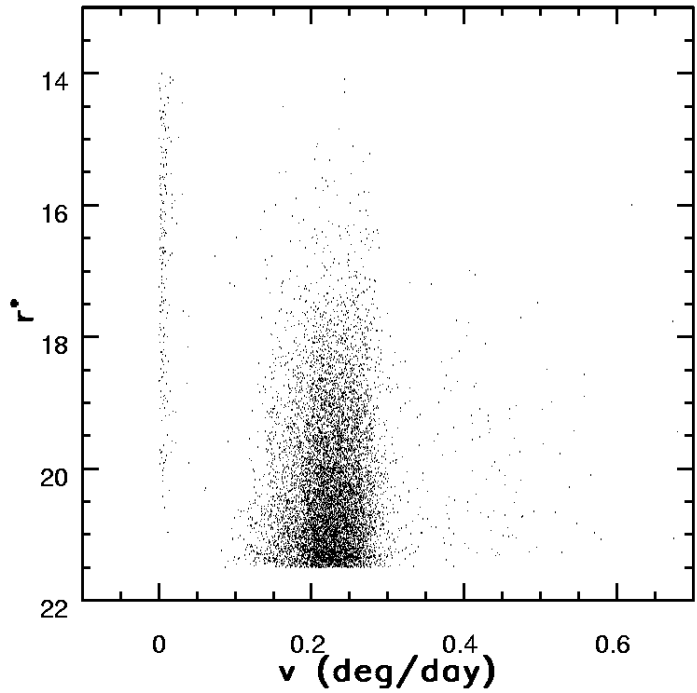}{10cm}{0}{80}{80}{-280}{-250}
\caption{The \r\ vs. $v$ distribution for the 11,216 moving object candidates.
The vertical strip of objects with $v \about 0$ are the rejected sources, and the large
concentration of sources to the right are 10,678 selected asteroids. Note the strikingly
empty gap between the two groups, which suggests that most detections with $v \ga 0.03$
deg day$^{-1}$ are real.
\label{r-v}
}
\end{figure}

The resulting sample is not very sensitive to the value of adopted cut, $v > 0.03$ \dd. Figure
\ref{r-v} shows the \r\ vs. $v$ distribution for the 11,216 moving object candidates.
The vertical strip of objects with $v \about 0$ are the rejected sources, and the large
concentration of sources to the right are 10,678 selected asteroids. Note the strikingly
clean gap between the two groups indicating that the false moving object detections are
probably well confined to a small velocity range. It is noteworthy that the objects
with $v < 0.03$ \dd\ are not concentrated towards the faint end, i.e. they do not
represent evidence that the moving object algorithm deteriorates for faint sources.
The number of $\about$ 500 objects appears consistent with the expected scatter due to
the large number of processed objects ($\about 5 \times 10^6$).

\begin{figure}
\plotfiddle{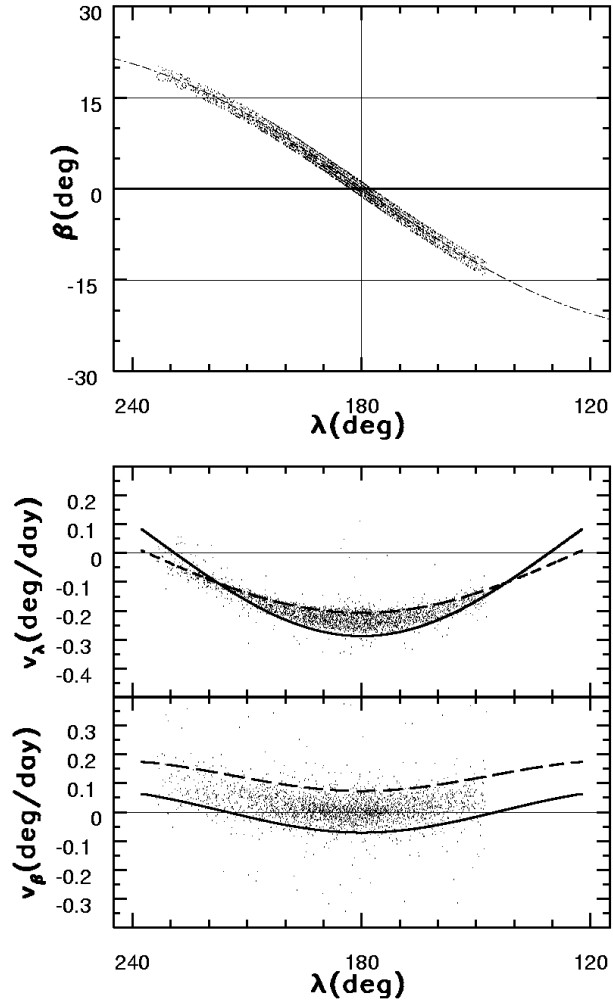}{11cm}{0}{65}{65}{-220}{-50}
\caption{The top panel shows the positions in ecliptic coordinate system of 2757
asteroids, marked by dots, selected from run 752. The dash-dotted line shows the
Celestial Equator. The bottom two panels show the longitude dependence of the two ecliptic
components of the measured asteroid velocity (corrected for diurnal motion). The
antisun is at $\lambda = 180^\circ$. The curves show predictions for varying
heliocentric distance and inclination (see text).
\label{phi752}
}
\end{figure}

The SDSS scans are very long, and some asteroids are observed at large angles
from the antisun direction. As discussed by Jedicke (1996), the observed velocity distribution
is therefore significantly changed from the one that would be observed near opposition.
Figure \ref{phi752} shows the 2757 asteroids selected from run 752. The top panel
displays each object as a dot in the ecliptic coordinate system. The overall
scan boundaries, and to some extent the six column boundaries (the strip is made
of six disjoint columns since the data from only one night are displayed), are outlined
by the object distribution. Note that the density of objects decreases with the distance
from the Ecliptic, as expected. The dash-dotted line shows the Celestial Equator. The lower
two panels show the dependence of the two ecliptic components of the measured asteroid
velocity, corrected for diurnal motion (i.e. the proper motion component due to the
Earth's rotation). These data were obtained on March 21, and thus the antisun is at
$\lambda = 180^\circ$. A strong correlation between the measured velocity and distance
from the antisun, $\phi = \lambda - 180^\circ$, is evident.

Following Jedicke (1996), we derive expressions for $v_\lambda(\phi,\beta,R,i)$ and
$v_\beta(\phi,\beta,R,i)$ for a circular orbit (see Appendix A). The
expected errors due to nonvanishing orbital eccentricity are 5-10\%, depending
on the heliocentric distance, $R$ and inclination, $i$ (e.g. Jedicke \& Metcalfe 1998).
The two lines in the upper part of the bottom panel show the predicted
$v_\lambda(\phi)$ curves for $R$=2 AU (lower, solid) and $R$=3 AU (upper, dashed).
The dependence of these curves on inclination is much smaller than on $R$ (the curves
are computed for $i$ = 0$^\circ$). Note that the curves cross for $|\phi| \about 40^\circ$.
The two lines in the lower part of the bottom panel show the predicted $v_\beta(\phi)$
curves for $i$ = -10$^\circ$ (lower, solid) and $i$ = 10$^\circ$ (upper, dashed).
The dependence of these curves on $R$ is much smaller than on inclination (the curves
are computed for $R$=2.5 AU).

The plotted curves bracket the observed distribution, and show that it is
impossible to accurately estimate the heliocentric distance for large $\phi$ since
the $v_\lambda$ curves cross. Since the heliocentric distance is crucial for a
significant part of the analysis presented here, we further limit the sample to
$|\phi| < 15^\circ$, except when studying the ecliptic latitude distribution
(Section 4.2.1) and the dependence of colors on the phase angle (Section 6.2).
This constraint produces a sample of 6666 asteroids. We estimate the heliocentric
distance and orbital inclination for these asteroids as described in \S 3.4.2.

\subsection{       The Sample Completeness and Reliability   }
\label{comrel}
The sample completeness (the fraction of moving objects in the data
recognized as such by the moving object algorithm) and reliability
(the fraction of correctly recognized moving objects) are important
factors which can significantly affect the conclusions derived in
the subsequent analysis. It is not possible to determine the completeness
and reliability by comparing the sample with catalogued asteroids because
existing catalogs are complete only to \r \about 15-16 (Zappala \& Cellino 1996,
Jedicke \& Metcalfe 1998), while this sample extends to \r = 21.5 (less
than 1/3 of moving objects observed by SDSS can be linked to previously
documented asteroids, see Appendix B). We estimate the sample completeness
and reliability by analyzing data for 99.5 deg$^2$ of sky observed twice
1.9943 days apart (runs 745 and 756, see Table 1). These two runs were
obtained during SDSS commissioning and overlap almost perfectly.

We estimate the sample reliability by matching 2474 objects selected
by the moving object algorithm in run 756 to the source catalog for run
745. The probability that two moving objects would be found at
an identical position in two runs is negligible, and such a matched pair
indicates a stationary object that was erroneously flagged as moving.
We find 62 matches within 1 arcsec implying a sample reliability of 97.5\%.
Visual inspection shows that the majority of false detections are either
associated with saturated stars, or are close to the faint limit for
detectability. Objects with very unusual velocities are more likely
to be false detections; we discuss further the reliability of such
candidates in Section 5.4.

The sample completeness can be obtained by comparing 2412 reliable
detections to the ``true" number of moving objects in the data.
This ``true" number can be estimated by using the fact that the true
moving objects observed in one epoch will not have a positional counterpart
in the other epoch. There are 4808 unsaturated point sources
brighter than \r=21.5 from run 756 that do not have counterparts within
3 arcsec in run 745 (the number of matched objects is $\about 10^6$).
Not all 4808 unmatched objects are moving objects, as many are not matched
due to instrumental effects (e.g. diffraction spikes, blended sources, etc.).
The hard part is to select moving objects among these 4808 objects without
relying on the moving object algorithm itself. We select the probable moving objects
by using the difference between the point-spread-function (PSF) and ``model"
magnitudes in the $g'$ band, and taking advantage of a bug in the code,
since fixed.

The PSF magnitudes are measured by fitting a PSF model, and model magnitudes
are measured by fitting an exponential and a de Vaucouleurs profile convolved
with the PSF, and using the formally better model in the $r'$ band to evaluate
the magnitude (Lupton {\em et al.} 2001).
Model magnitudes are designed for galaxy photometry and become equal to
the PSF magnitudes for unresolved sources, if they are not moving.
The difference between the PSF and model magnitudes for moving unresolved
sources is due to different choices of centroids. For PSF magnitudes the local
centroid in each band is used, while for model magnitudes the $r'$ band
center is used\footnote{This has been changed in more recent versions of
the photometric pipeline, for which local centroids are used for model magnitudes,
as well.}.
The moving objects thus have fainter model magnitudes than PSF magnitudes
in all bands but $r'$ (the point sources have small $m_{psf}(\r)-m_{mod}(\r)$
by definition). This difference is maximized for the $g'$ band and
we find that 4808 unmatched sources show a bimodal distribution of
$m_{psf}(\g)-m_{mod}(\g)$ with a well-defined minimum at about $-1$.
There are 2424 sources with $m_{psf}(\g)-m_{mod}(\g) < -1$, and they can
be considered as ``true" moving objects (that is, we are essentially
comparing two different detection algorithms). In the same area
there are 2412 objects flagged by the moving object algorithm, and
2377 of these are included in the above 2424, implying that the sample
completeness is 98\% (this is indeed a lower limit on the sample completeness
since it is possible that some of the 2424 sources with $m_{psf}(\g)-m_{mod}(\g)
< -1$ are not truly moving).

\subsection{        The Accuracy of the Measured Velocities       }

\begin{figure}
\plotfiddle{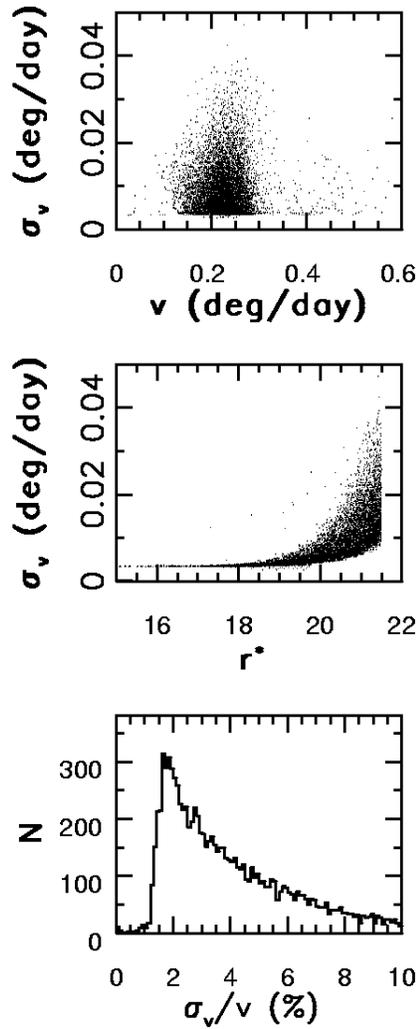}{11cm}{0}{80}{80}{-250}{-85}
\caption{The velocity errors quoted by the moving object algorithm implemented in
SDSS photometric pipeline. The top panel displays errors for 6,666 selected asteroids
vs. the asteroid velocity magnitude, the middle panel shows errors vs.
\r\ magnitude, and the bottom panel shows a histogram of fractional errors (expressed
in percent).
\label{photoErrors}
}
\end{figure}

While the sample completeness and reliability are sufficiently high
for robust statistical analysis of the moving objects, inaccurate
velocity measurements may contribute significant uncertainty because
the heliocentric distance is determined from the asteroid rate of motion.
Figure \ref{photoErrors} gives an overview of the achieved accuracy, as
quoted by the photometric pipeline. The top panel shows the velocity
error vs. velocity, and the middle panel shows the velocity
error vs. \r\ magnitude, where each of the 6666 asteroids is shown as
a dot. It is evident that the errors are dominated by photon statistics
at the faint end. The histogram of the fractional velocity errors
displayed in the bottom panel shows that, for the majority of objects,
the accuracy is better than 4\%.

Of course, it is not certain whether the quoted errors are realistic.
A test requires independent velocity measurements.
We estimate the accuracy of the measured velocity by comparing
two adjacent runs (752 and 756, see Table 1) obtained one day
apart. These two runs can be interleaved to make a full stripe.
Thanks to a fortuitous combination of the column width (13.5 arcmin)
and the time delay between the two scans, many asteroids observed
in the first run move into the area scanned by the second run
the following night. Out of 1626 asteroids with $|\phi| < 15^\circ$
observed in run 752, 693 were expected to be reobserved in run 756.
For reobserved objects the velocity can be obtained about 300 times
more accurately than from single run data, due to longer time
baseline (\about 24 hours vs. \about 5 minutes).

We positionally match within 60 arcsec these 693 asteroids to the
asteroids observed in run 756, and find 476 matches. By rematching
a random set of positions within the same matching radius, we estimate
that about 18 are random associations, implying a true matching
rate of 66\% (this matching rate is substantially lower than the
reliability of the sample because of the velocity errors).
The association of the two samples with a larger matching radius
increases the number of close pairs but also increases the fraction
of random associations. We find that matched and unmatched objects have
similar magnitude distributions, showing that the algorithm is robust
at the faint end.

\begin{figure}
\plotfiddle{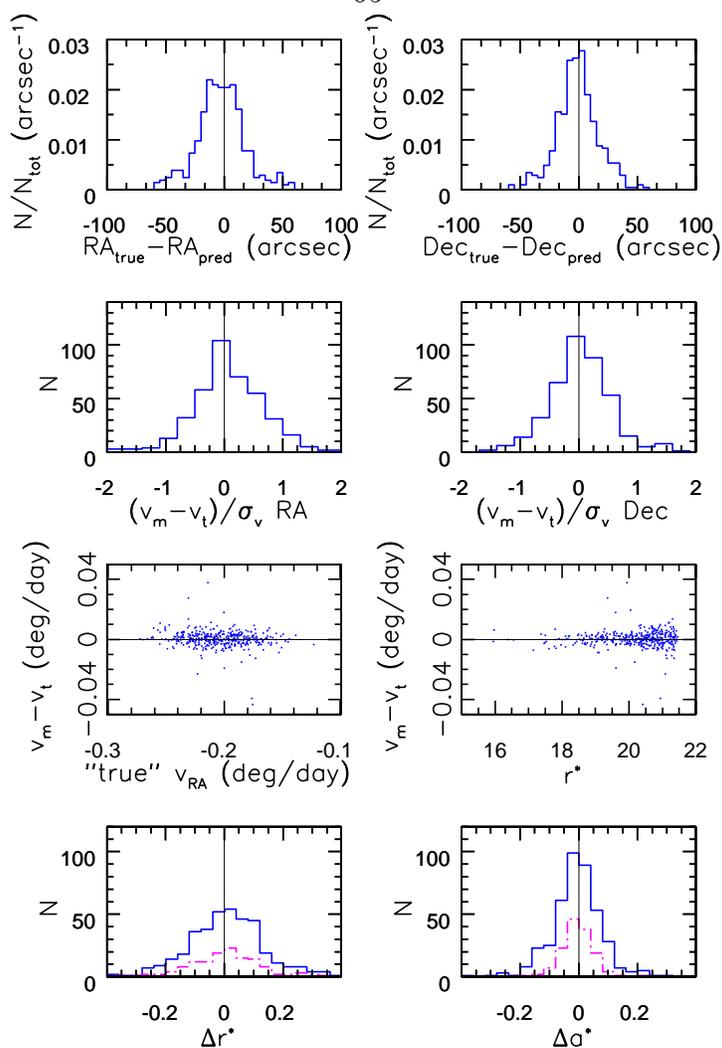}{12cm}{0}{60}{60}{-180}{-60}
\caption{The comparison of asteroid velocities and quoted errors measured in a
single observing run with the ``true'' values determined by reobserving objects
the following night for 476 asteroids observed twice in runs 752 and 756.
In the top two rows the left column corresponds to the right
ascension component and the right column to the declination component. The top two
panels show histograms of the positional difference, and the panels in the second
row show the velocity difference normalized by the quoted error. The two panels
in the third row show the velocity difference vs. the true velocity (left),
and the velocity difference vs. the mean \r\ magnitude. The bottom two panels show
the histograms of the magnitude and color differences between the two epochs for
the asteroids observed twice. The solid lines show histograms for all objects,
and the dot-dashed line show histograms for objects with $\r < 20$. Note that the
magnitude difference histogram is significantly wider than the color difference
histogram, presumably due to variability caused by the rotation of asteroids.
\label{matchesX}
}
\end{figure}

The detailed matching statistics are presented in Figure \ref{matchesX}.
The histograms of the differences between the predicted and observed
positions in the second epoch are shown in the top row. As discussed earlier,
the matched position can be used to determine the asteroid velocity to a much
better precision than from single run data. The two panels in the second
row in Figure \ref{matchesX}
show the histograms of the difference between this ``true" velocity
and the measured velocity in run 756, normalized by the quoted errors.
The quoted errors are overestimated by factor \about 2, presumably due
to overestimated centroiding errors in the photometric pipeline (whose
computation has been meanwhile improved). The two panels in the third row
show that the velocity difference is correlated with neither the velocity
nor the object's brightness (if comparing these two panels with the middle
panel in Figure \ref{photoErrors}, note that here only 476 matched objects
are included, while there all 6666 objects are shown).

The same matched data can be used to test the photometric
accuracy. The photometric errors determined by comparing measurements
for stars observed in two epochs are about 2-3\% for objects at the
bright end, and then start increasing due to photon counting noise
(for more details see Ivezi\'{c} {\em et al.} 2000). The bottom two panels
in Figure \ref{matchesX} show the histograms of the observed differences in
\r\ magnitude (lower left panel) and asteroid $a^*$ color (to be defined in
the next section, lower right panel). The histogram for all 476 matched asteroids
is shown by a solid line, and for 165 asteroids with $\r < 20$ by a dot-dashed
line. From the interquartile range we estimate that the equivalent Gaussian
width of the \r\ histogram is 0.11 mag, and the width of the $a^*$ histogram is
0.07 mag, independent of the magnitude limit. The color histogram is narrower
than the magnitude histogram, although its width is expected to be between
1 and $\sqrt{2}$ times wider\footnote{The exact value depends on the level of correlation
between the two measurements.} than the width of the magnitude difference
histogram, based on the statistical considerations. This may be interpreted
as the variability due to asteroid rotation which affects the brightness
but not the color. Such an interpretation was first advanced by Kuiper
{\em et al.} (1958).

In the subsequent analysis we consider only the sample with 5253
unique sources, formed by excluding sources from runs 94 and 752 whose
positions and velocities imply that they are also observed in runs 125
and 756, respectively. The uncertainty of measured velocities will result
in some sources being incorrectly excluded, and some sources being counted
twice. However, the excluded sources represent only $\about 24\%$ of the
full sample, and thus even an uncertainty of 20\% in the sample of excluded
sources corresponds to less than 5\% of the final sample. More importantly,
no significant bias with respect to brightness, color, position, and velocity
is expected in this procedure, as shown above.

\section{                   The Asteroid Colors                           }

\subsection{            The Colors of Main Belt Asteroids                }

\begin{figure}
\plotfiddle{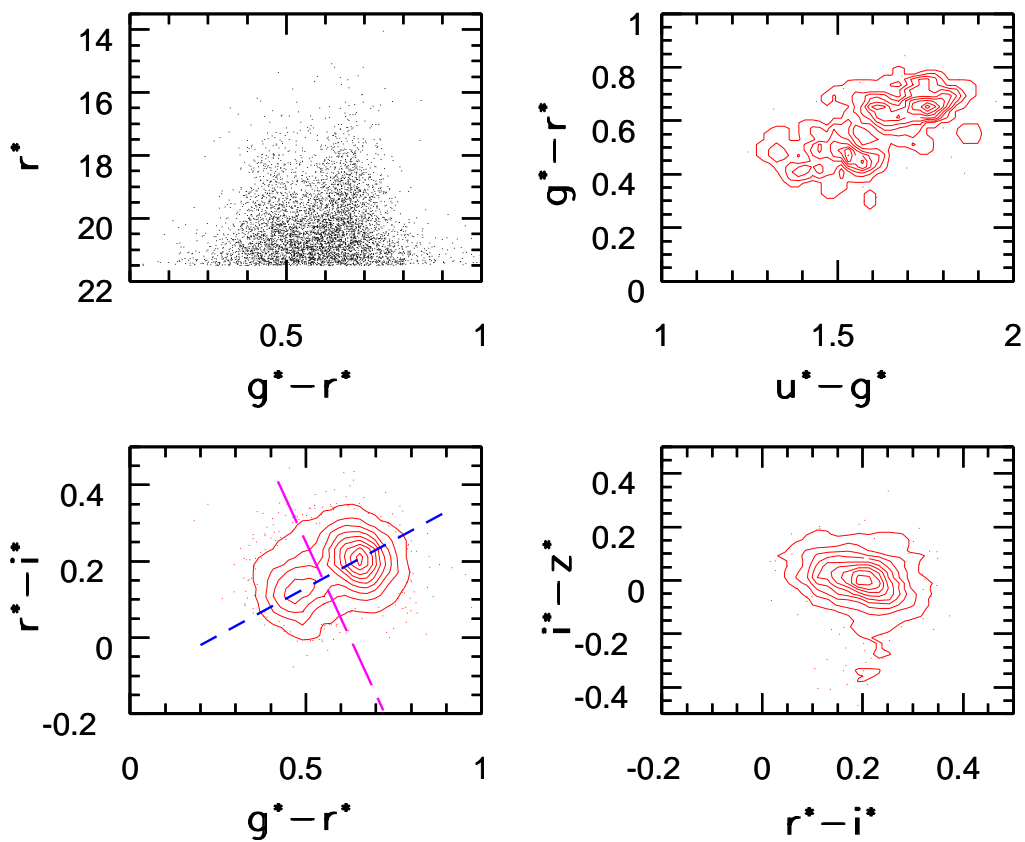}{10cm}{0}{90}{90}{-280}{-250}
\caption{The color-magnitude and color-color diagrams for 5,125 main belt asteroids.
The color-magnitude diagram in the upper left panel shows all objects as dots.
The three color-color diagrams in other panels show the distributions of objects
with photometric errors less than 0.05 mag. as linearly spaced isodensity contours,
and as dots below the lowest level. Note the bimodal distribution in the \gr\ vs. \ug\
and \ri\ vs. \gr\ diagrams. The two dashed lines in the \ri\ vs. \gr\ diagram show a
rotated coordinate system which defines an optimal asteroid color, named $a^*$.
\label{cmdMB}
}
\end{figure}

The color-magnitude diagram for the 5,125 main belt asteroids (selected by their
velocity vectors as described in \S 5.1 below) is displayed in the upper
left panel in Figure \ref{cmdMB}. The other three panels display color-color
diagrams\footnote{The color transformations between the SDSS and other photometric
systems can be found in Fukugita {\em et al.} (1996) and Krisciunas, Margon \&
Szkody (1998). For a quick reference, here we note that $\ug = 1.33(U-B)+1.18$
and $\gr = 0.96(B-V)-0.23$, accurate to within \about 0.05 mag.}. For clarity,
in the last three diagrams the asteroid distribution is shown by linearly spaced
density contours in the regions of high density and as dots outside the lowest
level contour. It is evident that the asteroid color distribution is bimodal,
in agreement with previous studies of smaller samples (Binzel 1989 and references
therein). In particular, the two color types are clearly separated in the \ri\ vs.
\gr\ color-color diagram which shows two well defined peaks.

The comparison of the observed color distributions with the colors of known asteroids
observed in the SDSS bands by Krisciunas, Margon \& Szkody (1998) shows that the
``blue" asteroids in the \ri\ vs. \gr\ diagram (lower left peak) can be associated
with the C type (carbonaceous) asteroids, while the redder class (upper right
peak) corresponds to the S type (silicate, rocky) asteroids. However, note that
not all ``blue" asteroids are C type asteroids, and not all ``red" asteroids are
S type asteroids, but rather they contain other taxonomic classes as well. For
example, Tedesco {\em et al.} (1989) defined 11 families among 357 asteroids,
based on IRAS photometry and three wideband optical filters. The IRAS photometry is used
to determine the albedo which, together with the two colors, defines the taxonomic
classes. Nevertheless, the same work shows that a large majority of all asteroids
belong to either C or S type, and we find that SDSS photometry clearly differentiates
between the two types.

\begin{figure}
\plotfiddle{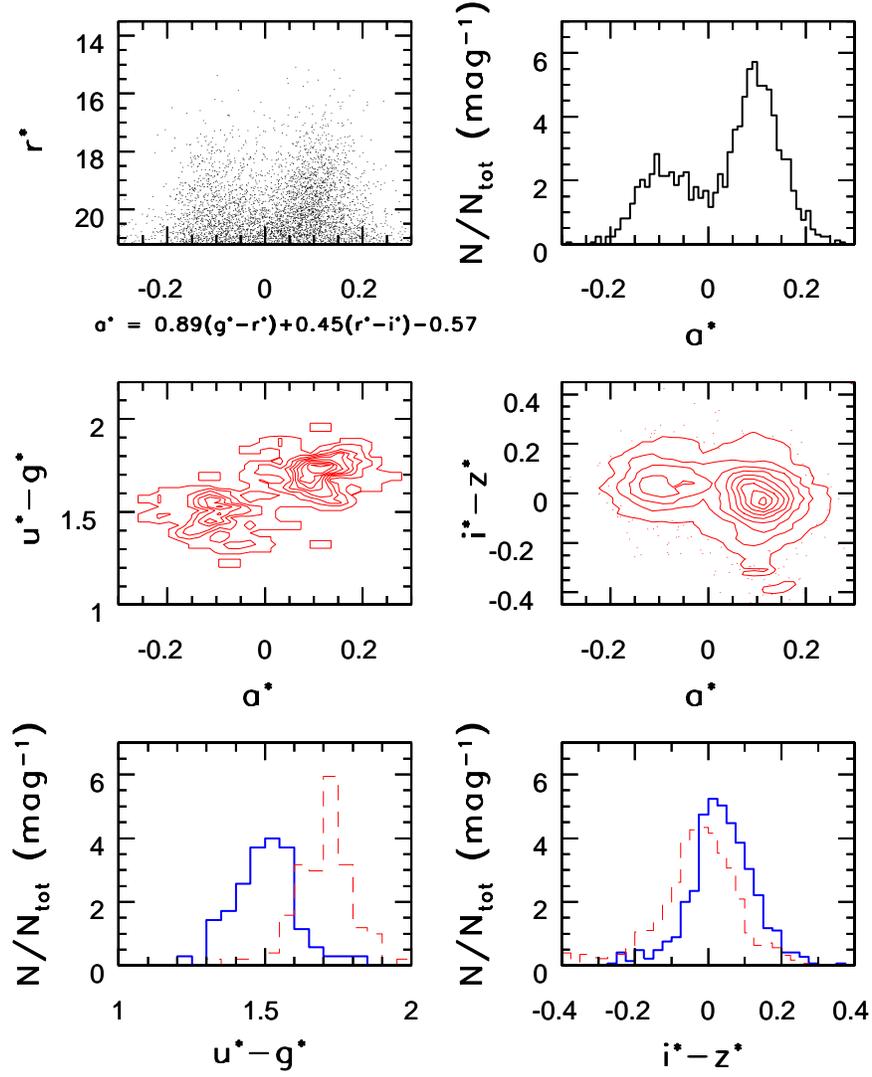}{10cm}{0}{75}{75}{-230}{-70}
\caption{The $a^*$ color based color-magnitude and color-color diagrams and various
histograms for 5,125 main belt asteroids. The bottom panels show color histograms
for the two classes of asteroids, separated by their $a^*$ color ($a^* < 0$, thick
solid line; $a^* \ge 0$, thin dashed line).
\label{cmdMBb}
}
\end{figure}

Tedesco {\em et al.} noted that much of the difference in the asteroid spectra
between 0.3 \mic\ and 1.1 \mic\ is due to two strong absorption features, one bluer
than 0.55 \mic\ and one redder than 0.70 \mic. The SDSS $r'$ filter lies
almost entirely between these two absorption features, which may explain why the
$\g-\r$ and $\r-\i$ colors provide good separation of the two color types. Moreover,
the \ri\ vs. \gr\ color-color diagram is constructed with the most sensitive SDSS bands.
We use this diagram to define an optimized color that can be used to quantify the
correlations between asteroid colors and other properties (e.g. heliocentric distance,
as discussed in the next section). We rotate and translate the \ri\ vs. \gr\ coordinate
system such that the new x-axis, hereafter called \a, passes through both peaks, and
define its value for the minimum asteroid density between the peaks to be 0 (i.e.
we find the principal components in the \ri\ vs. \gr\ color-color diagram). We obtain
\eq{
                 \a = 0.89 (\g - \r) + 0.45 (\r - \i) - 0.57.
}
Asteroids with \a $<$ 0 are blue in the \ri\ vs. \gr\ diagram, and those with \a $>$ 0
are red. Figure \ref{cmdMBb} shows various diagrams constructed with this optimized color.
The top left panel displays the \r\ vs. \a\ color-magnitude diagram.  The two color types
are much more clearly separated than in the analogous diagram using the \gr\ color
(displayed in the top left panel in Figure \ref{cmdMB}). The top right panel shows the
\a-histogram for asteroids brighter than \r=20. The number of main belt asteroids with
$\a > 0$ is 1.88 times larger than the number of asteroids with $a < 0$ (1.47 times for
\r $<$ 21.5). However, note that this result does not represent a true number
ratio of the two types because the same flux limit does not correspond to the same size
limit due to different albedos, and because of their different radial distributions
(see Section 6).

The two middle panels show the color-color diagrams constructed with the new color.
These diagrams suggest that the distribution of the SDSS colors for the main belt
asteroids reveals only two major classes: the asteroids with \a $<$ 0 have bluer \ug\
color and slightly redder \iz\ color than asteroids with \a $>$ 0. These color differences
are better seen in the color histograms shown in the two bottom panels, where
thick solid lines correspond to asteroids with $a^* \le 0$, and the thin dashed lines
to those with $a^* > 0$. In the remainder of this work we will refer to ``blue" and
``red" asteroids as determined by their \a\ color, but note that the \iz\ colors
are reversed (the subsample with bluer \iz\ color is redder in other SDSS colors).

\subsubsection{ Does SDSS Photometry Differentiate More Than Two Color Types? }

The large number of detected objects and accurate 5-color SDSS photometry
may allow for a more detailed asteroid classification. For example, some objects
have very blue \iz\ colors
($< -0.2$, see Figure \ref{cmdMBb}) which could indicate a separate family.
There are various ways to form self-similar classes\footnote{Here ``self-similar
class" means a set of sources whose measurement distribution is smooth and does
not indicate substructure.}, well described by
Tholen \& Barucci (1989). We decided to use the program AutoClass in an unsupervised
search for possible structure. AutoClass employs Bayesian probability analysis to
automatically separate a given data base into classes (Goebel {\em et al.} 1989).
This program was used by Ivezi\'c \& Elitzur (2000) to demonstrate that the
IRAS PSC sources belong to four distinct classes that occupy separate regions
in the 4-dimensional space spanned by IRAS fluxes, and the problem at hand
is mathematically equivalent.

AutoClass separated main-belt asteroids into 4 classes by using 5 SDSS magnitudes
(not the colors!) for objects brighter than \r=20. The largest 2 classes include
more than 98\% of objects and are easily recognized in color-color diagrams as
the two groups discussed earlier. The remaining 2\% of sources are equally split
in two groups. One of them is the already suspected group with $\iz < -0.2$. The
inspection of color-color diagrams for these sources shows that the blue $\iz$ color
is their only clearly distinctive characteristic. The remaining group is similar
to the ``blue" group but has \about 0.2 mag. redder \iz\ color and \about 0.1 mag.
bluer \ri\ color. Since the number of asteroids in the two additional groups proposed
by AutoClass is only 2\%, we retain the original manual classification into two
major types in the rest of this work.

\subsubsection{ Comparison With Independent Taxonomic Classification }

\begin{figure}
\plotfiddle{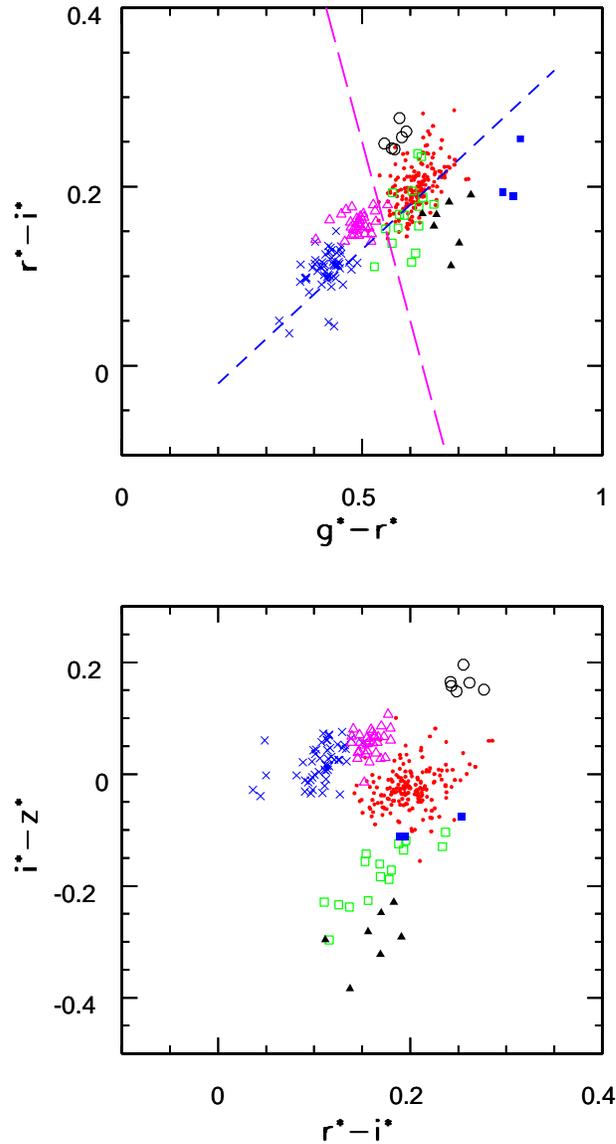}{14cm}{0}{70}{70}{-230}{-70}
\caption{The synthetic color-color diagrams for the 316 asteroids whose spectra
were obtained by the SMASS Survey, and convolved with the SDSS response functions
(see text). The taxonomic class is shown by different symbols: crosses
for the C type, dots for S, circles for D, solid squares for A, open squares for
V, solid triangles for J, and open triangles for the E, M and P types (which are
indistinguishable by their colors). The dashed lines in the upper panel show
the principal axes from Figure 8.
\label{SMASS}
}
\end{figure}

The number of known asteroids observed through SDSS filters (Krisciunas, Margon
\& Szkody 1998) is too small for a robust statistical analysis of the
correlation between the taxonomic classes and their color distribution.
In order to investigate whether the known asteroids with independent taxonomic
classification (which also includes the albedo information, not only the colors)
segregate in the SDSS color-color diagrams, we synthesize their colors
from spectra obtained by Xu {\em et al.} (1995). Their Small Main-belt
Asteroid Spectroscopic Survey (SMASS) includes spectra for 316 asteroids
with wavelength coverage from 0.4 \mic\ to 1.0 \mic, with the resolution
of the order 10 \AA. We convolved their spectra with the SDSS response
functions for the $g', r', i',$ and $z'$ bands (the spectra do not extend
to sufficiently short wavelengths for synthesizing the $u'$ band flux).
The results are summarized in the color-color diagrams displayed in Figure
\ref{SMASS}. The taxonomic classification (also adopted from Xu {\em et al.})
is shown by different symbols: crosses for the C type, dots for S, circles
for D, solid squares for A, open squares for V, solid triangles for J,
and open triangles for the E, M and P types (which are indistinguishable
by their colors). The dashed lines in the upper panel show the principal
axes discussed above.

These diagrams show that the blue asteroids ($a^* < 0$) include the C, E, M and
P classes, while the red asteroids ($a^* \ge 0$) include the S, D, A, V, and J
classes. Their distribution in the \ri\ vs. \gr\ diagram seems to be consistent with
the bimodal distribution reported here. The segregation of the classes in the
\iz\ vs. \ri\ diagram is evident. It may be that the D class asteroids
could be separated by their red \iz\ color ($> 0.15$), and that the
V and J classes could be separated by their blue \iz\ color ($< -0.2$).
However, note that the number of sources in these classes shown in
Figure \ref{SMASS} is not representative of the SDSS sample due to a
different selection procedure employed by the SMASS. Restricting the analysis
to asteroids with \r$<$20, we find that \about 6\% of the sample have
$\iz < -0.2$, and another 6\% have $\iz > 0.15$ (these fractions are somewhat
higher than obtained by AutoClass because its Bayesian algorithm is
intrinsically biased against overclassification, Goebel {\em et al.} 1989).
We leave further analysis of such classification possibilities for future work,
and conclude that the synthetic colors based on the SMASS data agree well with the
observed color distribution.

\subsubsection{   Albedos }

\begin{figure}
\plotfiddle{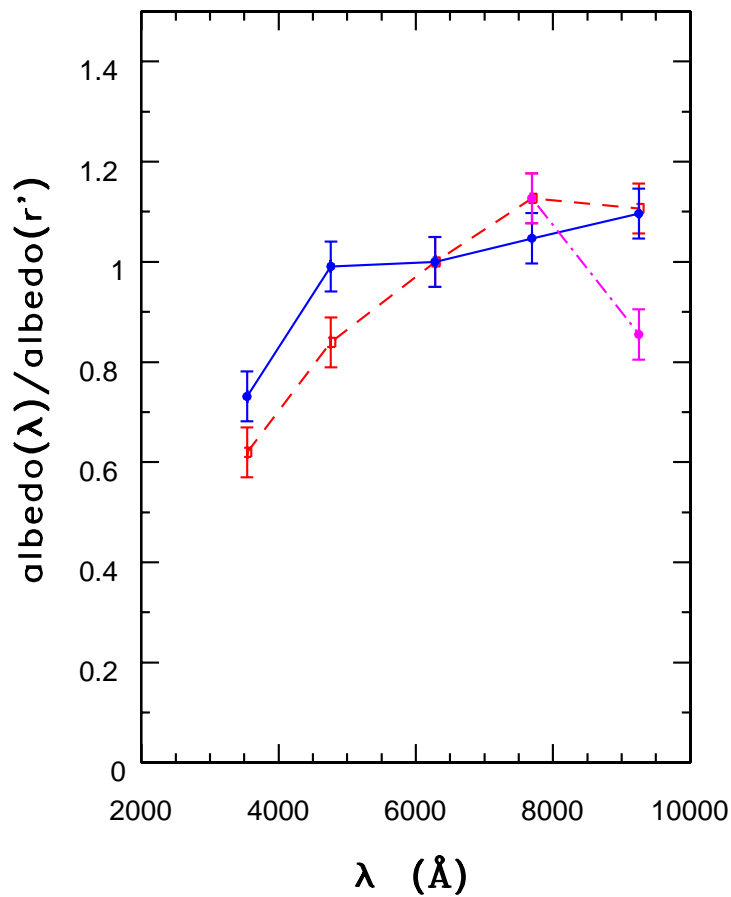}{10cm}{0}{80}{80}{-240}{-120}
\caption{The mean albedo for the two color types of main belt asteroids normalized to
its \r-band value. The solid line is for 1841 asteroids with $a^* < 0$, and
the dashed line is for 3302 asteroids with $a^* \ge 0$. The mean albedo for
a subset of the former, selected by \iz $<$ -0.20, is shown by the dot-dashed line.
The error bars show the distribution width for each subsample (equivalent Gaussian
width) -- they are not errors of each point, whose uncertainty is smaller than the
symbol size.
\label{albedo}
}
\end{figure}

The observed differences in the color distributions reflect differences in
asteroid albedos. Although the SDSS photometry cannot be used to estimate
the absolute albedos (which would require the measurements of the thermal
emission, see e.g. Tedesco {\em et al.} 1989), the spectral shape of the albedo
can be easily calculated since the colors of the illuminating source are
well known ($(\ug)_\odot$ = 1.32, $(\gr)_\odot$ = 0.45, $(\ri)_\odot$ = 0.10,
$(\iz)_\odot$ = 0.04). Figure \ref{albedo} shows the albedos obtained for the
median color for the two color-selected types normalized to the $r'$-band value.
The error bars show the (equivalent Gaussian) distribution width for each
subsample. The solid curve corresponds to asteroids with $\a \ge 0$, and the
dashed curve to those with $\a < 0$. Note the local maximum for the $\a \ge 0$
type, and that the \ug\ and $\r-\z$ colors are almost identical for the
two types. The dotted curve shows the mean albedo for
asteroids selected from the $\a \ge 0$ group by requiring \iz $< -0.2$.
The displayed wavelength dependence of the albedo is in good agreement with
available spectroscopic data (e.g. compare to Figure 5 in Tholen \& Barucci
1989, see also Figure 2 in Xu {\em et al.} 1995).

The subsequent analysis of the asteroid size distribution requires the
knowledge of the absolute albedo for each color type. The typical absolute values
of albedos for the two major asteroid types can be estimated from data presented
by Zellner (1979). Following Shoemaker {\em et al.} (1979), we adopt 0.04 (in the
$r'$ band) for the C-like asteroids ($a^* < 0$) and 0.14 for the S-like asteroids
($a^* \ge 0$). The intrinsic spread of albedo for each class is of the order 20\%,
in agreement with the range obtained by using IRAS data (Tedesco {\em et al.} 1989).
This difference in albedos implies that a C-like asteroid is $\sim$1.4 magnitudes
fainter than an S-like asteroid of the same size and at the same observed position,
and that a C-like asteroid with the same apparent magnitude and observed at the
same position is twice as large as an S-like asteroid.

\subsection{          The Asteroid Counts vs. Color       }

\subsubsection{ The Ecliptic Latitude Distribution }

\begin{figure}
\plotfiddle{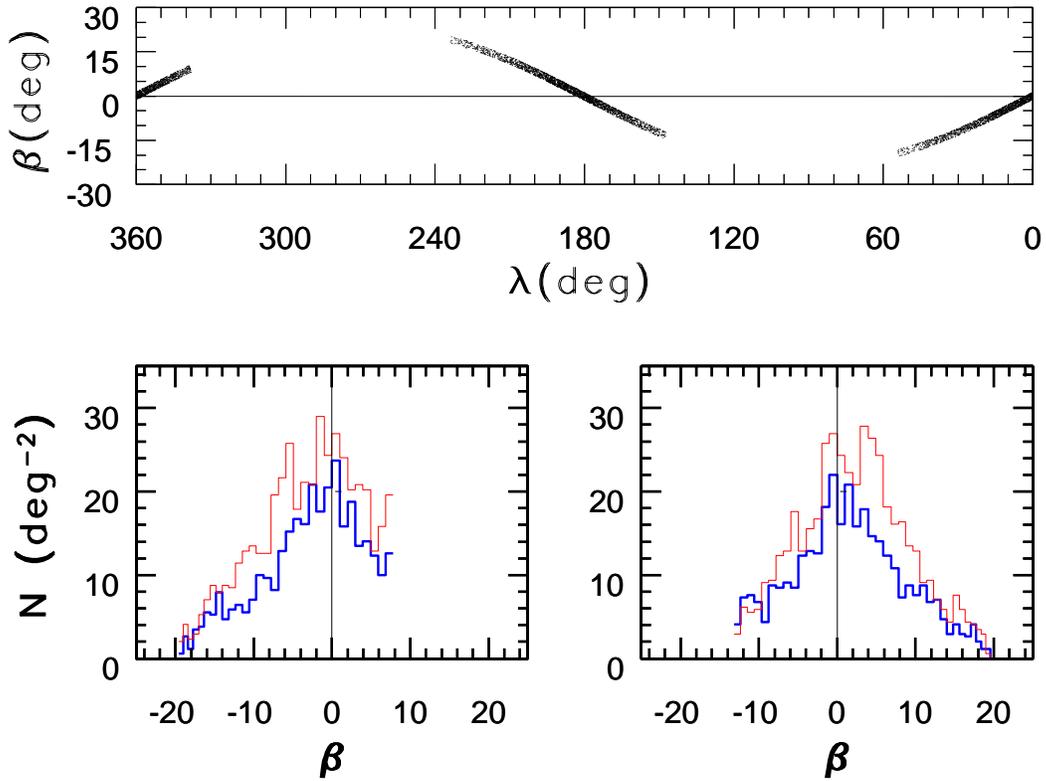}{8cm}{0}{90}{90}{-270}{-330}
\caption{The ecliptic latitude distribution for 10,678 asteroids with \r $<$ 21.5.
The top panel shows the observed angular (sky) distribution. The two bottom panels
show the dependence of the observed surface density on ecliptic latitude for
the Fall (left, $\lambda \about 0^\circ$) and Spring (right, $\lambda \about 180^\circ$)
samples. The two curves in each panel correspond to asteroids with
$a^* < 0$ (thick line) and with $a^* > 0$ (thin line).
\label{betadist}
}
\end{figure}

Figure \ref{betadist} shows the dependence of the observed asteroid surface (sky)
density on ecliptic latitude. The top panel shows the distribution of the 10,678
asteroids with \r $<$ 21.5 in the ecliptic coordinate system, where each object
is shown as a dot. The bottom two panels show the surface density vs. ecliptic
latitude where the thick lines correspond to $a^* < 0$ and thin lines to $a^* >0$.
The bottom left panel shows the results for the Fall sample ($\lambda \about 0^\circ$)
and the bottom right panel shows the results for the Spring sample $\lambda \about
180^\circ$).

It is evident that the ecliptic latitude distribution of the main belt asteroids is
not very dependent on their color. The distribution of the Spring sample is centered
on $\beta \about  2.0^\circ$, and the distribution of the Fall sample is centered
on $\beta = -2.0^\circ$, in agreement with the distribution of catalogued asteroids
(see Appendix B). The density of asteroids decreases rapidly with increasing ecliptic
latitude and drops to below 1 deg$^{-2}$ for $\beta > 20^\circ$. We do not detect
a single moving object in 16.2 deg$^2$ of sky with $\beta \about$80$^\circ$ (run 1336,
see Table 1).

The highest density of objects brighter than $r^* = 21.5$ is $46 \pm 2$ deg$^{-2}$
(including both color types). The mean density integrated over ecliptic latitudes
is 780 asteroids per degree of the ecliptic longitude (determined by counting observed
asteroids and accounting for all incompleteness effects). Assuming that this result
is applicable to the entire asteroid belt (the observed numbers of asteroids agree to
within 1\% between the Fall and Spring subsamples), we estimate that there are \about
280,000 asteroids brighter than $r^* = 21.5$ (observed near opposition). This estimate
implies that SDSS will observe $\about 100,000$ asteroids by its completion (see Figure
\ref{SDSSfoot}). We note that a fraction of these observations may be measurements of
the same asteroids. This fraction depends on the details of which areas of sky are
observed when, and cannot be estimated beforehand.

\subsubsection{    The Apparent Magnitude Distribution }

The brightness distribution of asteroids can be directly transformed into
their size distribution if all asteroids have the same albedo, and either have the
same heliocentric distance, or the size distribution is a scaleless power
law. While none of these conditions is true, we discuss the counts of asteroids
as a function of apparent magnitude because they clearly differ for the two
color types. The relationship between the distribution of apparent magnitudes
and heliocentric and size distributions is discussed in more detail in
Section 6 below.

\begin{figure}
\plotfiddle{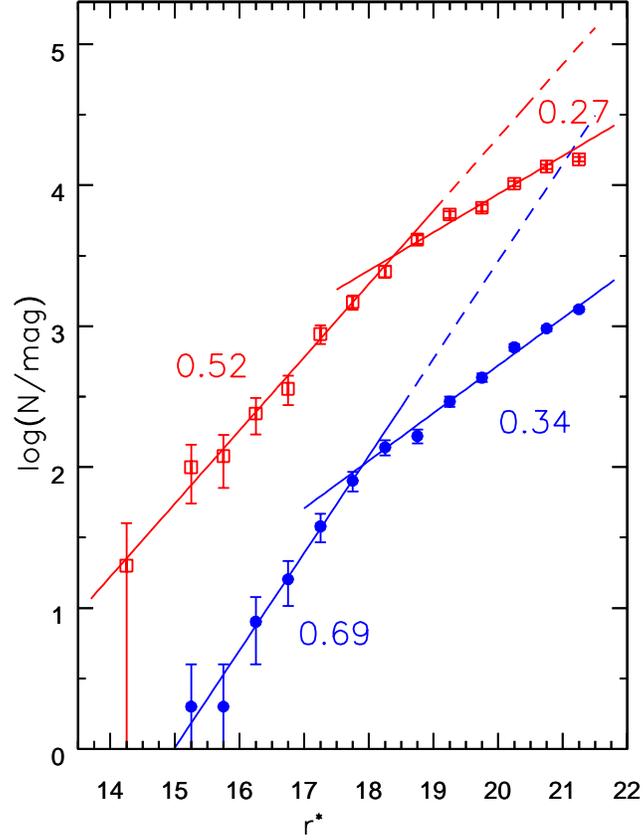}{10cm}{0}{70}{70}{-220}{-100}
\caption{The differential counts for the two types of main-belt
asteroids separated by their $a^*$ color. The open squares correspond to asteroids with
$a^* >0$ and the large dots to asteroids with $a^* <0$. The former are shifted
up by a factor of 10 for clarity. The error bars are computed by assuming Poisson
statistics. The lines show the best fit broken power laws. The numbers
show the best fit power-law indices for each magnitude range; other parameters are
listed in Table 3.
\label{countsMB}
}
\end{figure}

The measured counts for  main-belt asteroids, separated by their \a\ color,
are shown in Figure \ref{countsMB} (here we do not apply the phase angle
correction, see Section 6.1).
The circles correspond to asteroids with $\a < 0$, and the squares correspond
to the $\a > 0$ asteroids. The latter are shifted upwards by 1 dex for clarity.
A striking feature visible in both curves is the sharp change of slope around
\r \about 18-19. We find that for both color types the counts vs. magnitude
relation can be described by a broken power law
\eq{
             \log(N) \propto C_B +  k_B \r
}
for \r $< r^*_b$, and
\eq{
             \log(N) \propto C_F +  k_F \r
}
for \r $> r^*_b$, where $r^*_b$ is the magnitude for the power law break. The best
fits are shown by lines in Figure \ref{countsMB}, and the corresponding parameters
are listed in Table 3.

The changes in the slope of the counts vs. magnitude relations could be caused by
systematic effects in the detection algorithm. The dashed lines in Figure
\ref{countsMB} show the extrapolation of the bright end power-law fit, and
indicate that a decrease of detection efficiency by a factor of \about 10 is required
to explain the observed counts at \r \about 21.5.  However, such a significant
decrease of detection efficiency is securely ruled out (see Section 3.2).
In summary, the number of asteroids with \r \about 21 is roughly ten
times smaller than would expected from extrapolation of the power-law
relation observed for \r $\la$ 18.

The changes in the slope of the counts vs. magnitude relations disagree
with a simple model proposed by Dohnanyi (1969), which is based on an equilibrium
cascade in self-similar collisions, and which predicts a universal slope
of 0.5. We will discuss this discrepancy further in Sections 6 and 9.

\section{ The Ecliptic Velocity and  Determination of the Heliocentric Distance}

\subsection{   The Velocity-Based Classification of Asteroids  }

\begin{figure}
\plotfiddle{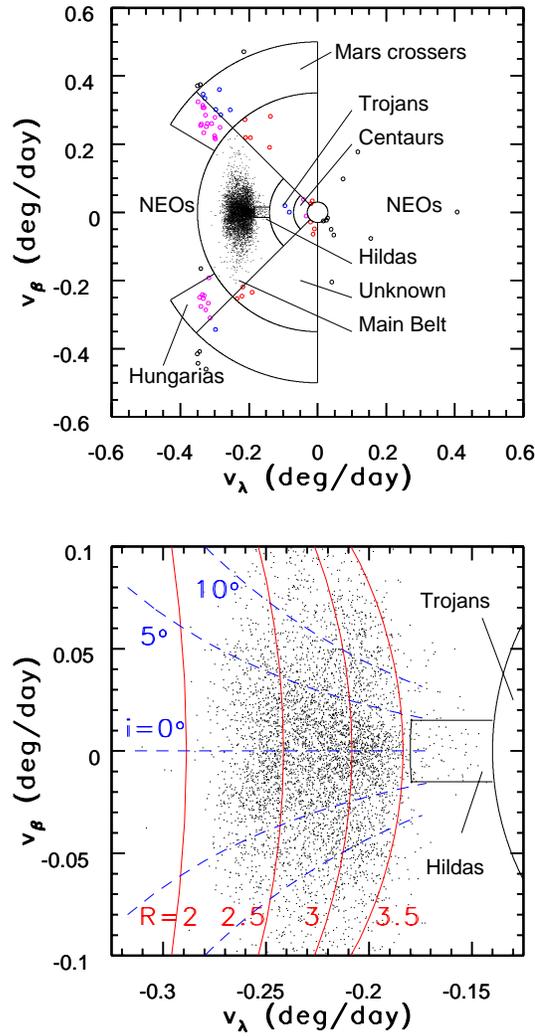}{10cm}{0}{60}{60}{-200}{-20}
\caption{The velocity distribution in ecliptic coordinates for the 5,253 reliably detected
moving objects. The lines in the top panel show the adopted separation of the objects into
main belt asteroids, Trojans, Centaurs, Hungarias, Hildas, Mars crossers, near Earth objects
(NEO), and Unknown. The lines in the bottom panel show the predictions for varying
heliocentric distance ($R$) and inclinations ($i$), as marked (see text for details).
\label{vLambdavBeta}
}
\end{figure}

Figure \ref{vLambdavBeta} shows the velocity distribution in ecliptic coordinates
for the sample of 5,253 unique asteroids. Its overall morphology is in agreement with other
studies (e.g. Scotti, Gehrels \& Rabinowitz 1991, Jedicke 1996) and shows a large
concentration of the main belt asteroids at $v_\lambda \about$ -0.22 \dd\ and
$v_\beta \about 0$. The sharp cutoff in their distribution at $v_\lambda \about$
-0.28 \dd\ is not a selection effect and corresponds to objects at a heliocentric
distance of $\about$ 2 AU. The lines in the top panel of Figure \ref{vLambdavBeta}
show the boundaries adopted in this work for separating asteroids into different
families, including the main belt asteroids, Hildas, Hungarias and Mars crossers, Trojans,
Centaurs, as well as near Earth objects (NEOs). This separation in essence reflects
different inclinations and orbital sizes of various asteroid families (for more details
see e.g. Gradie, Chapman \& Williams 1979, Zellner, Thirunagari \& Bender 1985).
These regions are modeled after the Spacewatch boundary for distinguishing NEOs from
other asteroids (Rabinowitz 1991), information provided in Jedicke (1996), and taking
into account the velocity distribution observed by SDSS.

There is some degree of arbitrariness in the proposed boundaries, for example the regions
corresponding to Hungarias and Mars crossers could be merged together. The boundary
definitions and asteroid counts for each region are listed in Table 2 (as indicated in
the table, the visual inspection shows that some objects are spurious; see \S 5.4 below).
These subsamples are used for comparative analysis of their colors and
spatial distributions in the following sections. We emphasize that this separation
cannot be used as a definitive identification of an asteroid with a particular family.
In particular, the main belt asteroids may cause significant contamination of other
regions due to the measurement scatter and their large number compared to other families.

\subsection{ The Correspondence between the Velocity and Orbital Elements }
\label{secFormulae}
Six orbital elements are required to define the motion of an asteroid. Since
the SDSS observations determine only four parameters (two sky coordinates and two
velocity components) the orbit is not fully constrained by the available data.
There are various methods to obtain approximate estimates for the orbital
parameters from the asteroid motion vectors (e.g. Bowell, Skiff, Wasserman
\& Russell 1989, and references therein). These methods provide an accuracy of
about 0.05-0.1 AU for determining the semimajor axes, and 1-5 deg. for
the inclination accuracy. As shown by Jedicke \& Metcalfe (1998), similar accuracy
can be obtained by assuming that the orbits are circular.

We follow Jedicke (1996) and derive the expressions for observed ecliptic
velocity components in terms of $R$, $i$ and $\phi$ listed in Appendix A.
We show in Appendix B that these expressions can be used to estimate the
heliocentric distance at the time of observation with an accuracy of 0.28
AU (rms). The uncertainty in the estimated heliocentric distance is significantly
larger than the uncertainty in the estimated semimajor axes (0.07 AU) because
asteroid orbits in fact have considerable eccentricity (\about 0.10-0.15).
However, we emphasize that the estimates for heliocentric distance and
semimajor axes are simply proportional to each other (for more details
see Appendix B). The uncertainty of 0.28 AU in the estimate of heliocentric
distance contributes an uncertainty of 0.5-0.8 mag. in the absolute
magnitude (see eq. \ref{distmod}).

The bottom panel in Figure \ref{vLambdavBeta} magnifies the part of
the top panel which includes the main belt and Hilda asteroids. The
dashed lines show the loci of points with $i$ ranging from $-15^\circ$
to $15^\circ$ in steps of $5^\circ$, and the solid lines show loci of
points with $R$ = 2, 2.5, 3 and 3.5 AU (the inclination is a positive
quantity by definition; here we use negative values as a convenient
way to account for different orbital orientations). They are computed by using
eqs. \ref{veq1} and \ref{veq2} listed in Appendix A with $\beta = 0$ and
$\phi = 0$. By using equations $2-6$ with the proper $\beta$ and $\phi$,
we compute $R$ and $i$ for all 5,125 main belt asteroids in our sample.

\subsection{ The Relation between Color and Heliocentric Distance }

\begin{figure}
\plotfiddle{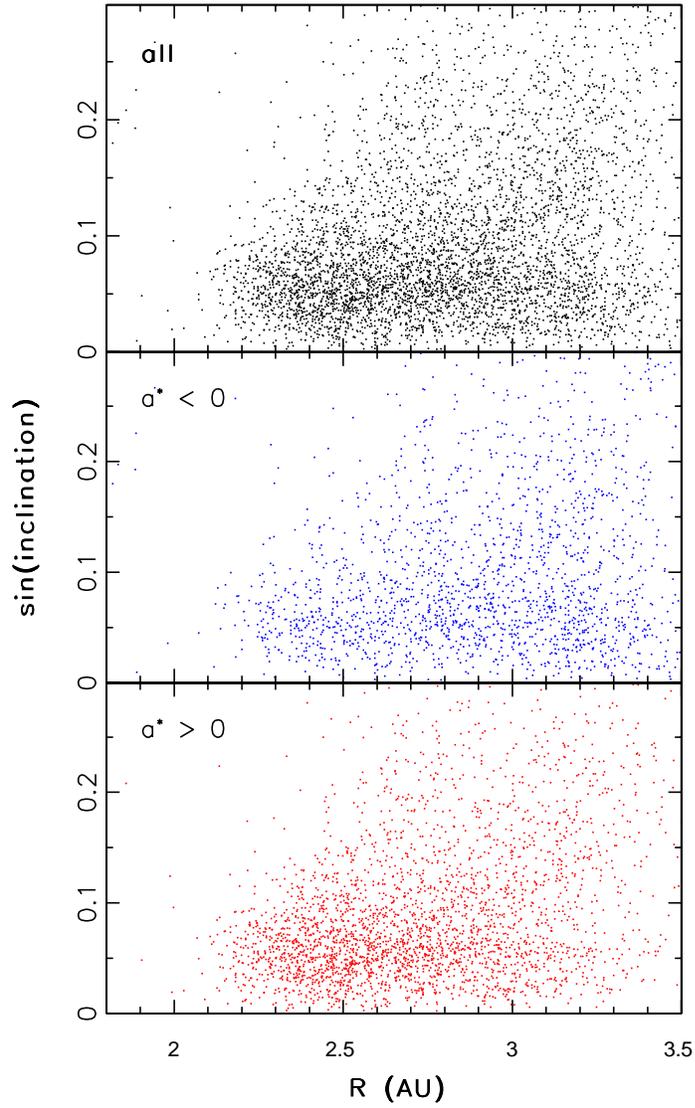}{11cm}{0}{60}{60}{-190}{-20}
\caption{The distribution of 5,125 main belt asteroids in the sin($i$) vs. $R$ plane.
The top panel shows the whole sample, while the other two panels show each color type
separately, as marked. Note that the red asteroids (bottom panel) tend to have
smaller heliocentric distances than do the blue asteroids (middle panel).
\label{iR}
}
\end{figure}

\begin{figure}
\plotfiddle{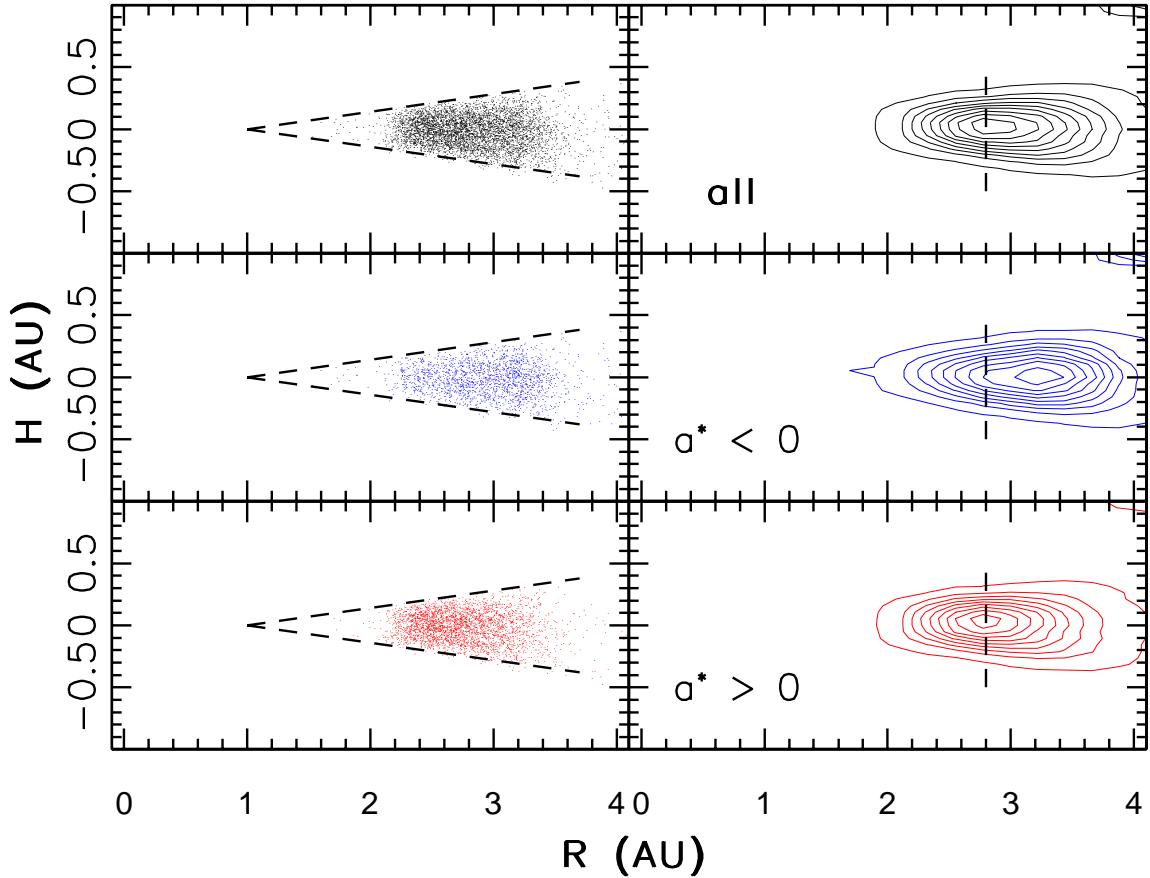}{11cm}{0}{90}{90}{-280}{-300}
\caption{The cross section of the asteroid belt. In the left column each asteroid is
shown as a dot, and in the right column the distribution is outlined by isodensity
contours. The top panels show the whole sample, and the other panels show each color
type separately, as marked. The two dashed lines in the left column are drawn at
$\beta = \pm 8^\circ$ and show the observational limits. The vertical dashed lines
in the right column show the position of the highest asteroid density for the red
subsample and are added to guide the eye. Note that the distributions of blue and red
asteroids are significantly different.
\label{HR}
}
\end{figure}

The $i-R$ distribution of the 5,125 main belt asteroids
is shown in the top panel in Figure \ref{iR}. The overall morphology
is in agreement with other studies (e.g. see figure 1 of Zappala
{\em et al.} 1990). The multi-color SDSS data allow the separation
of asteroids into two types,
and the middle and bottom panels show the distribution of each color type
separately. Because the heliocentric distance estimates have an accuracy
of 0.28 AU, these data cannot resolve narrow features\footnote{Note that,
e.g., the Kirkwood gaps would not be resolved even if we were to use true
$R$ values, since they are gaps in the distribution of asteroid semi-major
axis, not $R$. The $R$ distribution is smeared out because the orbits
are randomly oriented ellipses.}. Nevertheless, it is evident that
the red asteroids tend to be closer to the Sun. Another illustration
of the differences in the distribution of the two color types is shown
in Figure \ref{HR} which displays the cross-section of the asteroid belt:
the horizontal axis is the heliocentric distance and the vertical axis
is the distance from the ecliptic plane. The top panel shows the distribution
of all asteroids and the middle and bottom panels show the distribution of
each color type separately. The dashed lines in the left column are drawn
at $\beta = \pm 8^\circ$\, and show the observational limits.
The dashed lines in the right column mark the position of the maximum
density and are added to guide the eye. Note that the maximum
density for blue asteroids (middle panel) lies about 0.4 AU further out
than for red asteroids.

\begin{figure}
\plotfiddle{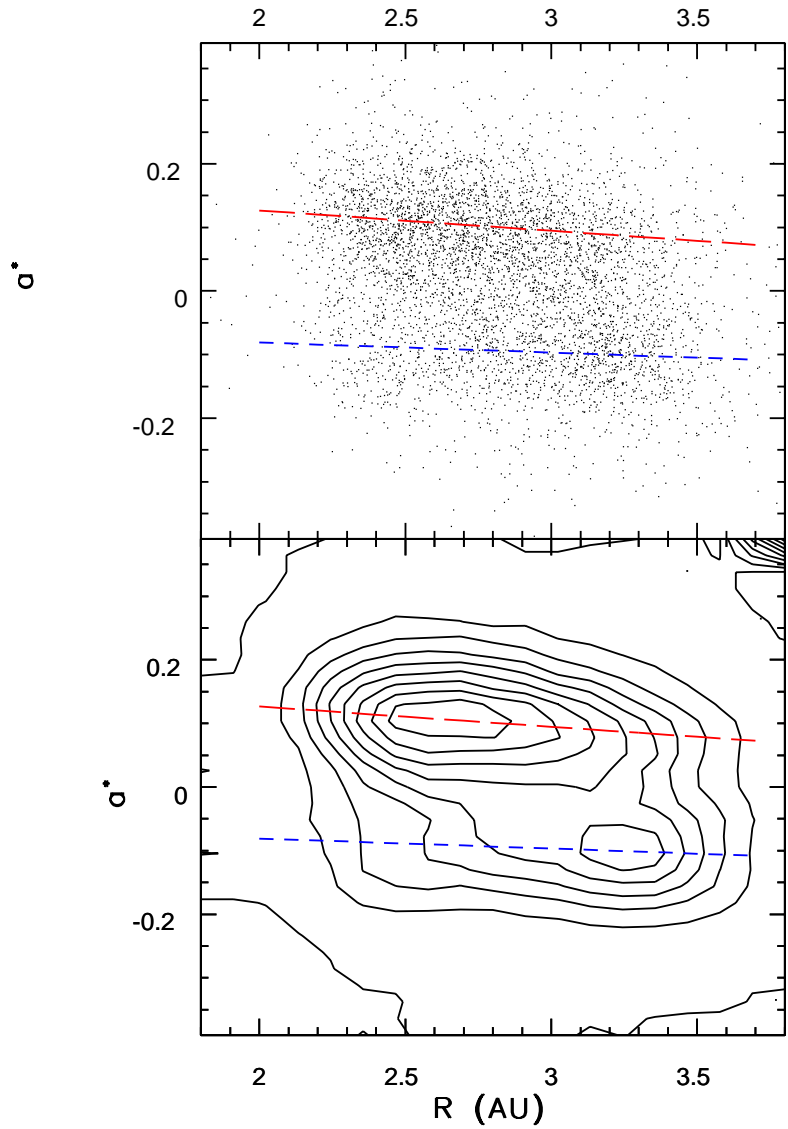}{11cm}{0}{70}{70}{-230}{-100}
\caption{The color--heliocentric distance dependence in asteroid belt.
In the top panel each asteroid is shown as a dot, and in the bottom panel the distribution
is outlined by isodensity contours. Two dashed lines are fitted separately for
$a^* < 0$ and $a^* \ge 0$ subsamples. Note that each subsample tends to become slightly
bluer as the heliocentric distance increases.
\label{avsR}
}
\end{figure}

If the mean color varies strongly with heliocentric distance, splitting
the sample by color would split the sample by heliocentric distance as
well, explaining the shifting maxima in Figure \ref{HR}. However, the
dependence of color on the heliocentric distance shown in
Figure \ref{avsR} indicates that this is not the case. In the top panel
each asteroid is marked as a dot, and the bottom panel shows the isodensity
contours. It is evident that the bimodal color distribution persists
at all heliocentric distances. The distributions of both types have a well
defined maximum, evident in the bottom panel, which may be interpreted as
the existence of two distinct belts: the inner, relatively wide belt dominated
by red (S-like) asteroids and the outer, relatively narrow belt, dominated
by blue (C-like) asteroids. We emphasize that the detailed radial distribution
of asteroids in this diagram is strongly biased by the heliocentric distance
dependent faint limit in absolute magnitude, and these effects will be discussed
in more quantitative detail in Section 6. Nevertheless, the {\em difference} between
the two types is not strongly affected by this effect, and the remarkable
split of the asteroid belt into two components is a robust result.

\begin{figure}
\plotfiddle{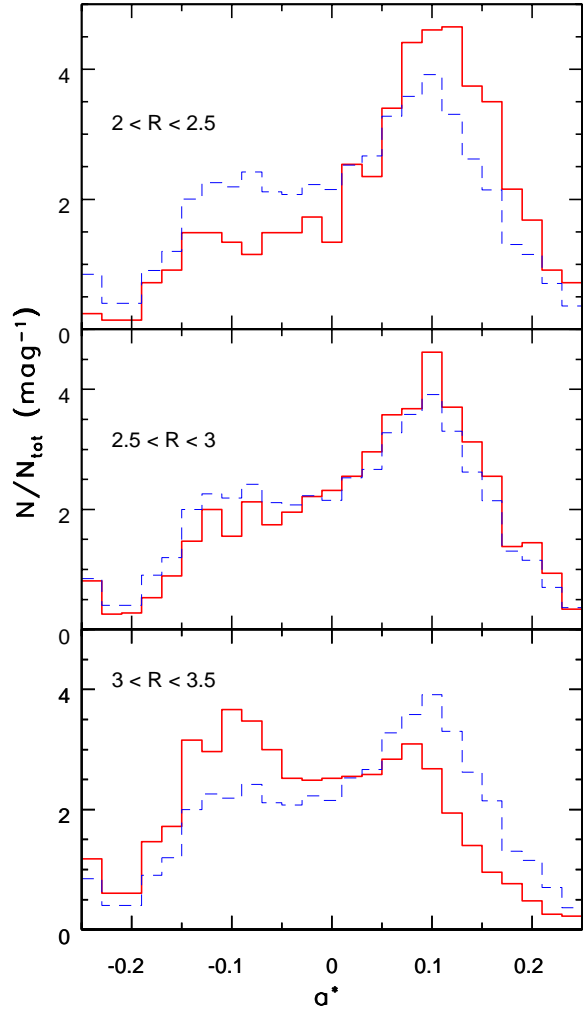}{11cm}{0}{60}{60}{-200}{-50}
\caption{The color distributions  for three main belt asteroid subsamples selected by
their heliocentric distance, as indicated in the panels. Each panel compares the color
distribution of a given subsample, shown as the solid line, to the color distribution
of all main belt asteroids, shown as the dashed line.
\label{histavsR}
}
\end{figure}

Another notable feature displayed by the data is that the median color for
each type becomes bluer with the heliocentric distance\footnote{A similar
result was obtained for the S asteroids by Dermott, Gradie \& Murray (1985)
who found that the mean U-V color of 191 S asteroids becomes bluer with $R$
by about 0.1 mag across the asteroid belt.}. The dashed lines
in Figure \ref{avsR} are linear fits to the color-distance dependence,
fitted for each color type separately. The best fit slopes are (0.016$\pm$0.005)
mag AU$^{-1}$ and (0.032$\pm$0.005) mag/AU for blue and red types, respectively.
Due to this effect, as well as to the varying number ratio of the asteroid
types, the color distribution of asteroids depends strongly on heliocentric
distance. Figure \ref{histavsR} compares the \a-color histograms for three
subsamples selected by heliocentric distance: $2 < R < 2.5, 2.5 < R < 3$, and
$3 < R < 3.5$. The thick solid lines show the color distribution of each
subsample, and the thin dashed lines show the color distribution of the whole
sample.

\subsection{ The Reliability of Asteroids with Unusual Velocities  }

While the sample reliability is estimated to be 98.5\% in Section \ref{comrel},
it is probable that it is much lower for asteroids with unusual velocities.
To estimate the fraction of reliable detections for such asteroids, we visually
inspect images for the 128 asteroids which are not classified as main belt
asteroids. As a control sample, and also for an additional reliability estimate,
we inspect 50 randomly selected main belt asteroids, and the 50 brightest
and 50 faintest main belt asteroids. While the visual inspection of 278 images
for the motion signature may seem a formidable task, it is indeed quite simple
and robust. Since asteroids move they can be easily recognized by their peculiar
colors in the $g'-r'-i'$ color composites. Moving objects appear as aligned
green--red--blue ``stars'', with the blue-red distance three times
larger than the green-red distance (due to filter spacings, see Section
\ref{DMflag}).

We find that neither of the two candidates for Centaurs are real. About 63\% of
NEOs (12/19), 57\% of Mars crossers (4/7), 50\% of Trojans (1/2) and 43\% of
Unknown (6/14) are not real. The contamination of the remaining subsamples is
lower; only 5\% for Hildas (3/59), while all 25 Hungarias are real. The majority
of false detections are either associated with saturated stars, or are close to
the faint limit. For all 3 subsamples with main belt asteroids the contamination
is 2\% (there is 1 instrumental effect per 50 objects in each class). Since the
fraction of objects not classified as main belt asteroids is very small (\about 2\%),
these results are consistent with estimates described in \S 3.2.

\subsection{  The Colors of Asteroid Families Other than Main Belt  }

The color distributions of various families are particularly useful for
linking them to other asteroid populations and constraining theories
for their origin. For example, one possible source of the NEOs is the
main belt asteroids near the 3:1 mean motion resonance with Jupiter.
These asteroids have chaotic orbits whose eccentricities increase until
they become Mars crossing, after which they are scattered into the inner
solar system (Wisdom 1986). If this scenario is true, the colors of
both the NEOs and Mars crossers should resemble the colors of main belt
asteroids, but show more similarity to the C-type asteroids than to the whole
sample due to the radial color gradient. An alternative source of NEOs
is extinct comet nuclei; this hypothesis predicts a wider color range
than observed for asteroids. Yet another hypothesis was forwarded by
Bell {\em et al.} (1989) who predicted that NEOs should be more similar
to S type than to C type asteroids. While there seems to be more evidence supporting
the first hypothesis (Shoemaker et al. 1979), the analyzed samples are small.

We find that \about 70\% of the visually confirmed NEOs (5/7), Unknown (6/8)
Hungarias (17/25), and Mars crossers (2/3) belong to the blue type ($a^* < 0$).
More than half of the remaining 30\% are typically borderline red ($a^* < 0.05$).
This is in sharp contrast with the overall color distribution of main belt asteroids,
where the fraction of blue asteroids is only \about 40\%, and the fraction
of asteroids with $a^* < 0.05$ is \about 47\%. The result for NEOs implies
that their source is the asteroid belt, rather than extinct comet nuclei.
Furthermore, it supports the hypothesis they originate in the outer part of
the asteroid belt\footnote{We show in Section 6.5 that the fraction of blue
asteroids in the outer belt is approaching 75\% (see also Figure \ref{histavsR}).},
contrary to the prediction by Bell {\em et al.}. The colors of Hungarias
and NEOs are similar, which is in agreement with the notion that Hungarias are
an intermediate phase for the main belt asteroids on their route to becoming
the NEOs. While it is not clear what the origin of asteroids from the
``Unknown'' region is, they are as blue as are Hungarias and NEOs.

The only confirmed Trojan candidate is distinctively red ($a^* = 0.29$).
Such a red color seems to agree with the colors of some Centaurs (Luu \& Jewitt 1996)
though it is hard to judge the significance of this result. We note that the Kuiper
Belt object candidate discussed in Section 8 is also significantly redder
($a^* \about 0.6$) than the main belt color distribution.

\section{     The Heliocentric Distance and Size Distributions    }

The size distribution of asteroids is one of most significant observational
constraints on their history (e.g. Jedicke \& Metcalfe 1998, and references
therein). It is also one of the hardest quantities to determine observationally
because of strong selection effects. Not only that the smallest observable
asteroid size in a flux limited sample strongly varies with the heliocentric
distance, but the conversion from the observed magnitude to asteroid size
depends on the, usually unknown, albedo. Assuming a mean albedo for all asteroids
may lead to significant biases since the mean albedo depends on the
heliocentric distance due to varying chemical composition. Because of its
multi-color photometry, SDSS provides an opportunity to disentangle these effects
by separately treating each of the two dominant classes, which are known to
have rather narrow albedo distributions (see Section 4.1.2).

For a fixed albedo, the slope of the counts vs. magnitude relation,
$k$, is related to the power-law index, $\alpha$, of the asteroid differential
size distribution, $dN/dD = n(D) \propto D^{-\alpha}$, via
\eq{
\label{k2alpha}
                         k = 0.2 (\alpha-1).
}
This simple relation assumes that the size distribution is a scale-free power law
independent of distance, which results in a linear relationship between log(counts)
and apparent magnitude. However, the observed change of slope around
\r \about 18 in the differential counts of asteroids (discussed in Section
4.2.3) introduces a magnitude scale which prevents a straightforward
transformation from the counts to size distribution.

In the limit where a single power law is a good approximation to the observed
counts, the above relation can still be used and shows that the power law index
$\alpha$ of the asteroid size distribution is \about 4 at the bright end and
\about 2.5 at the faint end for both color families (see Table 3). Since these
values may be somewhat biased by the phase angle effect, and the strong dependence of
the faint cutoff for absolute magnitude on heliocentric distance, in this section
we perform a detailed analysis of the observed counts. By applying techniques
developed for determining the luminosity function of extragalactic sources,
we estimate unbiased size and heliocentric distance distributions for the two
asteroid types.

\subsection{          The Correction for the Phase Angle         }

The observed apparent magnitude of an asteroid strongly depends on the phase
angle, $\gamma$, the angle between the Sun and the Earth as viewed from the
asteroid
\eq{
       \gamma = {\rm arcsin} \left( {\sin(\phi) \over R} \right).
}
We first determine the ``uncorrected" absolute magnitude from
\eq{
\label{distmod2}
                       \r(1,0) = \r - 5 \log(R\rho),
}
where $\rho$ is the Earth-asteroid distance expressed in AU,
\eq{
           \rho = - \cos(\phi) + \left(\cos(\phi) + R^2 - 1\right)^{1/2},
}
and correct it for the phase angle effect by subtracting
\eqarray{
      \Delta \r(1,0) &=& 0.15 |\gamma|, \,\,\, {\rm for} \,\, |\gamma| \le 3^\circ
                         \non
      \Delta \r(1,0) &=& 0.45 + 0.024(|\gamma|-3), \non
                         && {\rm for} \,\, |\gamma| > 3^\circ.
}
The adopted correction is based on the observations of asteroid 951 Gaspra
(Kowal 1989) and may not be applicable to all asteroids. In particular,
the true correction could depend on the asteroid type (Bowel \& Lumme
1979). However, as shown by Bowel \& Lumme the differences between
the C and S types for $\gamma \la 6^\circ$ are not larger than
\about 0.02 mag. The strong dependence of the correction on $\gamma$ for
small $\gamma$ is known as the ``opposition effect"; the adopted value
agrees with contemporary practice (e.g. Jedicke \& Metcalfe 1998). Note that
although the phase angle correction is somewhat uncertain, it is smaller
than the error in $\r(1,0)$ due to uncertain $R$ (0.5-0.8 mag) even for
objects at the limit of our sample ($\phi=15^\circ$, corresponding to
$\gamma \about 6^\circ$).

\subsection{ The Dependence of Color on Phase Angle and Absolute Magnitude }

The dimming due to a non-zero phase angle need not be the same at all
wavelengths, and in principle the colors could also depend on the observed
phase angle due to so-called differential albedo effect (Bowel \& Lumme
1979, and references therein). While such an effect does not have direct
impact on the determination of size and heliocentric distance distributions,
it would provide a strong constraint on the reflectance properties of
asteroid surfaces.

\begin{figure}
\plotfiddle{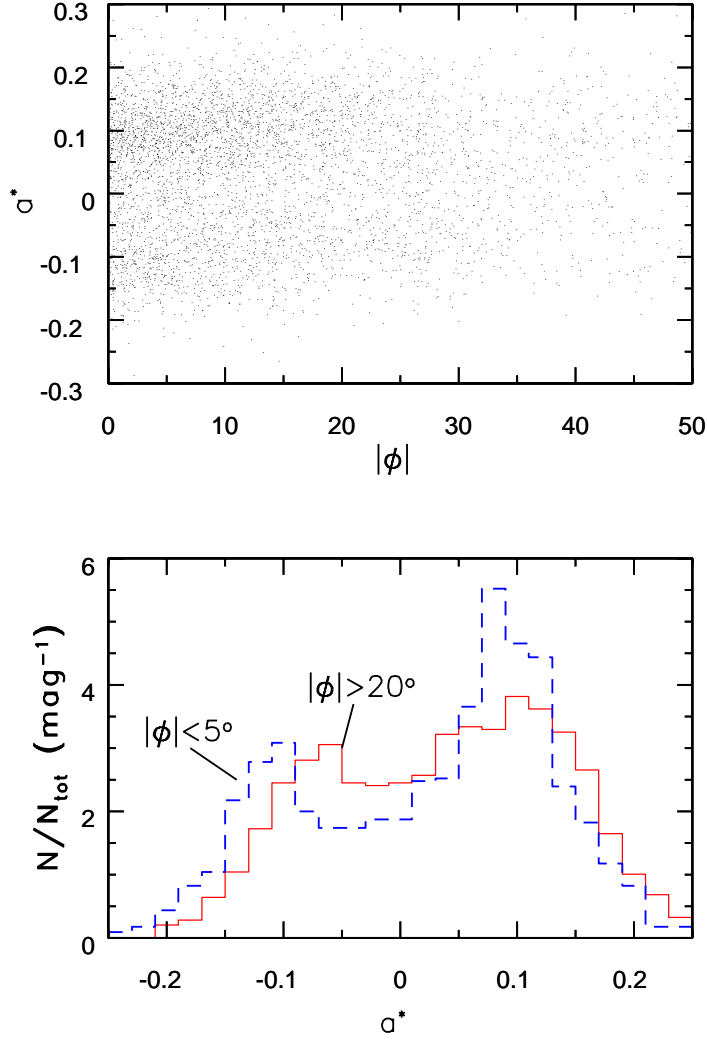}{13cm}{0}{70}{70}{-240}{-120}
\caption{The dependence of asteroid color on the phase angle for 4591 asteroids
with photometric errors less than 0.05 mag. in the $g'$, $r'$, and $i'$ bands
is shown in the top panel. The bottom panel shows the color distribution for
1150 asteroids observed close to opposition by the dashed line, and the
color distribution for 1244 asteroids observed at large angles from the
antisun by the solid line. The blue asteroids ($a^* < 0$) are redder by
$\sim 0.05$ mag when observed at large phase angles, while the color
distribution for red asteroids ($a^* \ge 0$) widens for large
phase angles, without a change in the median value.
\label{oposition}
}
\end{figure}

We investigate the dependence of color $a^*$ on angle $\phi$ by
comparing the color histograms for objects observed close to opposition
and for objects observed at large $\phi$. The top panel in Figure \ref{oposition}
shows the $a^*$ color for 4591 asteroids with photometric errors less than
0.05 mag. in the $g'$, $r'$, and $i'$ bands, as a function of the opposition
angle $\phi$. The bottom panel shows the color distribution for 1150 asteroids
observed close to opposition by the dashed line, and the color distribution
for 1244 asteroids observed at large $\phi$ by the solid line. We find a
significant difference in the $a^* - \phi$ relation for the two types of
asteroids: the blue asteroids ($a^* < 0$) are redder by $\sim 0.05$ mag when
observed at large phase angles, while the color distribution for red asteroids
($a^* \ge 0$) widens for large phase angles, without a corresponding change
of the median value. The same analysis applied to other colors shows that the
dependence on $\phi$ is the strongest for the \gr\ color, and insignificant for
other colors. While the magnitude of this effect seems to agree with the value
of 0.0015 mag deg$^{-1}$ observed for the Johnson $B-V$ color (Bowel \& Lumme 1979),
the difference in the behavior of the two predominant asteroid classes has not
been previously reported.

\begin{figure}
\plotfiddle{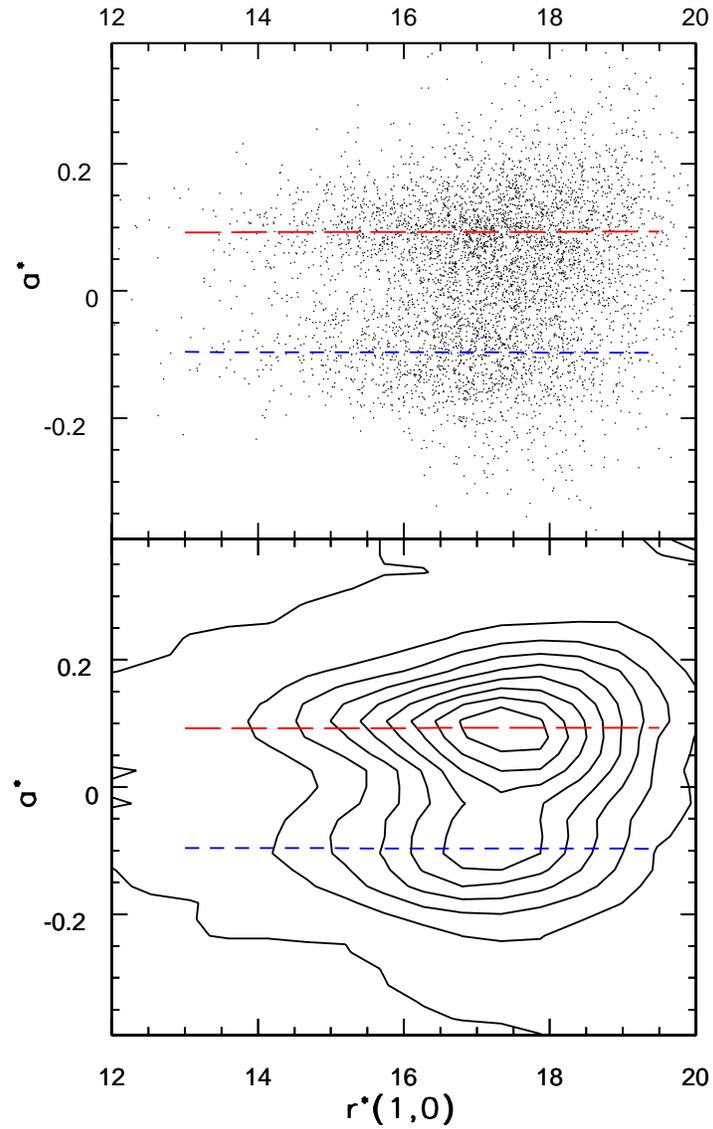}{11cm}{0}{70}{70}{-230}{-100}
\caption{The color--absolute magnitude dependence for main belt asteroids.
In the top panel each asteroid is shown as a dot, and in the bottom panel
the distribution is outlined by isodensity contours. Two dashed lines are fitted
separately for the $a^* < 0$ and $a^* \ge 0$ subsamples. There is no significant
correlation between the asteroid color and absolute magnitude.
\label{avsr10}
}
\end{figure}

Another effect which could bias the interpretation of the asteroid counts
is the possible dependence of the asteroid color on absolute magnitude.
Since for a given albedo the absolute magnitude is a measure of asteroid size,
the dependence of the asteroid surface properties on its size could be
observed as a correlation between the color and absolute magnitude.
Figure \ref{avsr10} shows the distribution of asteroids in the color vs. absolute
magnitude diagram for each color type separately (the top panel shows asteroids
as dots, and the bottom panel shows isodensity contours). The absolute magnitude
was calculated by using eq.\ref{distmod2}, and the heliocentric distance as
described in Section 5.2. The two dashed lines
are fitted separately for the $a^* < 0$ and $a^* \ge 0$ subsamples. There is no
significant correlation between the asteroid color and absolute magnitude
(the slopes are consistent with 0, with errors less than 0.005 mag mag$^{-1}$).

\subsection{The Absolute Magnitude and Heliocentric Distance Distributions}

The differential counts of asteroids can be used to infer their size
distribution if all asteroids have the same albedo. However, while the
albedo within each of the two major asteroid types has a fairly narrow
distribution (the scatter is \about 20\% around the mean), the mean
albedo for the two classes differs by almost a factor of 4 (Zellner 1979,
Tedesco {\em et al.} 1989). The fact that the composition of the asteroid belt,
and thus the mean albedo, varies with heliocentric distance has been a significant
drawback for the determination of the size distribution (e.g. Jedicke \&
Metcalfe 1998). SDSS data can remedy this problem because the colors are
sufficiently accurate to separate asteroids into the two classes.
We assume in the following analysis that each color type is fairly well
represented by a uniform albedo. We determine various distributions
for each type separately, but also for the whole sample in order to allow
comparison with previous work.

\subsection{  Determination of the Heliocentric Distance and Size Distributions }

For a given apparent magnitude limit (here \r = 21.5), the faint limit on
absolute magnitude is a strong function of heliocentric distance (see
e.g. eq.~\ref{distmod}). For example, at $R$=2 AU, \r =21.5 corresponds to
\r(1,0) = 19, while at $R$=3.5 AU, the limit is only 16.8. When interpreting
the observed distribution of objects in the \r(1,0) vs. $R$ plane, this
effect must be properly taken into account. This problem is mathematically
equivalent to the well studied case of the determination of the luminosity
function for a flux limited sample (e.g. Fan {\em et al.} 2001b and references
therein).

\begin{figure}
\plotfiddle{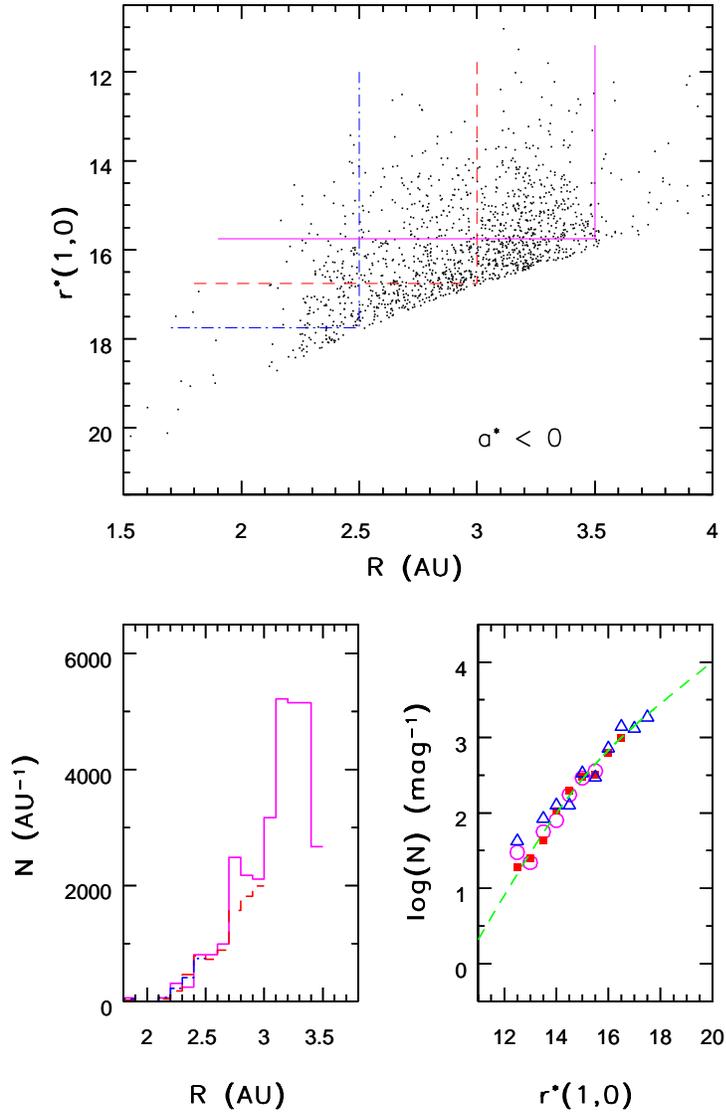}{11cm}{0}{60}{60}{-190}{-30}
\caption{The top panel shows the dependence of absolute magnitude on
heliocentric distance for blue asteroids. The three sets of lines
mark boundaries of regions used to compute the heliocentric distance
and absolute magnitude distributions shown in the bottom two panels
(for details see text).
\label{figLFblue}
}
\end{figure}

\begin{figure}
\plotfiddle{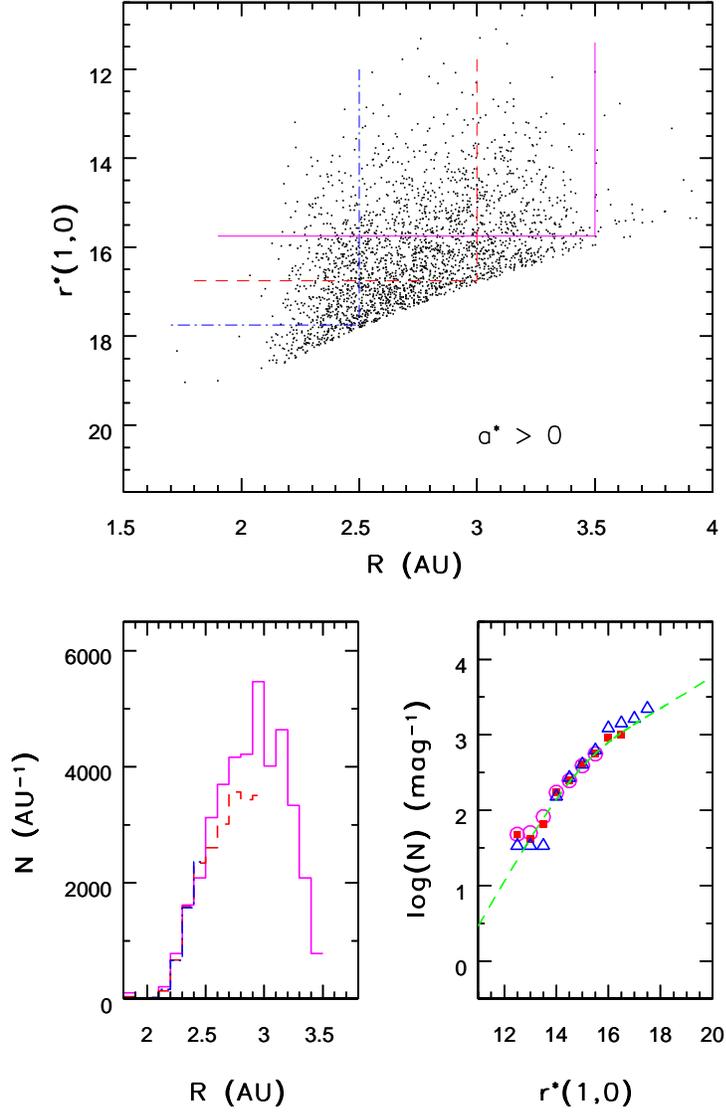}{11cm}{0}{60}{60}{-190}{-30}
\caption{The same as previous figure except for red asteroids.
\label{figLFred}
}
\end{figure}

The top panels in Figures \ref{figLFblue} and \ref{figLFred} show the
\r(1,0) vs. $R$ distribution of the two color-selected samples. We seek to
determine the marginal distributions of sources in the \r(1,0) and $R$ directions.
It is usually assumed that these two distributions are uncorrelated (i.e.
the \r(1,0) distribution is the same everywhere in the belt, and the radial
distribution is the same for all \r(1,0)). The sample discussed here is
sufficiently large to test this assumption explicitly, and we do so
using two methods.

First, we simply define three regions in the \r(1,0) vs. $R$ plane, outlined
by different lines in the top panels, and compare the properly normalized
histograms of objects for each of the two coordinates. If the distributions
of \r(1,0) and $R$ are uncorrelated, then all three histograms must agree.
The results are shown in the bottom panels; the left panels show $R$ histograms,
and the right panels show \r(1,0) histograms (shown by points, the dashed lines
will be discussed further below). In all four panels, the three histograms are
consistent within errors (not plotted for clarity). That is, there is no
evidence that the two distributions are correlated.

Another method to test for correlation between the two distributions
is based on Kendall's $\tau$ statistic (Efron \& Petrosian 1992), and
is well described by Fan {\em et al.} 2001b (see \S 2.2). For uncorrelated
distributions the value of this statistic should be much smaller than unity.
We find the values of 0.11 and 0.07, for the blue and red samples respectively,
again indicating that \r(1,0) and $R$ are uncorrelated.

The $R$ and \r(1,0) distributions plotted in the bottom panels of Figures
\ref{figLFblue} and \ref{figLFred} are not optimally determined because
they are based on only small portions of the full data set, and also
suffer from binning. Lynden-Bell (1971) derived an optimal method to determine
the marginal distributions for uncorrelated variables that does not require
binning and uses all the data.  We implemented this method, termed
the $C^-$ method, as described in Fan {\em et al.} (2001b). The output is the
cumulative distribution of each variable, evaluated at the measured value
for each object in the sample. One weakness of this method is that the uncertainty
estimate for the evaluated cumulative distributions is not available.
We determine the differential distributions by binning the sample in
the relevant variable, and assume the Poisson statistics based on the number
of sources in each bin to estimate the errors.

\subsection{      The Asteroid Heliocentric Distance Distribution          }

\begin{figure}
\plotfiddle{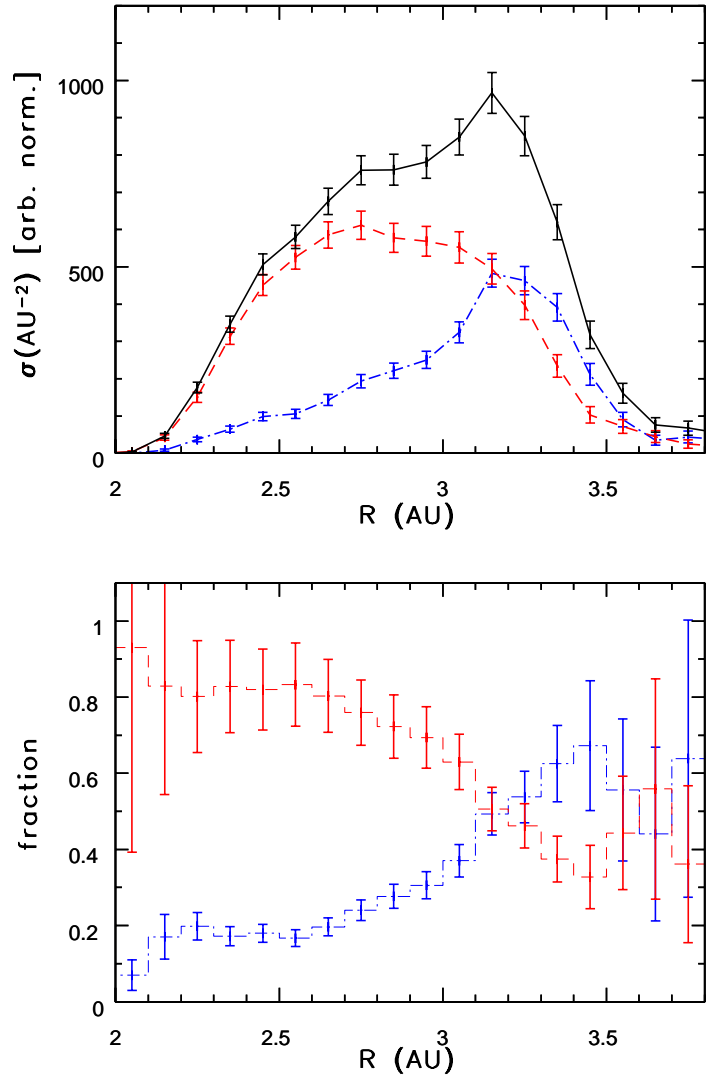}{11cm}{0}{60}{60}{-190}{-30}
\caption{The top panel shows the relative surface density of asteroids as
a function of heliocentric distance. The solid line corresponds to all asteroids,
and the dashed and dot-dashed lines correspond to red and blue asteroids,
respectively. The bottom panel shows the fractional contribution of each
type in an absolute magnitude limited sample to the total surface density.
\label{sigmaR}
}
\end{figure}

The top panel in Figure \ref{sigmaR} shows the surface density of asteroids
as a function of heliocentric distance. The surface density is computed
as the differential marginal distribution in the $R$ direction, divided
by $R$. The overall normalization is arbitrary and will be discussed in
the next Section. The solid line corresponds to the full sample, the dashed
line to red asteroids, and the dot-dashed line to blue asteroids. The relative
normalization of the two types corresponds to samples limited by absolute
magnitude. The distribution for the full sample is in good agreement
with the results obtained by Jedicke \& Metcalfe (1998, figure 3).
In particular, the depletion of asteroids near $R = 2.9$ spanning the
range from the 5:2 to 7:3 mean motion resonance with Jupiter is clearly
visible.

Figure \ref{sigmaR} confirms earlier evidence (see e.g. Figure
\ref{avsR}) that the heliocentric distance distributions of the two types
of asteroids are remarkably different. This difference may be interpreted
as two distinct asteroid belts: the inner belt dominated by red (S-like)
asteroids, centered at $R$ \about 2.8 AU, with a FWHM of 1 AU, and the outer
belt, dominated by blue (C-like) asteroids, centered at $R$ \about 3.2 AU,
with a FWHM of 0.5 AU. This distinction is also motivated by the strikingly
different {\em radial shapes} of the surface density. It may be possible to
derive strong constraints on the asteroid history by modeling the curves
shown in Figure \ref{sigmaR}, but this is clearly beyond the scope of this paper.

The distribution of red asteroids is fairly symmetric around $R \about 2.8$ AU,
while the distribution of blue asteroids is skewed and extends all the way
to $R \about 2$ AU. The overall shape indicates that it may represent a sum
of two roughly symmetric components, with the weaker component centered at
$R$ \about 2.5 AU, and the stronger component centered at $R$ \about 3.2 AU.
Gradie \& Tedesco (1982) showed that the distribution of M type asteroids
has a local maximum around $R \about 2.5$ AU. Furthermore, these asteroids have
colors similar to C type asteroids (see Figure \ref{SMASS}) and can be
distinguished only by their large albedo. Thus, it may be possible that
$\la 20\%$ of blue asteroids found at small $R$ ($\la 3$ AU) are dominated by 
M type asteroids.

The bottom panel shows the fractional contribution of each type to the total
surface density. The number ratio of the red to blue type changes from 4:1 in
the inner belt to \about 1:3 in the outer belt (note that the number ratio
for the outer belt has a large uncertainty). It should be emphasized that
the amplitudes of the distributions of asteroids shown in the top panel, and
thus the resulting number fractions, are strongly dependent on the sample
definition. The plotted distributions, and the change of number fractions from 4:1
to 1:3, correspond to an equal cutoff in absolute magnitude. However, it could
be argued that the cutoff should correspond to the same size limit. In such
a case the red sample should have a brighter cutoff due to its larger albedo,
which decreases its fractional contribution. For example, adopting a 1.4 mag
brighter cutoff (see Section 4.1.2) for the red asteroids (\r=20.1 instead of
\r=21.5) changes their fraction in the observed sample from 60\% to 38\%.
We further discuss the number ratios of the two dominant asteroid types
in the next two sections.

\subsection{     The Asteroid Absolute Magnitude Distribution           }

\subsubsection{   The Cumulative Distribution and Normalization         }

\begin{figure}
\plotfiddle{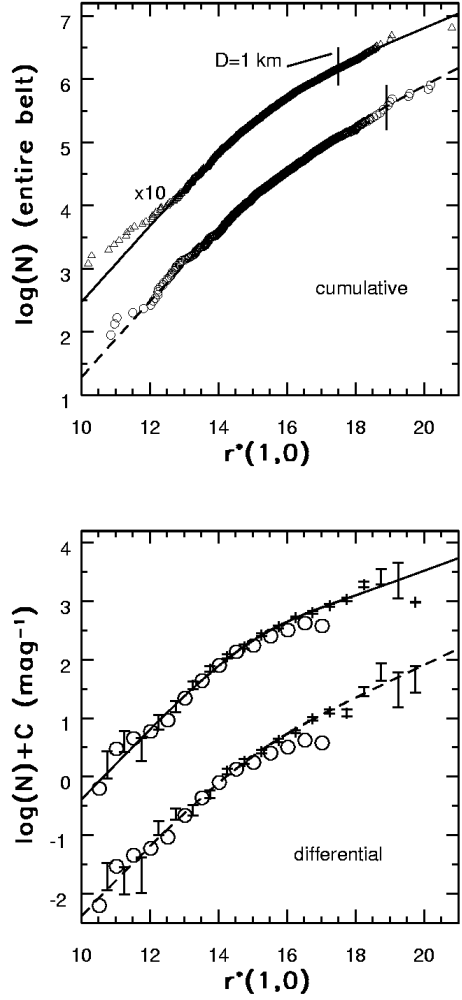}{12.5cm}{0}{60}{60}{-190}{-30}
\caption{The top panel shows the cumulative luminosity functions for
main belt asteroids, where the symbols correspond to the nonparametric estimates,
and the lines are analytic fits to these estimates (see text). The solid lines and
triangles correspond to asteroids with $a^* \ge 0$, and the dashed lines and circles
to asteroids with $a^* < 0$. The results for asteroids with $a^* \ge 0$ are
multiplied by 10 for clarity. The two vertical lines close to \r(1,0) = 18 correspond
to asteroid diameter of 1 km. The bottom panel displays the differential luminosity
function for main belt asteroids determined in this work (lines and error bars for
analytic fits and nonparametric estimates, respectively), and the result by Jedicke
\& Metcalfe (1998) shown as open circles.
\label{LFs}
}
\end{figure}

The top panel in Figure \ref{LFs} shows the cumulative \r(1,0) distribution
functions (the total number of asteroids brighter than a given brightness
limit) for main belt asteroids. The symbols (circles for asteroids with
$a^* < 0$ and triangles for asteroids with $a^* \ge 0$) show the nonparametric
estimate obtained by the $C^-$ method (every asteroid contributes to the
estimate and is represented by one symbol). The results for red asteroids
are multiplied by 10 for clarity.

The counts are re-normalized to correspond to the entire asteroid belt by
assuming that the belt does not have any longitudinal structure. This assumption
is well supported by the data because the counts of main belt asteroids observed
in the Spring and Fall subsamples agree to within their Poisson uncertainty
(\about 1\%). The normalization is determined by the counts of asteroids with
\r(1,0) $<$ 15.3, which is the faintest limit that is not affected by the
apparent magnitude cutoff of the sample (see Figures \ref{figLFblue} and
\ref{figLFred}). There are 337 blue and 519 red asteroids with \r(1,0) $<$ 15.3,
corresponding to 18,000 blue and 27,700 red asteroids in the entire belt.
The multiplication factor (53.4) includes a factor of 2.03 to account for
the $\phi$ cut (see Section 3.1) and the ``uniqueness" cut (see Section 3.3),
that decreased the initial sample of 10678 asteroids to the final sample of
5253 unique asteroids with reliable velocities. The remaining factor of 26.3
reflects the total area covered by the four scans analyzed here\footnote{Effectively,
each run covers 3.42 degree wide region of the Ecliptic; about 2.5 times more
than the width of six scanlines (1.35 degree) because of the inclined direction
with respect to the ecliptic equator.}, that is, the data cover 3.8\% of the
Ecliptic.

The normalization uncertainty due to Poisson errors is about 5\%. An additional
similar error in the normalization may be contributed by the uncertainty in the
estimated heliocentric distance, another comparable term by the uncertainty
of absolute photometric calibration ($\la 0.05$ mag), and yet another one
by the uncertainty associated with defining a unique sample (see Section 3.3).
Thus, the overall normalization uncertainty is about 10\%.

Since the blue and red asteroids have different albedo, a \r(1,0) limit does
not correspond to the same size limit. To illustrate this difference, the two
vertical lines close to \r(1,0) = 18 mark asteroid diameter of 1 km; the
cumulative counts are 370,000 for the blue type, and 160,000 for the red
type.

The normalization obtained here is somewhat lower than recent estimates
by Durda \& Dermott (1997) and Jedicke \& Metcalfe (1998). Durda \& Dermott
used the McDonald Survey and Palomar-Leiden Survey data (van Houten
{\em et  al.} 1970) and found that the number of main-belt asteroids
with $H < 15.5$ (absolute magnitude $H$ is based on the Johnson V band,
$H$-\r(1,0) \about 0.2) is 67,000. Jedicke \& Metcalfe used the Spacewatch
data and found a value of 120,000. As discussed above, the SDSS counts imply
that there are 45,700 asteroids with $H < 15.5$, or about 1.5 times less than
the former, and 2.6 less than the latter estimate. Taking the various
uncertainties into account, the normalization obtained in this work is
marginally consistent with the results obtained by Durda \& Dermott.
We point out that the methods employed here to account for various selection
effects are significantly simpler that those used in other determinations due
to the homogeneity of the SDSS data set.

The nonparametric estimate shows a change of slope around \r(1,0)\about 15-16
for both subsamples and suggests the fit of the following analytic function
\eq{
\label{LFcum}
            N_{cum} = N_o {10^{ax} \over 10^{bx} + 10^{-bx}},
}
where $x = \r(1,0) - r_C$, $a = (k_1+k_2)/2$, $b = (k_1-k_2)/2$, with $k_1$
and $k_2$ the asymptotic slopes of log(N)-$r$ relations. This function
smoothly changes its slope around \r \about $r_C$. The best fits for each
subsample are shown by lines, and the best-fit parameters (also including
the whole sample) are listed in Table 4 (note that the sum of the $N_o$ values
for the blue and red subsamples is not exactly equal to the $N_o$ for the
whole sample due to slightly varying $r_C$).

The red sample shows marginal evidence that the power-law index is smaller 
for \r(1,0)$\la$ 12.5 than for 12.5 $\la$ \r(1,0)$\la$ 15. A best-fit power-law 
index for the 20 points with 10 $<$ \r(1,0) $<$ 12.5 is 0.39$\pm$0.03. A few points 
at the bright end of the blue sample that could perhaps be interpreted as evidence 
for a similar change of slope have no statistical significance.

\subsubsection{       The Differential Distribution      }

The bottom panel in Figure \ref{LFs} displays the differential luminosity
distribution for main belt asteroids determined from the cumulative luminosity
distribution by two different methods (the curves are shifted for clarity; the
proper normalization can be easily reproduced using the best-fit parameters
from Table 4). The lines show the analytic derivative of the best fit to the
cumulative luminosity function (see eq.~\ref{LFcum}). The nonparametric
estimates are determined by binning the sample in \r(1,0) and piecewise fitting
of a straight line to the cumulative distribution. The points are not plotted
for clarity, and the displayed error bars are computed from Poisson statistics
based on the number of objects in each bin. There is no significant difference
between the results obtained by the two methods. The analytic curves are also
shown in the bottom right panel of Figures \ref{figLFblue} and \ref{figLFred}.

The recent result for the asteroid differential luminosity function
by Jedicke \& Metcalfe (1998) is shown as open circles. Their estimate
corresponds to the total sample because they did not have color information.
Overall agreement between the two results is encouraging, given that the
observing and debiasing methods are very different, and that they were forced
to assume a mean asteroid albedo due to the lack of color information. Our result 
indicates that the turn over at the faint end that they find is probably not real,
while the sample discussed here does not have enough bright asteroids (only
\about 20) to judge the reality of the bump at the bright end. Nevertheless, it seems 
that the two determinations are consistent, and, furthermore, appear in qualitative 
agreement with results from the McDonald Asteroid Survey and the Palomar-Leiden 
Survey (van Houten {\em et al.} 1970).

\subsection{            The Asteroid Size Distribution                }

\begin{figure}
\plotfiddle{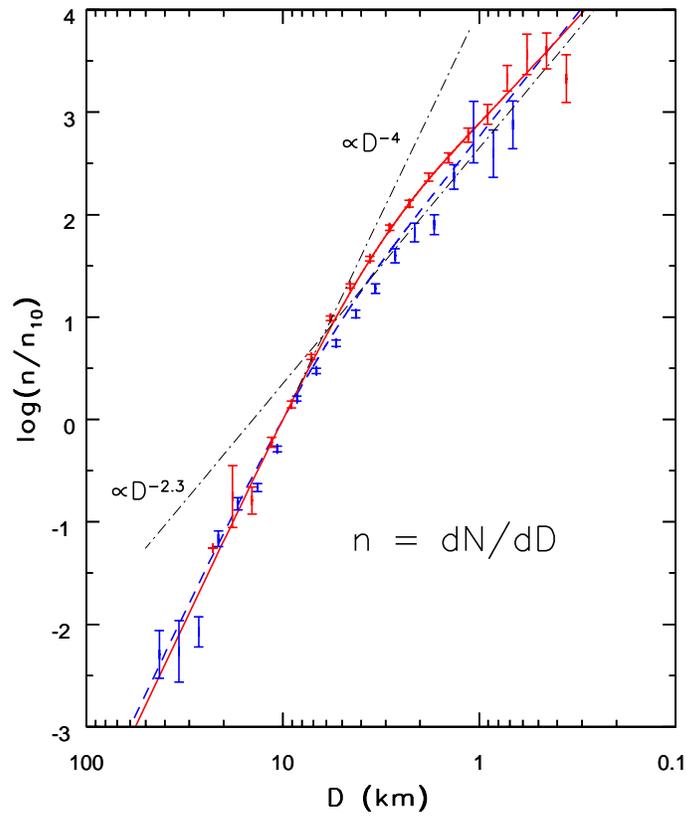}{12.5cm}{0}{60}{60}{-190}{-30}
\caption{The differential size distribution normalized by its value for $D$=10 km
(solid and dashed lines for analytic estimate, and error bars for nonparametric
estimate, for red and blue asteroids respectively). The dot-dashed lines are
added to guide the eye and correspond to power-law size distributions with
index 4 and 2.3 (see text for discussion).
\label{sizedist}
}
\end{figure}

Assuming that all asteroids in a given sample have the same albedo, the
differential luminosity function can be transformed to the size distribution
with the aid of eq.~\ref{absmag} (note that the best-fit parameters listed in
Table 4 fully describe the size distribution, too). Figure \ref{sizedist} shows
the differential size distributions ($dN/dD$) normalized by the value for $D$=10 km
(solid and dashed lines, for red and blue asteroids respectively). We have assumed
albedos of 0.04 and 0.16 for the blue and red asteroids (see Section 4.1.3). Since
the distributions are renormalized, these values matter only slightly in the
region with $D$ \about 5 km where the size distributions change slope. The dot-dashed
lines are added to guide the eye and correspond to power-law size
distributions with index 4 and 2.3.

The agreement between the data for D $\ga$ 5 km and a simple power-law size
distribution with index 4 is remarkable. From eq.~\ref{k2alpha} and the
best-fit values from Table 4, we find that the implied best-fit power-law index
is 4.00$\pm$0.05 for both types. The power-law index of both asteroid types
for D $\la$ 5 km is close to 2.3, although there may be significant differences.
The best fit values are 2.40$\pm$0.05 and 2.20$\pm$0.05 for the blue and red
types, respectively. However, these error estimates do not account for possible
hidden biases in the analysis. For example, a linear bias in the estimates
of heliocentric distance (e.g. $R_{est} = 0.95*R_{true} + 0.1$, see Appendix B)
would change the power-law index by about 0.03. Nevertheless, it is unlikely
that such a bias would be different for the two types, and thus the difference
in their power-law indices of 0.20 for D $\la$ 1 km may be a real (4 $\sigma$)
effect in the explored size range.

\section{       The Search for Asteroids by Combining SDSS and 2MASS Data        }

\subsection{                       2MASS Data                           }

The Two Micron All Sky Survey (2MASS, Skrutskie {\em et al.} 1997) is surveying
the entire sky in near-infrared
light\footnote{http://www.ipac.caltech.edu/2mass/overview/about2mass.html}.
The observations are done simultaneously in the $J$ (1.25 $\mic$), $H$ (1.65 $\mic$),
and $K_s$ (2.17 $\mic$) bands.  The detectors are sensitive to point sources brighter
than about 1 mJy at the $10\sigma$ level, corresponding to limiting (Vega-based) magnitudes
of 15.8, 15.1, and 14.3, respectively.  Point-source photometry is repeatable to
better than 10\% precision at this level, and the astrometric uncertainty for these
sources is less than 0.2 arcsec. The 2MASS has observed $\sim$300 million stars,
as well as several million galaxies, and covered practically the whole sky.

Moving objects cannot be recognized in 2MASS data without additional information.
Such information is available for known asteroids and comets, and they are identified
as a part of routine processing of the data (Sykes {\em et al.} 2000, hereafter S2000).
The initial catalogs for about 3,000 deg$^2$ contain observations of 1054 asteroids
and 2 comets. S2000 describe color-color diagrams for these asteroids and show that
they span quite narrow ranges of $J-H$ and $H-K_s$ colors (about 0.2-0.3 mag. around
$J-H$ \about 0.5 and $H-K$ \about 0).

\subsection{         Matching SDSS and 2MASS Data                    }

We use SDSS data to find candidate asteroids in 2MASS data. SDSS can be used
to detect moving objects in 2MASS catalogs because the two data sets were obtained at
different times and thus SDSS should not detect positional counterparts to asteroids
observed by 2MASS\footnote{While the converse is also true, 2MASS data could
be used to find asteroids in SDSS data only at the bright end, as it does not go
as deep as SDSS.}.
The aim of this search is to detect asteroids with very red colors
which may have been missed in optical surveys due to faint optical fluxes.
For example, the unusual R-type (red) asteroid 349 Dembowska has a K-band albedo about
5 times larger than its $r'$-band albedo (Gaffey, Bell \& Cruikshank 1989).
Since the solar $r'$-K color is about 2 (Finlator {\em et al.} 2000, hereafter F00),
a similar asteroid with $r'$=15, which is roughly the completeness limit for cataloged
asteroids, would have K=11 (here we used catalogued SDSS and 2MASS magnitudes;
note that SDSS magnitudes are calibrated on an AB system, while 2MASS magnitudes
are calibrated on Vega system, for details see F00). This is about 3 magnitudes
brighter than the 2MASS faint cutoff in the K band, and shows that any significant
population of red asteroids would be easily detected by combining SDSS and 2MASS data.

The positional matching of the sources observed by SDSS and 2MASS is described by
F00 and Ivezi\'c {\em et al.} (2001). Here we briefly summarize their results and
extend these studies to search for asteroids among 2MASS Point Source Catalog (PSC)
sources without optical counterparts.

F00 positionally matched the 2MASS PSC from the recent 2MASS Second Incremental Data
Release\footnote{http://www.ipac.caltech.edu/2mass/releases/second/index.html}
to data from two SDSS fall commissioning runs (94 and 125, see Table 1)
along the Celestial Equator extending from $\alpha_{J2000}$ = 0$^h$ 24$^m$
to $\alpha_{J2000}$ = 3$^h$ 0$^m$ and with $-1.2687^\circ <$ $\delta_{J2000}$
$< 0^\circ$ (2MASS data for $\delta > 0$ are not yet publicly available). The region
with 1$^h$ 51$^m$ $<$ $\alpha_{J2000}$ $<$ 1$^h$ 57$^m$ is missing from the
2MASS catalog. The resulting overlapping area (47.41 deg$^2$) includes 64,695
2MASS sources, and 97.9\% of them are matched to a catalogued SDSS source to better
than 2 arcsec. The inspection of SDSS images at the positions of unmatched 2MASS
sources shows that in about 30\% of cases (0.6\% of the total) there is an SDSS
source not listed in the catalog. They are usually associated with a saturated or
very bright star (\about 1/2), with a complex blend of sources (\about 1/3), or
with a diffraction spike or satellite trail. The remaining 1.4\% unmatched 2MASS
sources can be either
\begin{enumerate}
\item
Moving objects
\item
Extremely red sources (e.g. stars heavily obscured by circumstellar dust)
\item
Spurious 2MASS detections
\end{enumerate}

Since we are trying to maximize the number of candidate asteroids, we rematched
the 2MASS PSC and SDSS catalogs in a larger region centered on the Ecliptic
and observed in two SDSS spring commissioning runs (752 and 756, see Table 1)
along the Celestial Equator. This region extends from $\alpha_{J2000}$ = 10$^h$ 00$^m$
to $\alpha_{J2000}$ = 14$^h$ 00$^m$ with $-1.2687^\circ <$ $\delta_{J2000}$ $< 0^\circ$.
The resulting overlapping area (76.07 deg$^2$) includes 115,076 2MASS sources,
and 98.57\% of them are matched to a catalogued SDSS source to better than 2 arcsec.
The inspection of SDSS images at the positions of 1641 unmatched 2MASS
sources shows that in about 50\% of cases (817, or 0.7\% of the total) there
is an SDSS source not listed in the catalog, in agreement with F00, and indicating
that the completeness of the SDSS catalog at the bright end is \about 99.3\%.
The fraction of 2MASS sources without an optical counterpart (824, or 0.7\%)
is lower by about a factor of 2 for this sample than for the sample discussed by F00 (1.5\%).

\subsection{ Analysis of 2MASS sources without Optical Counterparts }

The sample of 824 2MASS sources without optical counterparts is a promising source
of faint, unusual asteroids. To minimize the number of spurious 2MASS detections,
we first analyze the statistics of 2MASS processing flags.
In each 2MASS band, $rd\_flg$ indicates the quality of profile extraction,
$bd\_flg$ indicates whether a source is blended, and $cc\_flg$ indicates whether
the source's photometry and position may be affected by artifacts of nearby bright
stars or by confusion with other nearby sources. By using
these flags we divide the sample into 5 categories listed in Table 5. We find
that only 5.3\% of sources have impeccable detections in all three 2MASS bands, while
more than 80\% of sources are detected only in J, or only in H. Because the objects
do not
have SDSS counterparts, the latter are probably spurious detections\footnote{A
high-redshift quasar with a very
strong emission line in the detection band could be undetected in all other seven
SDSS and 2MASS bands. However, the plausible numbers of such sources are incompatible
with the observed density of \about 9 deg$^{-2}$.}. We point out that the fraction
of J-only and H-only detections in the 2MASS catalog is only 2.3\%, as shown in the
table. Thus, although the fraction of these, presumably spurious, detections is
above 80\% for the 2MASS sources without optical counterparts, these listings
represent only 25\% of the total number of such J-only and H-only detections
(for H-only detections considered alone this fraction is 57\%). Consequently, the
2MASS processing software was appropriately tuned to reach a reasonable compromise
between the reliability and completeness of the J-band only detections.

\begin{figure}
\plotfiddle{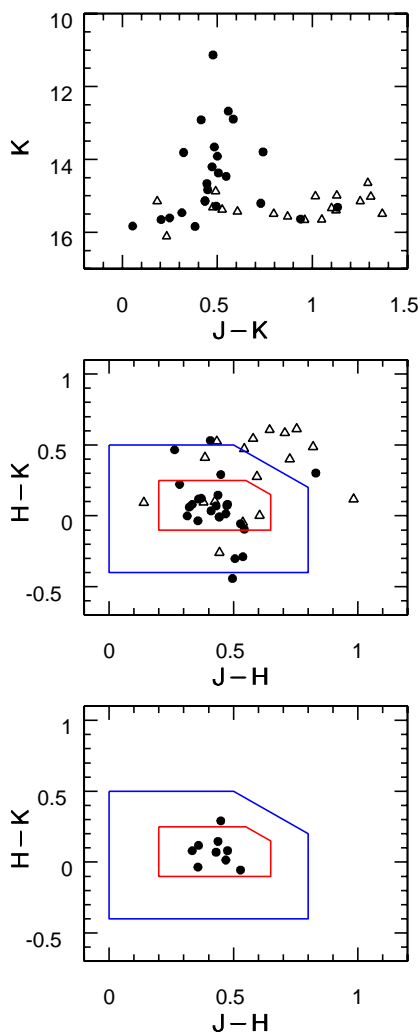}{12cm}{0}{60}{60}{-200}{-30}
\caption{The color-magnitude and color-color diagrams for the candidate asteroids
selected in 2MASS data. The top panel shows the K vs. J-K color-magnitude diagram for
42 reliable 2MASS PSC objects in the analyzed region without an SDSS counterpart
within 3 arcsec. The 24 objects flagged as minor planets in the 2MASS database are
shown as large dots, and the rest of objects are shown as open triangles. The middle
panel shows the H-K vs. J-H color-color diagram for the same objects. The two contours
outline the distribution of known asteroids, and enclose approximately 2/3 and 95\%
of sources from Sykes et al. (2000). The bottom panel shows H-K vs. J-H color-color
diagram for the 9 objects with $K_s < 14.3$ (10$\sigma$ limit).
\label{2MASSdiags}
}
\end{figure}

We select the 42 objects with good detections in all 3 2MASS bands as plausible
asteroid candidates. The $K_s$ vs. $J-K_s$ color-magnitude diagram and the $H-K_s$
vs. $J-H$ color-color diagram for these objects are shown in the top two panels in
Figure \ref{2MASSdiags}. The 24 objects flagged by 2MASS as known minor
planets\footnote{There
are 38 objects flagged as minor planets in the starting 2MASS sample of 115,076 sources,
and 25 have good detections in all 3 2MASS bands. One object was erroneously matched
to a nearby (1.9 arcsec away) faint source in SDSS and 24 were uncovered by our
technique as 2MASS objects without SDSS counterparts.}
are shown as large dots, and the others as open triangles. The two contours in the
middle panel
outline the distribution of the known asteroids, and enclose approximately 2/3
and 95\% of sources from S2000.  The bright sources predominantly have
$J-K_s < 1$ as shown by S2000. The sources outside the contour outlining
the S2000 sample are predominantly faint and it is not clear whether they are scattered
out by measurement errors, whether they are asteroids with peculiar colors,
or whether they are asteroids at all. We find no strong correlation with the distance
from the Ecliptic for these sources, but note that the sample is very small.

We further limit the sample by requiring at least 10$\sigma$ detection in the $K_s$
band ($K_s < 14.3$), which automatically guarantees at least 10$\sigma$ detections
in the other two bands (because asteroids are bluer than the colors of 2MASS
N$\sigma$ limits). This requirement leaves 9 sources whose $H-K_s$ vs. $J-H$ color-color
diagram is shown in the bottom panel in Figure \ref{2MASSdiags}. All of these are
flagged as minor planets, showing that reliable 2MASS detections without SDSS
counterparts are heavily dominated by previously known asteroids. Furthermore, to
a magnitude limit of \r=16.3, there are 24 asteroids in the analyzed area. This
is in good agreement with the faint cutoff for SDSS-2MASS matched stars with \gr\
color similar to asteroids (see F00, fig. 4), and indicates that the catalogued
asteroids are complete to a similar depth, in agreement with Zappala \& Cellino
(1996). We conclude that there is no evidence for the existence of a significant
population of asteroids with extremely red optical-IR colors, and estimate an
upper limit of 10\% for the fraction of extremely red asteroids which escaped
detection at optical wavelengths.

It cannot be excluded that at least some asteroids have very
peculiar IR colors. We have attempted to reobserve two sources selected
from the sample of 2MASS sources without SDSS counterparts discussed by
F00. These (2MASS 0041309-002730 and 2MASS 0053480-003836) were selected to
have the most extreme near-infrared colors; the first has good detections in all 3 2MASS
bands, and the second was detected only in the K band and flagged as
an unreliable source. We used the United Kingdom Infrared Telescope UKIRT
and the infrared array camera UFTI in J band on the night of Sept 21, 2000.
We did not detect either object within UFTI's 50 arcsec field of view to a
limit of about 18 mag, implying that they are moving objects. Since the second
source was already flagged as unreliable by 2MASS processing software, we consider
only the first source as a good candidate for an asteroid with peculiar
infrared colors (it is not flagged by 2MASS as a minor planet). Its 2MASS
measurements are J=15.64$\pm$0.05, J-H=1.01$\pm$0.07, H-K=0.04$\pm$0.10.

S2000 discuss two asteroids with very peculiar colors and warn that they may be
blends with a background object. However, this explanation is probably not applicable
for this asteroid because there are no optical nor infrared sources at the position
of the 2MASS detection. In principle, there could exist a very red and very variable
source that was observed by 2MASS, and was not detected in UKIRT observation
because it was in its faint phase. However, the probability for this seems
very small.

\section{             The Search for the Kuiper Belt Objects            }

The Kuiper Belt Objects (KBOs, sometimes known as trans-Neptunian objects) are an ancient
reservoir of objects located beyond Neptune's orbit, which are believed to be the origin
of short-period comets (Fern\'andez 1980, Duncan, Quinn \& Tremaine 1988; for a detailed
review see Jewitt 1999). It is estimated that the number of the KBOs larger than $100$ km
in diameter (assuming geometric albedo of 0.04) is $\sim 10^5$ in the $30$ to $50$ AU distance
range (Jewitt 1999). The total mass in the KBOs is thought to be somewhat less than
one Earth mass, and the expected sky density of the KBOs brighter than $\r \about$ 21.5 mag
is about 0.01-0.04 deg$^{-2}$.

The first KBO was detected by Jewitt \& Luu (1992), and today there are close to 300
known objects\footnote{For an up-to-date list of known KBOs see
http://cfa-www.harvard.edu/cfa/ps/lists/TNOs.html}.
Most of these are very faint: a KBO with radius equal to $\about$100 km and albedo 0.04 has a
visual magnitude of 22 at a distance of $\about$30 AU from the Sun. Finding such objects
poses quite a challenging task; to date detections have usually been accomplished
by detecting their angular motion on the sky using large optical telescopes.

As discussed in Section 2.3, the KBOs move a bit too slowly to be detected in a single SDSS run.
However, they can be detected by comparing two scans of the same area, as long as they
are obtained within a week (otherwise a KBO would move to the region between two scanned
columns). Here we describe a method developed for finding KBOs in two-epoch data and apply it
to data for about 100 deg$^2$ of sky observed twice $\about$ 1.9943 days apart (runs 745
and 756, see Table 1). The lower limit for detectable angular motion range is 0.0004 \dd,
and is set by a matching radius of 1 arcsec. The upper limit is determined by the
maximum change in position for an object to be observed in both epochs and is about
0.08 \dd. As the expected KBO motion is well within these limits, the only practical
constraint is the brightness. The lower limit on KBO size to be detected with $\r=22$
during oposition at a heliocentric distance of 40 AU is about 400 km (assuming an
albedo of 0.04).

\subsection{                     The Search Method          }

Slow moving objects like KBOs can be detected in two-epoch data because at each of the
two observed positions they will not have a positional match in the other epoch.
The probability of a random association with a background object is very low because
the SDSS photometric pipeline deblends overlapping objects, unless they are too close
to one another. Assuming a lower limit of 2 arcsec for a pair of nearby sources to be
deblended, only $\about 10^{-3}$ of potential KBOs will be lost. The objects without
positional counterparts in other epoch, hereafter called {\em orphans}, can then be
matched within a much larger radius (a few arcmin) than the typical distance between
two SDSS sources ($\about$ 30 arcsec for \r $\about$ 22). Of course, there will also
be orphans caused by instrumental effects such as diffraction spikes, satellite trails,
seeing dependent deblending, etc. In addition to these, main-belt asteroids may be
considered as contaminants when searching for KBOs.

\begin{figure}
\plotfiddle{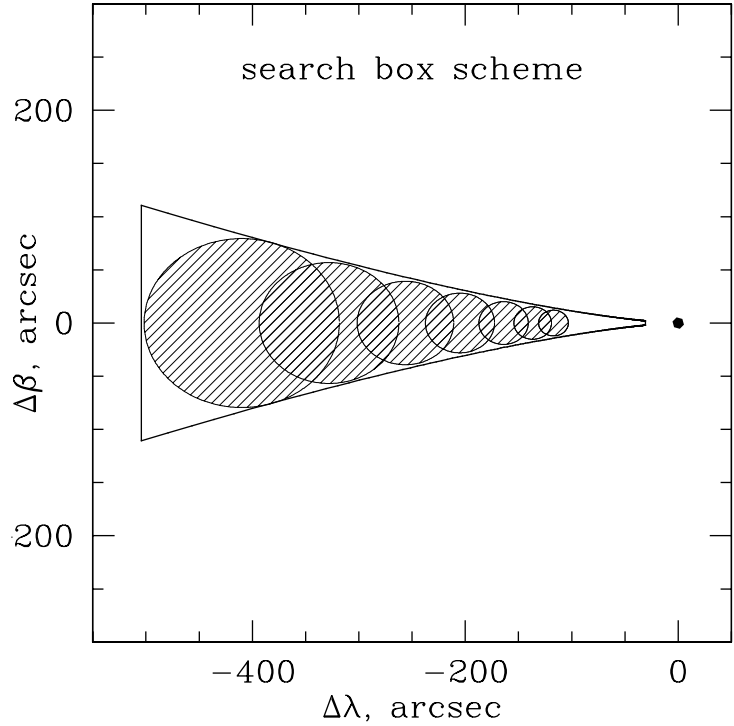}{8cm}{0}{50}{50}{-150}{-50}
\caption{The construction of the search box for matching moving objects observed
in two different runs. Axes show the change in ecliptic coordinates between the two
observations. An object at a given heliocentric distance would be found in one of the
ellipses, with the exact position depending on the orientation of its proper motion
vector. Different ellipses correspond to varying distances from 20 AU to 70 AU.
Note that the orbital parameters constrained by an observed position are degenerate
since any position can belong to more than one ellipse.
\label{KBObox}
}
\end{figure}

Given a position of an orphan (i.e. a KBO candidate) in the first epoch, its position
in the second epoch can be approximately predicted by assuming that the observed motion
is due to Earth reflex motion and its own Keplerian proper motion. The predicted position
depends on the time elapsed between the two observations, and object's unknown heliocentric
distance and orbital inclination. Because of the KBO's proper motion, the region in the
second epoch where a matching orphan would be found is an ellipse for each possible
heliocentric distance, with its major axis parallel to the Ecliptic. Assuming
observations close to opposition, the distance parallel to the Ecliptic between
the center of this ellipse and the first position is
\eq{
         \Delta \lambda = 0.9856 {\Delta t \over R - 1} \, \dd,
}
where $\Delta t$ is the elapsed time in days, and is determined by the Earth's
reflex motion. The major axis of this ellipse is given by
\eq{
         a = { \Delta \lambda(R) \over \sqrt{R}},
}
and its minor axis by
\eq{
         b = a \cos(\beta).
}
An example of such ellipses for several different $R$ is shown in Figure
\ref{KBObox}. Since the distribution of possible distances is continuous\footnote{We
do not exploit the fact that the observed KBO distance distribution is highly
bimodal with peaks at $\about$ 39-40 AU and $\about$ 43-44 AU, see Jewitt 1999.},
the overall search region assumes a wedge-like shape as shown in Figure \ref{KBObox}.
This shape is approximately described by
\eq{
        \Delta \beta = \pm {\cos(\beta) \Delta \lambda(R) \over \sqrt{R}}.
}

The pairs of orphans that are matched within such search box, determined
for a range of plausible $R$ (assumed to be 10-300 AU), can still be spurious
matches and we further require that each candidate pair of detections satisfy

\begin{enumerate}
\item
No significant motion should be detected within a single run to avoid spurious
matches with main-belt asteroids.
\item
Both detections must be unresolved sources brighter than \r=22, and
must not be blended (objects with more than one peak) or saturated, nor
cosmic rays.
\item
The measured \gr\ and \ri\ colors for the two epochs must agree within 1$\sigma$
(the agreement between the two sets of magnitudes is not required because
the brightness could vary if the object has irregular shape and rotates).
\end{enumerate}

\subsection{                      Results                         }

\begin{figure}
\plotfiddle{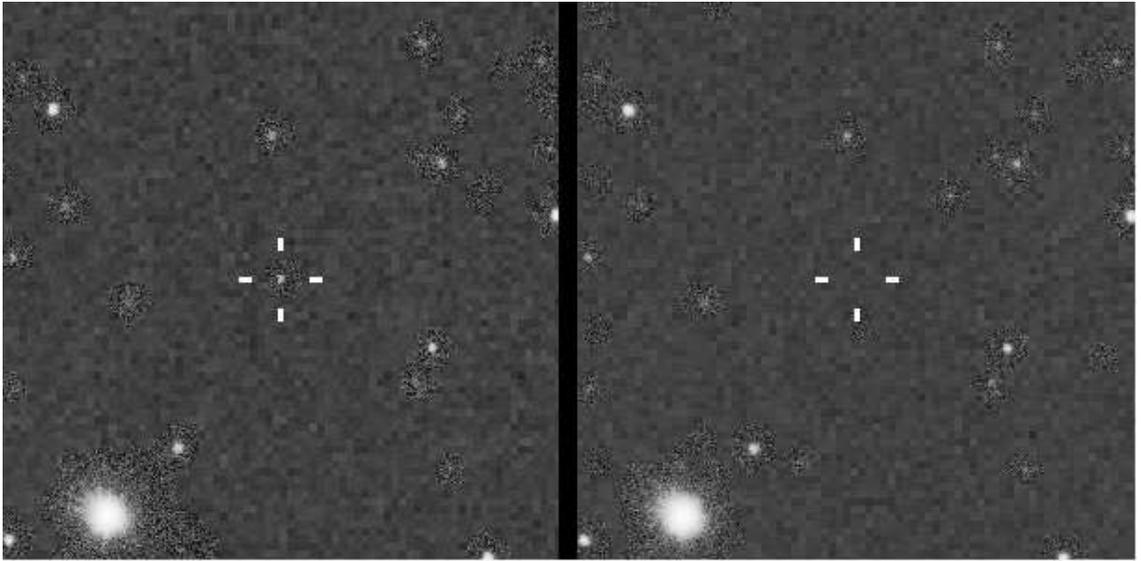}{10cm}{0}{70}{70}{-200}{-150}
\caption{The KBO candidate observed in the first epoch. The left panel
shows the 1x1 arcmin $r'$-band image from run 745 with the position of the KBO
marked by a cross. The right panel shows the same part of the sky observed in
in the second epoch (run 756) with the same position marked by a cross. 
The images are assembled from the postage stamps for all objects detected
by photometric pipeline.
\label{KBO1}
}
\end{figure}

\begin{figure}
\plotfiddle{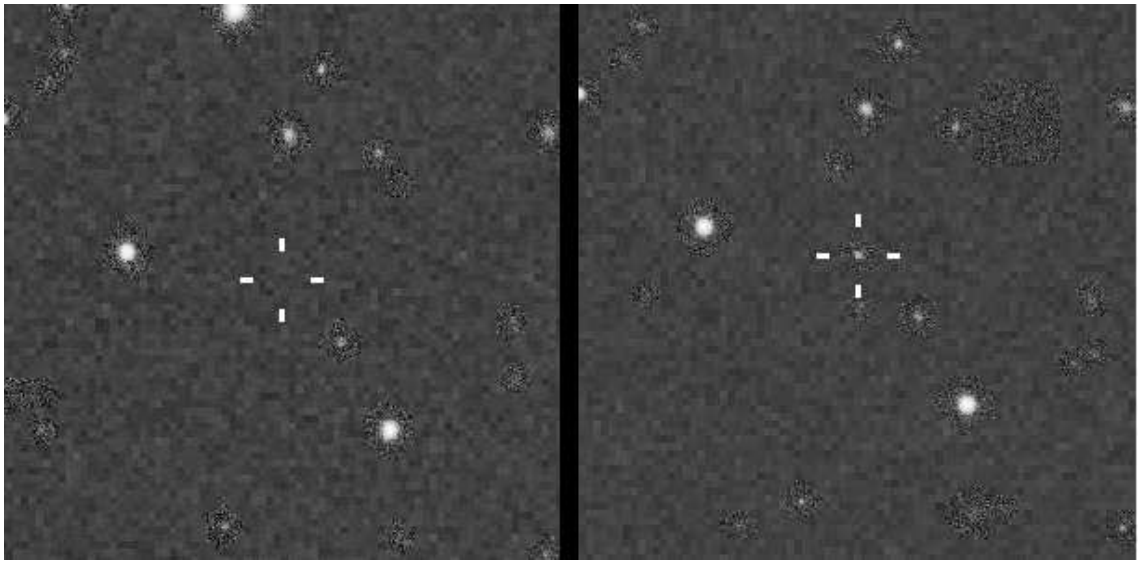}{10cm}{0}{70}{70}{-200}{-150}
\caption{Same as the previous figure, except that the KBO candidate is observed
in the second epoch.
\label{KBO2}
}
\end{figure}

We searched 99.5 deg$^2$ of sky observed twice 1.9943 days apart (runs 745
and 756, see Table 1) using the algorithm described in the previous section.
The application of the first two criteria results in 166 candidate pairs, which
are reduced to only 6 candidates by the application of the color constraint.
We inspected visually 24 images corresponding to these 6 candidates and found
that five candidates can be excluded as instrumental effects. The 4 images for
the sixth candidate are shown in Figures \ref{KBO1} and \ref{KBO2}.

The measured parameters for this KBO candidate are listed in Table 6, and
the derived properties are listed in Table 7. Note that the magnitudes are
the same within the measurement uncertainty, although this was not required
by the algorithm, indicating that this object may have quite a regular shape.
The candidate's distance is obtained from its angular motion; it is multi-valued
because the two-epoch data are insufficient to fully constrain the candidate's
orbital parameters. The ambiguity in distance propagates to the ambiguity in size.

We use the photometric transformations listed by Fukugita et al. (1996)
to translate the KBO colors measured by other groups (Luu \& Jewitt 1996;
Tegler \& Romanishin 1998) to the SDSS photometric system. We find that typical
KBO colors are $g'-r' \approx 0.8\pm0.3$ and $r'-i'\approx 0.4\pm0.2$.
The colors of the SDSS KBO candidate (\gr \about 1.0, \ri \about 0.5)
fall within these ranges. As noted earlier, these colors are significantly
redder than the colors of main-belt asteroids.

The search for KBOs is also sensitive to Centaurs, but we did not find any
reliable candidates in the searched area.

\subsection{         Completeness of the search for KBOs              }

The third search criterion described above requires that the \gr\ and \ri\
colors must agree within 1$\sigma$ between the two observations, and
decreases the number of candidates from 166 to 6. If this criterion is too
conservative, then the number of detected KBOs could be 28 times higher.

We have tested this possibility by removing the third criterion and instead
requiring that the observed \gr\ color falls in the range 0.7-1.2 in both
epochs (the left edge is 0.2 redder than discussed above due to the large
increase in the number of stars with bluer colors). We also abandon any
requirement on the orbit and simply match the orphans with required
colors (120 and 300 from runs 745 and 756, respectively) within a circle
with radius of 220 arcsec (corresponding to the motion of 0.03 deg day$^{-1}$).

We selected only one other candidate, and by visual inspection of images
found that it was an instrumental effect. Thus we conclude that it is
highly unlikely that our searching method missed any KBOs present in the
data. It is not easy to precisely estimate this probability, but assuming
that the most likely cause would be blending with another source, we place
an approximate upper limit of 0.001 on the probability that there is
another KBO which the search algorithm missed.

One detected source brighter than \r=22 in 99.5 deg$^2$ is in agreement
with expected density of 0.01-0.04 such objects per deg$^2$ (Jewit 1999).
The significance of this result is limited by the small-number statistics.

\section{                           Discussion                }

\subsection{  Moving Objects in SDSS  }

The detection of asteroids discussed in this work is based on a new algorithm, and
thus it is of concern whether these sources really are asteroids. Based on the
analysis presented here, the main arguments in favor of the reality of the
SDSS detections are:
\begin{itemize}
\item
Visual Inspection. But Yet They Move.
\item
The density of selected candidates correlates with the ecliptic latitude,
although the position of the Ecliptic is not known to the algorithm,
and the ecliptic latitude distribution width is consistent with observed
distribution of known asteroids (see Sections 2.3 and 4.2.1).
\item
The color distribution of the selected candidates spans a much smaller range
than that of stars, and corresponds to a G type star. Furthermore, it matches the
colors of known asteroids observed with SDSS filters, and agrees with the
synthetic colors derived from spectral data (see Section 4.1).
\item
The morphology of the velocity distribution for the selected candidates is in close
agreement with the known orbital distribution of asteroids (see Section 5.1).
\item
For a limited area we show that the majority of candidates are recovered
in a repeated scan (see Section 3.2).

\end{itemize}

The results presented here show that there are \about 530,000 objects larger
than 1 km in the asteroid belt. This is about four times less than the
previous estimates, e.g. the SIRTF planning for observing Solar System bodies
assumes that there are as many as $2\times10^6$ asteroids larger than 1 km
(see Figure 4 in Hanner \& Cruikshank 1999). As discussed in Section 6.6,
about half of that discrepancy can be attributed to different normalizations,
and the other half is probably due to smaller number of asteroids in the
previously unexplored 1--5 km diameter range.

Extrapolating the observed asteroid density to the whole survey, we predict
that by its completion SDSS will obtain about 100,000 near simultaneous five-color
measurements for a subset drawn from 280,000 asteroids brighter than r $<$ 21.5 at
opposition. While it is certain that a fraction of these observations will be multiple
measurements of the same objects, the fraction cannot be estimated beforehand because
it depends on the detailed observing strategy.

\subsection{ The Colors of Asteroids and the Heliocentric Distance Distribution }

The color distribution of observed asteroids is strongly bimodal. While
there is evidence for the existence of additional classes (see Sections
4.1.1 and 4.1.2), more than 98\% of the objects in the sample have colors consistent 
with a simple separation into two basic color types. We find a significant correlation
between asteroid colors and heliocentric distance, suggesting the existence
of two distinct asteroid belts: the inner rocky belt, about 1 AU wide (FWHM)
and centered at $R$ \about 2.8 AU, and the outer carbonaceous belt, about 0.5
AU wide and centered at $R$ \about 3.2 AU (see Sections 5.3 and 6.5). This
result represents a significant constraint for the theories of Solar System
formation (e.g. Ruzmaikina {\em et al.} 1989).

The correlation between the asteroid colors and their heliocentric distance
has been recognized since the earliest development of taxonomies (for a
historical overview see Gradie, Chapman \& Tedesco 1989). The first evidence
that various taxonomic types show distinct heliocentric distributions was
presented by Chapman, Morrison \& Zellner (1975) and later reinforced by using
larger samples (Gradie \& Tedesco 1982, Zellner, Tholen \& Tedesco 1985).
This study, which represents an increase in the sample size by more than a factor
of 10, confirms that the number of red main belt asteroids decreases, and
the number of blue main belt asteroids increases with the heliocentric distance. 
For samples defined by the same absolute magnitude cutoff,
the number fraction of the red to blue asteroids varies from 4:1 at $R$ \about 2 AU,
to about 1:3 at $R$ \about 3.5 AU. For sample defined by the same size cutoff,
the number fraction of the red to blue asteroids is 1:2.3 for $D > 1$ km (this 
ratio slightly depends on the size cutoff due to different size distributions).
Our study has shown in addition that the surface number density of each type 
has a well defined maximum in the radial direction, and that the median color of 
each subsample becomes systematically bluer as the heliocentric distance increases.

The colors of Hungarias, Mars crossers, and near-Earth objects are more similar
to the C-type than to S-type asteroids, suggesting that they originate in the
outer belt. However, there are less than 50 objects in this sample and thus
a larger data set is needed for a robust conclusion. We investigated the possibility
that optical surveys are biased against very red asteroids by positionally matching
the SDSS and 2MASS surveys. While we find one example of an asteroid with peculiar
colors, there is no evidence for a significant fraction of very red asteroids ($\la
10\%$).

\subsection{         The Asteroid Size Distribution       }

The size distribution discussed in Section 6.6 shows a striking
change of the power-law index from 4 to \about 2.3 around $D$\about5 km
for both asteroid types. The steep slope applies up to the sample limit
at $D$\about40 km, and the shallow slope down to $D$\about0.4 km. The 
dynamic range of \about100 is the largest one obtained with a single 
data set.

In the first detailed study of the asteroid size distribution,
Zellner (1979) found the power-law index $\alpha \about (1.5-2)$ for
asteroids larger than $\about$ 20 km, corresponding to $r^* \la 12$
where the counts analyzed in this study are limited by small-number
statistics. Thus, our results are not in direct conflict with those
obtained by Zellner, and may be considered the extension of the size
distribution to smaller asteroid sizes. While it is conspicuous
that the change from Zellner's shallower size distribution to the steeper
one obtained here occurs at the magnitude where the data samples barely
overlap, the flattening of the size distribution for $D>20$~km is
supported by more recent determinations. Using a sample of about
4000 asteroids, Cellino {\em et al.} (1991) found that the asteroid
size distribution resembles a power law with index $\alpha \about 3$ 
for $D \ga 150$~km, and becomes flatter for smaller sizes ($\alpha \about 
1$), down to their sample size limit of 44 km. 
The results presented here are roughly consistent with a recent
result for the absolute magnitude distribution by Jedicke \& Metcalfe (1998)
(shown as open circles in the bottom panel in Figure \ref{sizedist}).
A most notable difference in the region of overlap is that we do not detect
the turn-over in the counts at the faint end (we detect about 2-3
times more objects in the relevant range of magnitudes, \r(1,0)
\about 16-17, or \r $\la$ 19).

It may be of interest to note that the size distribution for KBOs larger than
$\about$100 km seems to follow the $D^{-4}$ power law (Jewitt 1999), while the
size distribution of comets appears to follow the $D^{-3}$ power law (Shoemaker
\& Wolfe 1982). The simulations of the asteroid belt evolution predict slopes similar
to those detected here which range from 2.5 to 4.0 (Gradie, Chapman \& Tedesco
1989), depending on detailed assumptions and modelled effects. For example,
 the $D^{-3.5}$ power law is expected to be produced by shattering collisions
(Dohnanyi 1969, Williams \& Wetherill 1994). The Jedicke \& Metcalfe result
for the absolute magnitude distribution was transformed into a size
distribution and modeled by Durda, Greenberg
\& Jedicke (1998). They join the resulting distribution with the data
for catalogued asteroids and find two ``humps": the first spans the
range 30-300 km and the second spans the range 3-30 km (see their figure
2). These ``humps" were explained by tailoring a detailed dependence of
the critical specific energy (energy per unit mass required to fragment
an asteroid and disperse the fragments to infinity) on asteroid size.

The size distributions derived here and shown in Figure \ref{sigmaR}
are smooth, and apart from the change in slope around D \about 5 km,
do not show any structure. The discrepancies in the region of overlap with
studies quoted above are probably due to different methods of correcting
for selection effects. They were forced to assume a mean albedo due to
the lack of color information, a problem that is alleviated by SDSS data.
Due to significant differences in the mean albedo and radial density distribution
of the two classes, this assumption may lead to biases, as was already pointed
out in those studies. The same explanation can be invoked to account for
the fact that, contrary to previous studies (e.g. Jedicke \& Metcalfe 1998),
we find no evidence that the size distribution varies with the heliocentric
distance.

The SDSS data indicate that the size distribution of main belt asteroids is well
described by a $D^{-4}$ power-law for 5 km $\la D \la$ 40 km, and by a $D^{-2.3}$
power-law for 0.4 km $\la D \la$ 5 km. Similar broken power laws have already
been proposed for the size distribution of various solar system objects
(e.g. Levison \& Duncan 1990, Tremaine 1990), and in particular for the main-belt
distribution (Cellino {\em et al.} 1991). While we don't yet have an understanding
of the processes responsible for the observed change of slope of size distribution
from 4 to 2.3, it seems plausible that they could be modeled as in e.g. Durda,
Greenberg \& Jedicke (1998).

\subsection{      The Asteroid-Earth Impact Rates               }

A decrease by a factor of 4 in the estimated number of asteroids with $D > 1$~km 
discussed here implies a similar decrease in estimates of various other quantities 
proportional to the number of asteroids, e.g. the contamination rate by asteroids 
when searching for stellar occultations by KBOs (C. Alcock, priv. comm.), and the 
frequency of  ``killer asteroids'', usually defined as objects larger than 1 km. 
The results presented here imply that the time until the next catastrophic asteroid 
collision with the Earth may be at least four times longer than previously thought 
(estimated to be $1-2 \times 10^5$ years for objects larger than 1 km, see e.g. 
Spaceguard Survey Workshop Report\footnote{Available at 
http://impact.arc.nasa.gov/reports/spaceguard/index.html}, Verschuur 1991,
Steel 1997, Ceplecha {\em et al.} 1998, Rabinowitz {\em et al.} 2000). 
Of course, the estimates of such impact rates based
on the number of asteroids and dynamical considerations are rather uncertain, so
it may be more robust to determine them from the historical impact records.
On the other hand, the latter method suffers greatly from the small number 
statistics and unknown sample completeness.

\begin{figure}
\plotfiddle{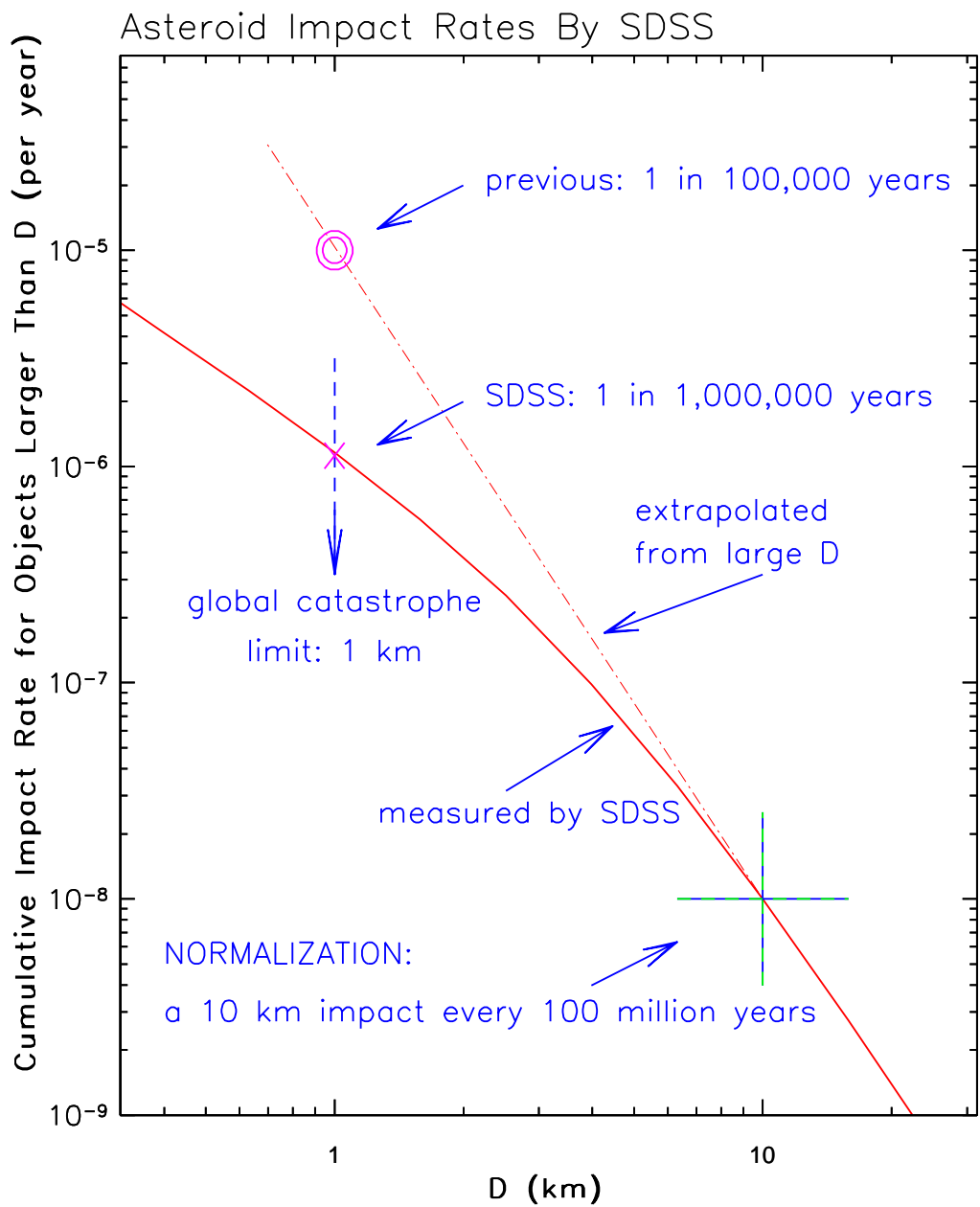}{13cm}{0}{70}{70}{-200}{-70}
\caption{The solid line shows the estimate for asteroid impact rates with the Earth 
based on the asteroid size distribution determined here. The overall normalization
is tied to the impact of a 10 km large body that caused the extinction of dinosaurs
65 million years ago. The dot-dashed line shows that extrapolation of 
the size distribution from the $D \about 10$ km end results in erroneous
impact rate estimate for smaller objects.
\label{impacts}
}
\end{figure}

The only known catastrophic event with a reliable estimate of the object size 
is the impact of a 10$\pm$4 km body that probably caused the extinction 
of dinosaurs 65$\pm$0.1 million years ago (Alvarez {\em et al.} 1980, van den 
Bergh 1994). Tying this impact rate (\about1 per 10$^8$ years for objects larger 
than 10 km) to the impact rate for objects larger than 1 km that are capable
of causing global catastrophes, was until now based on the extrapolations of 
the size distribution. SDSS measurements for the first time reliably show that
the number ratio of $D >$ 1 km asteroids and $D >$ 10 km asteroids is about 100,
implying an impact rate for ``killer asteroids'' of about 1 every 10$^6$ years, 
or at least several times less frequently than previous estimates. While still
dependent on rather uncertain normalization, this estimate no longer relies on 
the extrapolation of the asteroid size distribution. Figure \ref{impacts} shows
that such an extrapolation overestimates the number of $D >$ 1 km asteroids by
about a factor of 10.

\subsection{               Future  Work             }

This work analyzes only 10\% of the total number of asteroids to be observed by
SDSS. Although this sample represents more than an order of magnitude increase
in the number of asteroids with accurate multi-color photometry, another increase
by factor of ten will be possible soon. With such massive and detailed information,
it may be possible to develop a more sophisticated taxonomic scheme than attempted
here, as well as to perform other types of studies; for example detailed modeling
of the dependence of color on phase angle.

The main shortcoming of SDSS data for asteroid studies is insufficient information
to determine orbits. Although the completeness of the catalogued asteroids is
significant only at the bright end of the SDSS sample, the matching of known asteroids
to SDSS observations would increase the sample of objects with both known
orbits and accurate multi-color photometry by at least an order of magnitude.
Additionally, about 225 deg$^2$ of the sky along the Celestial Equator will
be surveyed dozens of times, and depending on the frequency of the observations,
it may be possible to determine the asteroid orbits. These data will also be
invaluable when searching for KBOs as described in Section 8.

We hope that this preliminary analysis will motivate further studies of the solar
system using SDSS data. In particular, most of the data discussed here are part
of the June 2001 SDSS Early Data Release.

\vskip 0.4in
\leftline{Acknowledgments}

We are grateful to Scott Tremaine, Mario Juri\'{c}, Dejan Vinkovi\'{c} and
Tim Knauer for their careful reading and insightful comments.

The Sloan Digital Sky Survey (SDSS) is a joint project of The University of Chicago,
Fermilab, the Institute for Advanced Study, the Japan Participation Group, The Johns
Hopkins University, the Max-Planck-Institute for Astronomy, New Mexico State University,
Princeton University, the United States Naval Observatory, and the University of Washington.
Apache Point Observatory, site of the SDSS telescopes, is operated by the Astrophysical
Research Consortium (ARC).

Funding for the project has been provided by the Alfred P. Sloan Foundation,
the SDSS member institutions, the National Aeronautics and Space Administration, the National
Science Foundation, the U.S. Department of Energy, Monbusho, and the Max Planck Society.
The SDSS Web site is http://www.sdss.org/.

This publication makes use of data products from the Two Micron All Sky
Survey, which is a joint project of the University of Massachusetts and
the Infrared Processing and Analysis Center/California Institute of Technology,
funded by the National Aeronautics and Space Administration and the National
Science Foundation.


\newpage

\appendix{\bf Appendix A: Determination of Asteroid Heliocentric Distance from
                          its Proper Motion }

Following Jedicke (1996), we derive the following expressions which relate the
observed ecliptic latitude, $\beta$, opposition angle $\phi$ (angular distance from
the antisun), and the ecliptic velocity components, $v_\lambda$ and $v_\beta$,
to the semimajor axis $a$ (or heliocentric distance $R$ because circular orbits are
assumed), orbital inclination $i$, the longitude of the ascending node $\Omega$,
and the rate of the radial motion $\dot{\Delta}$.

The heliocentric velocity vector of an asteroid is the composition of
its relative velocity with respect to the Earth and the reflex motion
of the Earth
\begin{eqnarray}
\label{veceq1}
{\bf v} = {\bf v}_{\rm rel} + {\bf v}_{\oplus},
\end{eqnarray}
where ${\bf v}_{\oplus}$ is the Earth's velocity ($\| {\bf v}_{\oplus} \|
= v_E = 0.986$ {\rm deg day$^{-1}$}). Since the orbits are assumed circular
(for the Earth's motion this is quite a decent approximation), the expressions
for three vector components plus the norm of Eq.~\ref{veceq1} yield a set of
four equations with four unknowns ($R$, $i$, $\Omega$ and $\dot{\Delta}$).
\begin{eqnarray}
\label{veceq2}
{\bf v} &=& {v_E \over \sqrt{R}} \left( \begin{array}{c}
-\cos b \sin l \cos i - \sin b \cos \Omega \sin i \\
\cos b \cos l \cos i - \sin b \sin \Omega \sin i \\
\cos b \sin i ( \cos l \cos \Omega + \sin l \sin \Omega )
\end{array} \right) \nonumber\\
&=& \left( \begin{array}{c}
-\Delta (v_\beta \sin \beta \cos \phi + v_\lambda \sin \phi) +
\dot{\Delta} \cos \beta \cos \phi \\
\Delta (v_\lambda \cos \phi - v_\beta \sin \beta \sin \phi) +
\dot{\Delta} \cos \beta \sin \phi + v_E \\
\Delta v_\beta \cos \beta + \dot{\Delta} \sin \beta
\end{array} \right) \\
\| {\bf v} \| &=& \Delta ^2 (v_\lambda^2 + v_\beta^2) +
\dot{\Delta}^2 + v_E^2 \nonumber \\
&& +2v_E \left[ \Delta \left( v_\lambda \cos \phi -
v_\beta \sin \beta \sin \phi \right) +
\dot{\Delta} \cos \beta \sin \phi \right] \nonumber \\
&=& {v_E \over \sqrt{R}},
\end{eqnarray}
where $(l,b)$ represent the heliocentric ecliptic longitude and
latitude and $\Delta$ is the distance between the Earth and the object
\begin{eqnarray}
\tan l &=& \frac{ \Delta \cos \beta \sin \phi}{1+\Delta \cos \beta
\cos \phi} \\
\tan b &=& \frac{\Delta \sin \beta (\sin l + \cos l)}{1+\Delta \cos
\beta (\sin \phi + \cos \phi)} \\
\Delta &=& \sqrt{\cos^2 \beta \cos^2 \phi + R^2-1} - \cos \beta \cos
\phi.
\end{eqnarray}

These equations are not invertible and we solve them by an
iterative procedure based on routine {\em newt.c} from Numerical Recipes
(Press {\em et al.} 1992).

The analysis of the accuracy of this method is described in Appendix B.
It is noteworthy that somewhat simpler set of equations involving only
two unknowns ($R$ and $i$) can be derived by utilizing the Cartesian
coordinates and introducing several additional assumptions (Jedicke 1996).

\eqarray{
\label{veq1}
    v_\lambda &=& \sin(\phi) {v_x \over \Delta} +
               \cos(\phi) {v_y - v_E \over \Delta},
    \non
    v_\beta &=& -\sin(\beta) \cos(\phi) {v_x \over \Delta} +
               \sin(\beta) \sin(\phi) {v_y - v_E \over \Delta} + \non
             &&  \cos(\beta) {v_z \over \Delta},
}
where
\eqarray{
\label{veq2}
     \theta &=& \phi - {\rm arcsin}\left({\sin(\phi) \over R} \right)
     \non
     \Delta &=& [R^2 + 1 - 2 R \cos(\theta)]^{1/2}
     \non
     v_x &=& {v_E \over R^{1/2}} \cos(i) \sin(\theta)
     \non
     v_y &=& {v_E \over R^{1/2}} \cos(i) \cos(\theta)
     \non
     v_z &=& {v_E \over R^{1/2}} \sin(i).
}

For small $\phi$ and $i$ ($\la 10^\circ$), we find that this simplified
set of equations performs as well as the first set of equations. For
larger $\phi$ and $i$ its performance deteriorates.

\appendix{\bf
Appendix B: Tests of the Algorithms By Using Database of Catalogued Asteroids }

Many of the results presented in this work involve the asteroid heliocentric
distance determined as described in Appendix A. Here we determine
errors for the estimated values of heliocentric distance
for a sample of asteroids with known orbits.

\begin{figure}
\plotfiddle{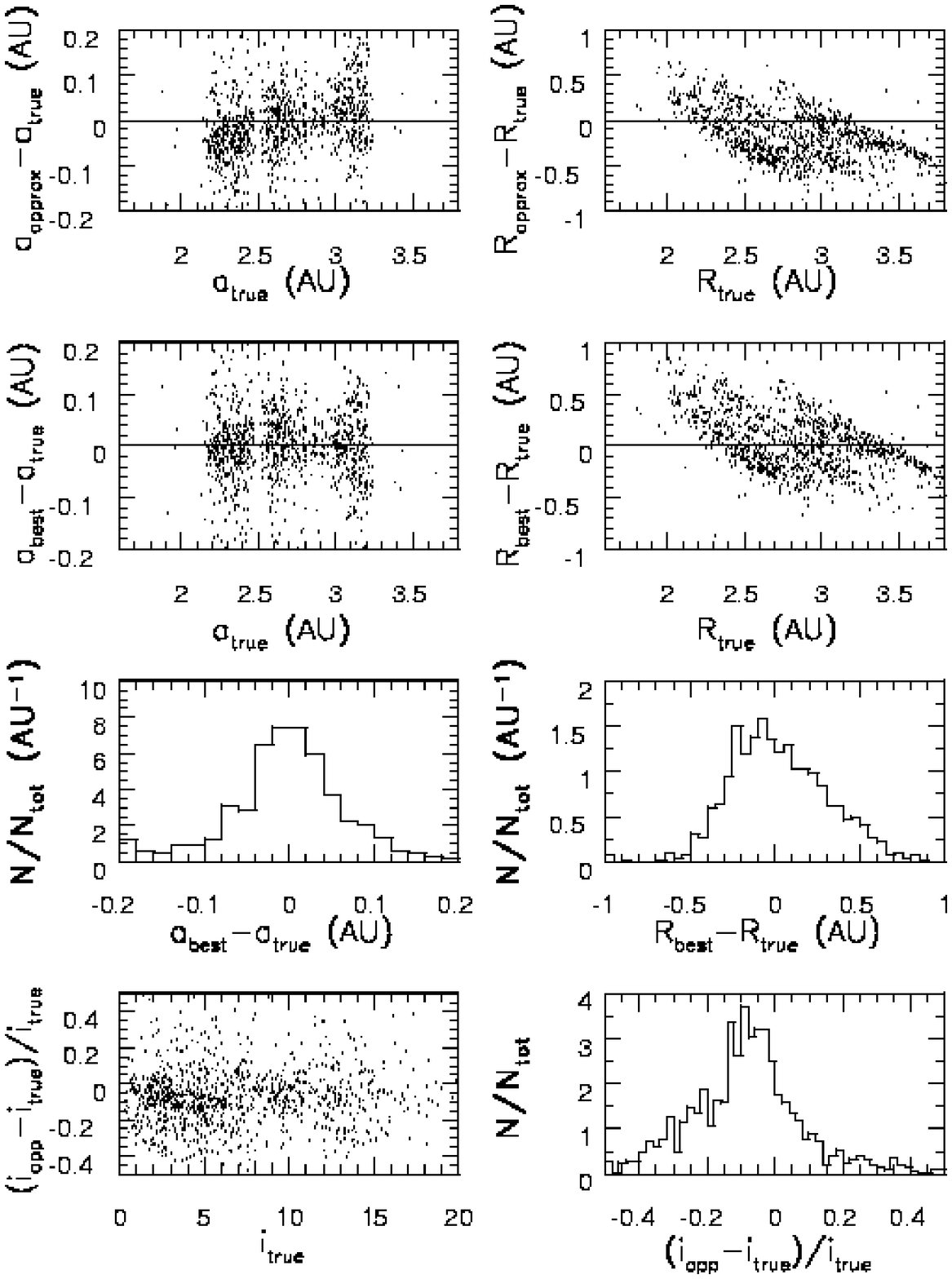}{12cm}{0}{70}{70}{-220}{-60}
\caption{The accuracy of orbital elements determined from proper motions.
For details see Appendix B.
\label{testR}
}
\end{figure}

We utilized the largest available database with orbital parameters for
114,539 asteroids provided by Bowell, Muinonen \& Wasserman (1994). The positions
of all asteroids were generated by a simple 2-body ephemeris generator for
September 21, 1998, and March 21, 1999, and 1729 asteroids were selected in
the observed area with $|\phi| < 15^\circ$. This number corresponds to 28\%
of the number density observed by SDSS in the same area and selected with the same
criteria. The asteroid proper motions were determined by computing the positions
288 seconds later and expressing the linear velocity in equatorial coordinates,
as was done with the SDSS observations. These simulated observations were then
processed by the same software as used in the analysis of the SDSS observations, producing
a list of estimated semi-major axes and heliocentric distances, $a$ and $R$ ($a = R$),
and orbital inclinations ($i$). Note that this procedure is an end-to-end test
of the analysis pipeline (i.e. test of all the steps {\em after} detecting asteroids
by the SDSS photometric pipeline). We determine the accuracy of the estimated values
by comparing them to the known true values. The results are summarized in
Figure \ref{testR}.

The top left panel shows the difference between the estimated and true
semi-major axis plotted vs. the true value, with each asteroid shown as a dot
(note that estimate for $a$ is the same as estimate for $R$).
The Kirkwood gaps are clearly visible in the distribution of semi-major axes.
The mean error is correlated with the semi-major axis and we correct for this
effect by fitting a straight line, producing the ``best" estimate
$a_{best} = a_{app} (1.085-0.03a_{app})$. The left panel in the second row
shows the difference between this new value and true semi-major axis plotted vs.
the true value, and the left panel in the third row shows the histogram of this
difference. Its equivalent Gaussian width is only 0.072 AU, showing that the
semi-major axis can be determined to within \about 3\% although the method
assumes circular orbits. This is the ``intrinsic" accuracy of the method based
on true proper motion. The errors in SDSS measurements will result in a wider
distribution and we model this effect by adding errors with random amplitude
of $\pm$5\% to the true velocity. The added errors increase the width of
$a_{best}-a_{true}$ distribution to 0.102 AU (the identical transformation from
the ``approximate" to ``best" value was used).

The assumption of a circular orbit results in identical estimates of semi-major
axis and heliocentric distance. The heliocentric distance is thus inevitably
underestimated due to the eccentric orbits of asteroids (the mean orbital value of $R/a$
is greater than unity for an eccentric Keplerian orbit). This effect is visible
in the upper right panel of Figure \ref{testR} which shows the difference between
the approximate and true heliocentric distance plotted vs. the true value,
where each asteroid is shown as a dot. We apply the same correction procedure
as for the major axis estimate, resulting in a best estimate of
$R_{best} = R_{app} (1.16 - 0.03R_{app})$ (which was used in the data analysis).
The right panel in the second row
shows the difference between this best value and the true value plotted vs.
the true value, and the right panel in the third row shows the histogram of this
difference. Its equivalent Gaussian width is 0.28 AU, about 4 times larger than
the width of the corresponding distribution for the semi-major axis estimate.

Although the inclination estimate is not used quantitatively here, we also
study its accuracy for completeness. The uncertainty of the inclination estimate
is 17\% with a bias of -7.8\% (note that these values refer to fractional errors).
The relevant diagrams are shown in the bottom row of Figure \ref{testR}.

It is somewhat counterintuitive that the semi-major axis can be determined
with about 4 times better accuracy than the heliocentric distance. After all,
the main component of the asteroid proper motion is due to the Earth's reflex
motion which depends only on the asteroid distance. To illustrate how this
apparent contradiction arises, it is sufficient to consider an asteroid observed
at opposition,  whose orbit has eccentricity $e$ and orbital inclination $i$=0$^\circ$.
In such a case, the observed angular velocity is
\eq{
\label{vL1}
                v_\lambda = { v_{ast} - v_E \over R - 1},
}
where
\eq{
                v_{ast} = {v_E \over R} \sqrt{a(1-e^2)}.
}
The semi-major axis and heliocentric distance are related by
\eq{
                R = {a(1-e^2) \over 1+ e \cos{\Psi}},
}
where $\Psi$ is the true anomaly, which specifies the orientation of the orbit 
relative to the observer. Two limiting cases are perihelion ($\Psi = 0^\circ$) and 
aphelion ($\Psi = 180^\circ$). For fixed $a$ and $e$, the observation in perihelion
corresponds to the smallest $R$ and the largest $v_{ast}$, while the opposite
is true for observation in aphelion. The changes of $a$ and $R$ have opposite
effects on $v_\lambda$ (note that $v_E > v_{ast}$), and this compensation results 
in a very minor dependence of $v_\lambda$ on unknown $\Psi$. We find that for the 
relevant range of parameters (2.2 $< a <$ 3.2, $e \la 0.2$) the change of $v_\lambda$ 
is $\la 3\%$ when $\Psi$ is varied from $0^\circ$ to 180$^\circ$. Or equivalently,
for a fixed (i.e. measured) $v_\lambda$, the uncertainty of $a$ is about 3\% due
to nonzero eccentricity (as found empirically). Note that even if the eccentricity
were known (and nonzero, of course), this uncertainty would not be smaller because
the problem is in unknown $\Psi$. Similarly, the larger uncertainty in $R$ is the
result of random orientations of eccentric asteroid orbits. Even if all asteroid
orbits had identical {\em known} eccentricity, the resulting uncertainty would be
similar because the heliocentric distance can be anywhere between
$R_{min}=(1-e)a$ and $R_{max}=(1+e)a$.

\newpage
\clearpage

\begin{scriptsize}
\begin{deluxetable}{ccccrrr}
\tablenum{1}
\tablecolumns{6}
\tablewidth{340pt}
\tablecaption{SDSS commissioning runs used in this study}
\tablehead
{
 Run$^a$  &  Date$^b$  & Min. R.A.$^c$ & Max. R.A.$^d$ & Min. $\beta^e$ & Max. $\beta^f$ & Area$^g$
}
\startdata
        94  & 1998-09-19 &   22$^h$ 25' &   03$^h$ 46' & -20   &   5  & 100.9    \\
       125  & 1998-09-25 &   23$^h$ 21' &   05$^h$ 07' & -20   &   5  &  97.5    \\
       752  & 1999-03-21 &   09$^h$ 40' &   16$^h$ 46' & -14   &  20  & 133.9    \\
       756  & 1999-03-22 &   07$^h$ 50' &   15$^h$ 44' & -14   &  20  & 149.3    \\
   1336$^h$ & 2000-04-04 &   16$^h$ 43' &   03$^h$ 29' &  73   &  86  &  16.2    \\
745-756$^i$ & 1999-03-20 &   10$^h$ 42' &   15$^h$ 46' &  -8   &  20  &  97.5    \\
\enddata
\tablenotetext{a}{Unique SDSS scan name.}
\tablenotetext{b}{The date of observation.}
\tablenotetext{c}{Starting right ascension.}
\tablenotetext{d}{Ending right ascension.}
\tablenotetext{e}{The minimum ecliptic latitude.}
\tablenotetext{f}{The maximum ecliptic latitude.}
\tablenotetext{g}{Total area (deg$^2$).}
\tablenotetext{h}{Not an equatorial run, $52^\circ < {\rm Dec} < 64^\circ$.}
\tablenotetext{i}{The overlap between the two runs (date given for run 745).}
\end{deluxetable}
\end{scriptsize}

\begin{scriptsize}
\begin{deluxetable}{cccrr}
\tablenum{2}
\tablecolumns{5}
\tablewidth{400pt}
\tablecaption{Classification regions in the $v_\lambda$ - $v_\beta$ diagram.}
\tablehead
{
  Region  &   velocity (\dd)  &  angle (deg)  &  N (\r$<$20.0)  & N (\r$<$21.5)
}
\startdata
         main belt &   0.14 -- 0.35  &  135 -- 225     &  1718  &  5125      \\
            Hildas &   0.14 -- 0.18  &  175 -- 185     &    11  &    59      \\
           Trojans &   0.07 -- 0.14  &  135 -- 225     &     1  &     2      \\
          Centaurs &   0.03 -- 0.07  &  135 -- 225     &     1  &     2$^a$  \\
         Hungarias &   0.35 -- 0.50  &   90 -- 135$^b$ &    13  &    25      \\
     Mars crossers &   0.35 -- 0.50  &  135 -- 150$^b$ &     0  &     7      \\
              NEOs &   0.35 -- 0.50  &  150 -- 210$^c$ &     8  &    19      \\
           Unknown &   0.35 -- 0.50  &   90 -- 135$^b$ &     2  &    14      \\
           Total   &        ---      &        ---      &  1754  &  5253      \\
\enddata
\tablenotetext{a}{Visual inspection of images shows that all are spurious.}
\tablenotetext{b}{Symmetric around $v_\lambda$ axis, see Figure \ref{vLambdavBeta}.}
\tablenotetext{c}{Also includes $v > 0.5$ and $v_\lambda > 0$.}
\end{deluxetable}
\end{scriptsize}

\begin{scriptsize}
\begin{deluxetable}{rrr}
\tablenum{3}
\tablecolumns{3}
\tablewidth{240pt}
\tablecaption{Parameters for broken power-law fits to main belt asteroid counts.}
\tablehead
{
   Parameter  &   $a^* < 0$  &   $a^* > 0$
}
\startdata
   $r_b$$^a$    &   18.0       &    18.5        \\
     k$_B^b$    &   0.69       &    0.52        \\
     k$_F^c$    &   0.34       &    0.27        \\
  $\alpha_B^d$  &   4.5        &    3.6         \\
  $\alpha_F^e$  &   2.7        &    2.4         \\
\enddata
\tablenotetext{a}{Magnitude for the power law break.}
\tablenotetext{b}{Coefficient k from log(N) = C + k \r\ at the bright end.}
\tablenotetext{c}{Coefficient k at the faint end.}
\tablenotetext{d}{The index $\alpha$ from $n(a) \propto a^{-\alpha}$ (=5*kB + 1)
 at the bright end.}
\tablenotetext{e}{The index $\alpha$ at the faint end.}
\end{deluxetable}
\end{scriptsize}

\begin{scriptsize}
\begin{deluxetable}{cccc}
\tablenum{4}
\tablecolumns{4}
\tablewidth{240pt}
\tablecaption{Parameters for fits to the cumulative distributions of absolute magnitudes.}
\tablehead
{
     Parameter$^a$  &   $a^* < 0$  &   $a^* > 0$ & all
}
\startdata
     $N_o$      &       44,300      &      74,000     &     111,600     \\
     $r_C$      &   15.5$\pm$0.01   &  15.6$\pm$0.01  & 15.5$\pm$0.01   \\
     k$_1$      &    0.61$\pm$0.01  &   0.61$\pm$0.01 &  0.61$\pm$0.01  \\
     k$_2$      &    0.28$\pm$0.01  &   0.24$\pm$0.01 &   0.25$\pm$0.01 \\
\enddata
\tablenotetext{a}{See eq. 12. The normalization is given for the entire
asteroid belt.}
\end{deluxetable}
\end{scriptsize}

\begin{scriptsize}
\begin{deluxetable}{rrrrrrrr}
\tablenum{5}
\tablecolumns{8}
\tablewidth{360pt}
\tablecaption{The flag statistics for 2MASS sources without optical matches.}
\tablehead
{
  Type$^a$  & rd\_flg & bl\_flg & cc\_flg & N$^b$ & \%$^c$ &  N$_{all}^d$ & \%$_{all}^e$
}
\startdata
  impeccable &  222  &  111   &   000  &   42 &   5.1  & 93050 &  80.86  \\
   J-only    &  200  &  100   &   000  &  461 &  56.0  &  2325 &   2.02  \\
   H-only    &  020  &  010   &   000  &  211 &  25.6  &   368 &   0.32  \\
   K-only    &  002  &  001   &   000  &   17 &   2.1  &    81 &   0.07  \\
  all other  & \dots & \dots  &  \dots &   93 &  11.2  & 19252 &  16.73  \\
\enddata
\tablenotetext{a}{The meaning of flags.}
\tablenotetext{b}{The number of unmatched 2MASS sources in the selected area (76.07 deg$^2$).}
\tablenotetext{c}{The fraction of total for unmatched sources.}
\tablenotetext{d}{The number of all 2MASS sources in the same area.}
\tablenotetext{e}{The fraction of total for all sources.}
\end{deluxetable}
\end{scriptsize}


\begin{scriptsize}
\begin{deluxetable}{rrr}
\tablenum{6}
\tablecolumns{3}
\tablewidth{200pt}
\tablecaption{Measured parameters for the KBO candidate.}
\tablehead
{
   Parameter  &   Run 745  &  Run 756
}
\startdata
         RA & $193.63749^\circ$ & $ 193.59462^\circ$ \\
        Dec & $-0.56884^\circ$  & $ -0.55040^\circ$  \\
 MJD-51,000 & 257.33775         & 259.33200          \\
  $\lambda$ & $192.77092^\circ$ & $192.72408^\circ$  \\
    $\beta$ & $-5.05787^\circ$  & $-5.05827^\circ$   \\
      u$^*$ &  $23.66 \pm 0.59$ & $23.49 \pm 0.54$   \\
      g$^*$ &  $22.47 \pm 0.11$ & $22.40 \pm 0.11$   \\
      r$^*$ &  $21.41 \pm 0.07$ & $ 21.42 \pm 0.06$  \\
      i$^*$ &  $20.94 \pm 0.07$ & $ 20.95 \pm 0.06$  \\
      z$^*$ &  $20.65 \pm 0.20$ & $ 20.90 \pm 0.21$  \\
g$^*$-r$^*$ &  $1.06$           & $0.98$             \\
r$^*$-i$^*$ &  $0.47$           & $0.47$             \\
      a$^*$ &  $0.60$           & $0.53$             \\
\enddata
\tablenotetext{a}{MJD = TAI/86400, where TAI is the number of seconds
since Nov 17 1858 0 (TAI-UT $\approx$ 30 sec.).}
\end{deluxetable}
\end{scriptsize}

\begin{scriptsize}
\begin{deluxetable}{ll}
\tablenum{7}
\tablecolumns{2}
\tablewidth{240pt}
\tablecaption{Derived parameters for the KBO candidate.}
\tablehead
{
   Parameter  &  Value
}
\startdata
   $v_\lambda$ &   -0.0235 \dd \\
   $v_\beta$   &    0.0002 \dd \\
distance       &  36-49    AU \\
radius for:    &              \\
albedo=0.04    &  210-390  km \\
albedo=0.25    &   80-150  km \\
\enddata
\end{deluxetable}
\end{scriptsize}

\newpage

\clearpage

\clearpage

\clearpage

\begin{thebibliography}{99}


\bibitem[Allen 1973]{Allen1973} Allen, C.W. 1973, Astrophysical Quantities
             (London: Athlone Press, 1973)

\bibitem[Alvarez {\em et al.} 1980]{Alvarez1980} Alvarez, L.W., Alvarez, W.,
         Asaro, F. \& Michel, H.V. 1980, Science, 208, 1095  

\bibitem[Bell 1989]{Bell89} Bell, J.F., Davis, D.R., Hartmann, W.K. \&
         Gaffey, M.J. 1989, in Asteroids II, eds. R.P. Binzel,
         T. Gehrels and M.S. Matthews, (Tucson: Univ. of Arizona Press), 921

\bibitem[Binzel 1989]{Binzel89} Binzel, R.P. 1989, in Asteroids II, eds. R.P. Binzel,
         T. Gehrels and M.S. Matthews, (Tucson: Univ. of Arizona Press), 3

\bibitem[Bowell \& Lumme 1979]{BL79} Bowell, E. \& Lumme, K. 1979, in Asteroids,
         ed. T. Gehrels, (Tuscon: Univ. of Arizona Press), 132

\bibitem[Bowell, Skiff, Wasserman \& Russell 1989]{BSWR89} Bowell, E.,
         Skiff, B.A., Wasserman, H. \& Russell K.S, 1989, in Asteroids, Comets,
         and Meteors III, eds. C.I. Lagerkvist, H. Rickman, B.A. Lindblad, and
         M. Lindgren (Uppsala University, Uppsala), 19

\bibitem[Bowell, Muinonen \& Wasserman 1994]{BMW94} Bowell, E., Muinonen, K.,
         \& Wasserman, H. 1994, A public-domain asteroid orbit database, In
         Asteroids, Comets, Meteors 1993, eds. Milani, A., {\em et al.},
         (Dordrecht: Kluwer Academic), 477

\bibitem[Cellino {\em et al.} 1991]{C91} Cellino, A., Zappala, V., Farinella, P.
          1991, MNRAS, 253, 561

\bibitem[Ceplecha {\em et al.} 1998]{C98} Ceplecha, Z., {\em et al.} 1998,
         Space Science Reviews, 84, 327.

\bibitem[Chapman, Morrison \& Zellner 1975]{CMZ75} Chapman, C.R., Morrison, D. \&
         Zellner, B. 1975, Icarus, 25, 104

\bibitem[Dermott, Gradie \& Murray 1985]{DGM85} Dermott, S.F., Gradie, J., \&
         Murray, C.D. 1985, Icarus 62, 289

\bibitem[Dohnanyi 1969]{D69} Dohnanyi, J. 1969, J. Geophys. Res., 74, 2531

\bibitem[Duncan, Quinn \& Tremaine 1988]{DQT88} Duncan, M., Quinn, T., \&
         Tremaine, S. 1988, ApJ, 328, L69

\bibitem[Durda \& Dermott 1997]{DD97} Durda, D.D., \& Dermott, S.F. 1997,
         Icarus, 130, 140

\bibitem[Durda, Greenberg \& Jedicke 1998]{DGJ98} Durda, D.D., Greenberg, R.,
         \& Jedicke, R. 1998, Icarus, 135, 431

\bibitem[Efron \& Petrosian 1992]{EP92} Efron, B. \& Petrosian, V. 1992, ApJ,
         399, 345

\bibitem[Fan 1999]{Fan99}Fan, X. 1999, AJ, 117, 2528

\bibitem[Fan {\em et al.} 2001a]{F01a}Fan, X., {\em et al.} 2001a, AJ, 121, 31

\bibitem[Fan {\em et al.} 2001b]{F01b}Fan, X., {\em et al.} 2001b, AJ, 121, 54

\bibitem[Fern\'andez 1980]{F80} Fern\'andez, J. 1980, MNRAS, 192, 481

\bibitem[Finlator 2000]{F2000} Finlator, K., {\em et al.} 2000, AJ, 120, 2615

\bibitem[Fukugita {\em et al.} 1996]{F96}Fukugita, M., Ichikawa, T., Gunn, J.E.,
         Doi, M., Shimasaku, K., \& Schneider, D.P. 1996, AJ, 111, 1748


\bibitem[Gaffey, Bell \& Cruikshank 1989]{GBK89} Gaffey, Bell \& Cruikshank 1989,
         in Asteroids II, eds. R.P. Binzel, T. Gehrels and M.S. Matthews, (Tucson: Univ.
         of Arizona Press), 117

\bibitem[Gehrels 1979]{Gehrels79} Gehrels, T. 1979, in Asteroids, ed. T. Gehrels,
         (Tuscon: Univ. of Arizona Press), 3

\bibitem[Gradie \& Tedesco 1982]{GT82} Gradie, J., \& Tedesco, E.F. 1982, Science, 216, 1405

\bibitem[Gradie, Chapman \& Tedesco 1988]{GCT89} Gradie, J., Chapman, C.R., \& Tedesco, E.F.
         1989, in Asteroids II, eds. R.P. Binzel, T. Gehrels and M.S. Matthews, (Tucson: Univ.
         of Arizona Press), 316

\bibitem[Gradie, Chapman \& Williams 1988]{GCTW79} Gradie, J., Chapman, C.R., \& Williams, J.G.
         1979, in Asteroids, ed. T. Gehrels, (Tuscon: Univ. of Arizona Press), 359

\bibitem[Gunn {\em et al.} 1998]{Gunnetal} Gunn, J.E., {\em et al.} 1998,
         AJ, 116, 3040

\bibitem[Hanner \& Cruikshank 1999]{HC99} Hanner, M.S. \& Cruikshank, D.P. 1999,
         The Solar System and Circumstellar Dust Disks: Prospects for SIRTF, Report of
         a Workshop, August 18-20, 1999, Dana Point, California, available as
         http://sirtf.caltech.edu/SciUser/A\_GenInfo/SSC\_A4\_DANA1999.html

\bibitem[Hapke 1989]{Hapke89} Hapke, B., Dominigue, D., Lumme, K., Peltoniemi, J. \&
         Harris, A.W. 1989, in Asteroids II, eds. R.P. Binzel,
         T. Gehrels and M.S. Matthews, (Tucson: Univ. of Arizona Press), 524

\bibitem[Ivezi\'c \& Elitzur 2000]{IE00} Ivezi\'c, \v Z., \& Elitzur, M. 2000,
         ApJ, 534, 93.

\bibitem[Ivezi\'c {\em et al.} 1999]{I00} Ivezi\'c, \v Z., {\em et al.}
         2000, AJ, 120, 963

\bibitem[Ivezi\'c {\em et al.} 2001]{I01} Ivezi\'c, \v Z., {\em et al.}
         2001, in prep.

\bibitem[Jedicke 1996]{Jedicke96}Jedicke, R. 1996, AJ, 111, 970

\bibitem[Jedicke \& Metcalfe 1998]{JM98}Jedicke, R., \& Metcalfe, T.S. 1998,
         Icarus 131, 245

\bibitem[Jewitt 1999]{Jewitt99} Jewitt, D. 1999, Annu. Rev. Earth. Planet. Sci, 27, 287

\bibitem[Jewitt \& Luu 1992]{JL92} Jewitt, D., \& Luu, J. 1992, Nature, 362, 730

\bibitem[Kowal 1989]{Kowal89} Kowal, C. 1989, Icarus, 77, 118

\bibitem[Krisciunas, Margon \& Szkody 1998]{KMS98} Krisciunas, K., Margon,
             B., \&  Szkody P. 1998, PASP 110, 1342

\bibitem[Kuiper et al. 1958]{K58} Kuiper G.P., {\em et al.} 1958, ApJS 3, 289.

\bibitem[Levison \& Duncan 1990]{LD90} Levison, H., \& Duncan, M. 1990, AJ, 100, 1669

\bibitem[Lupton, Gunn \& Szalay 1999]{LGS99} Lupton, R.H., Gunn, J.E., \&
             Szalay, A. 1999,  AJ, 118, 1406

\bibitem[Lupton {\em et al.} 2001]{Lupton01} Lupton, R.H., {\em et al.} 2001,
         in preparation

\bibitem[Luu \& Jewitt 1996]{LuuJewit96} Luu, J., \& Jewitt, D. 1996, AJ, 112, 2310

\bibitem[Lynden-Bell 1971]{LyndenBell71} Lynden-Bell, D. 1971, MNRAS, 155, 95

\bibitem[Muinonen, Bowell \& Lumme 1995]{MBL95} Muinonen, K., Bowell, E. \& Lumme, K.
        1995, A\&A, 293, 948

\bibitem[Oke \& Gunn 1983]{OG83} Oke, J.B., \& Gunn, J.E. 1983, ApJ 266, 713

\bibitem[Pier {\em et al.} 2000]{Pier01} Pier, J.R., {\em et al.} 2001, in
         preparation

\bibitem[Press {\em et al.} 1992]{Press92} Press, W.H., Teukolsky, S.A., Vetterling, W.T.,
         \& Flannery, B.P. 1992, Numerical Recipes in C, Second Edition, Cambridge
         University Press

\bibitem[Rabinowitz 1991]{Rabin91} Rabinowitz, D.L. 1991, AJ, 101, 1518

\bibitem[Rabinowitz {\em et al.} 2000]{Rabin00} Rabinowitz, D.L., Helin, E.,
         Lawrence, K., \& Pravdo, S. 2000, Nature, 403, 165

\bibitem[Richards {\em et al.} 2001]{R01} Richards, G.T., {\em et al.} 2001, in
         preparation

\bibitem[Ruzmaikina 1989]{Ruzma89} Ruzmaikina, T.V., Safronov, V.S., Weidenschilling,
         S.J. 1989, in Asteroids II, eds. R.P. Binzel, T. Gehrels and M.S. Matthews,
         (Tucson: Univ. of Arizona Press), 681

\bibitem[Scotti, Gehrels \& Rabinowitz 1991]{SGR91} Scotti, J.V., Gehrels, T.,
     \& Rabinowitz, D.L. 1991, in Asteroids, Comets, Meteors 1991, Lunar and
     Planetary Institute, Houston, 1992, p. 541

\bibitem[Shoemaker {\em et al.} 1979]{Setal79} Shoemaker, E., Williams, J.G.,
         Helin, E.F., \& Wolfe, R.F. 1979, in Asteroids,
         ed. T. Gehrels, (Tuscon: Univ. of Arizona Press), 253

\bibitem[Shoemaker \& Wolfe 1982]{SW82} Shoemaker, E., \& Wolfe, R. 1982, in
            Satellites of Jupiter (ed. Marrison, D.), 277

\bibitem[Skrutskie {\em et al.} 1997]{Skrutskie97} Skrutskie, M.F., {\em et al.} 1997,
        The Impact of Large-Scale Near-IR Sky Surveys, ed. F. Garzon et al. (Dordrecht:
        Kluwer), 25

\bibitem[Steel 1997]{Steel97} Steel, D. 1997, Irish Astronomical Journal, 24, 19

\bibitem[Sykes {\em et al.} 2000]{Sy2000} Sykes, M.V., {\em et al.} 2000, Icarus, 146, 161

\bibitem[Tedesco {\em et al.} 1989]{Tedesco89} Tedesco, E.F., Williams, J.G., Matson,
         D.L., Veeder, G.J., Gradie, J.C., \& Lebofsky, L.A. 1989, AJ, 97, 580

\bibitem[Tegler \& Romanishin 1998]{TR98} Tegler, S.C. \& Romanishin, W. 1998, Nature, 392, 49

\bibitem[Tholen \& Barucci 1989]{TB89} Tholen, D.J., \& Barucci, M.A. 1989, in Asteroids II,
         eds. R.P. Binzel, T. Gehrels and M.S. Matthews, (Tucson: Univ. of Arizona Press), 298

\bibitem[Tombaugh 1961]{Tombaugh61}Tombaugh, C.W. 1961, in Planets and Satellites,
         eds. G.P. Kuiper and B.M. Middlehurst (Chicago: University of Chicago Press), 12

\bibitem[Tombaugh 1996]{Tombaugh96}Tombaugh, C.W. 1996, in Completing the Inventory of
         the Solar System, eds. T.W. Rettig and J.M. Hahn (San Francisco: Astronomical Society
         of the Pacific), 157

\bibitem[Tremaine 1990]{T90} Tremaine, S. 1990, in Baryonic Dark Matter, (eds. D. Lynden-Bell
         and G. Gilmore) 37

\bibitem[van den Bergh 1994]{vdBergh1994} van den Bergh, S. 1994, PASP, 106, 689

\bibitem[van Houten {\em et  al.} 1970]{vanHouten1970} van Houten, C.J., {\em et al.}
         1970, A\&AS, 2, 339

\bibitem[Verschuur 1991]{Verschuur91} Verschuur, G.L. 1991, Air and Space, vol. 6, 
         Oct--Nov 1991, p. 88-94

\bibitem[Williams \& Wetherill 1994]{WW94} Williams, D.R. \& Wetherill, G.W. 1994,
          Icarus, 107, 117

\bibitem[Wisdom 1985]{Wisdom} Wisdom, J. 1986, Nature, 315, 731

\bibitem[Xu {\em et al.} 1995]{Xu95} Xu, S., Binzel, R.P., Burbine, T.H. \& Bus,
         S.J. 1995, Icarus, 115, 1

\bibitem[York  {\em et al.} 2000]{York} York, D.G., {\em et al.} 2000,
         AJ, 120, 1579

\bibitem[Zappala \& Cellino 1996]{ZC96} Zappala, V. \& Cellino, A. 1996, In Completing
        the Inventory of the Solar System, (eds. T.W. Rettig and J.M. Hahn), ASP
        Conference Series 107, 29

\bibitem[Zappala {\em et al.} 1990]{Zetal90} Zappala, V., Cellino, A., Farinella,
        P. \& Kne\v{z}evi\'{c}, Z. 1990, AJ, 100, 2030

\bibitem[Zellner 1979]{Zellner79} Zellner, B. 1979, in Asteroids, ed. T. Gehrels,
         (Tuscon: Univ. of Arizona Press), 783

\bibitem[Zellner {\em et al.} 1985]{Zellner85} Zellner, B., Tholen, D.J., \& Tedesco,
         E.F. 1985, Icarus, 61, 355

\bibitem[Zellner {\em et al.} 1985b]{Zellner85b} Zellner, B., Thirunagari, A., \&
         Bender, D. 1985, Icarus, 62, 505

\end{thebibliography}
\end{document}